\documentclass[%
reprint,
 amsmath,amssymb,
 aps,
prb,
floatfix,
]{revtex4-2}

\usepackage{multirow}
\usepackage{graphicx}
\usepackage{dcolumn}
\usepackage{bm}

\begin{document}

\title{Strictly localized states in the octagonal Ammann-Beenker quasicrystal}

\author{M.\"O. Oktel}
\email{oktel@bilkent.edu.tr}
\affiliation{Department of Physics, Bilkent University, Ankara, 06800, TURKEY}

\date{\today}

\begin{abstract}
Ammann-Beenker lattice is a two-dimensional quasicrystal with eight-fold symmetry, which can be described as a projection of a cut from a four-dimensional simple cubic lattice. We consider the vertex tight-binding model on this lattice and investigate the strictly localized states at the center of the spectrum. We use a numerical method based on the generation of finite lattices around a given perpendicular space point and QR decomposition of the Hamiltonian to count the strictly localized states. We apply this method to count the frequency of localized states in lattices of up to 100 000 sites. We obtain an orthogonal set of compact localized states by diagonalizing the position operator projected onto the manifold spanned by the zero energy states. We identify twenty localized state types and calculate their exact frequencies through their perpendicular space images. Unlike the Penrose lattice, all the localized state types are eight-fold symmetric around an eight edge vertex, and all vertex types can support localized states. The total frequency of these twenty types gives a lower bound of $f_{LS}=30796-21776\sqrt{2} \simeq 0.08547$ for the fraction of strictly localized states in the spectrum. This value is in agreement with the numerical calculation and very close to the recently conjectured exact fraction of localized states $f_{Ex}=3/2-\sqrt{2}\simeq 0.08579$. 
\end{abstract}

\maketitle

\section{Introduction}

The theory of elementary excitations in crystalline solids relies on the periodicity of the lattice through Bloch's theorem. Lattice periodicity is combined with the point group symmetries of the crystal, and the reciprocal space description of the 
 system reflects the orientational order in the system. The discovery of quasicrystals that have sharp X-ray spectra \cite{she84} with rotational symmetries impossible for a periodic system showed the link between orientational order and periodicity can be severed even in the absence of randomness. Alloys which have this non-periodic but rotationally symmetric quasicrystalline order can be routinely synthesised \cite{jan97}, and smaller-scale quasiperiodic order is recently engineered in synthetic surfaces\cite{col17}, photonic \cite{var13}, polaritonic \cite{tan14} and cold atom systems\cite{vie19,sin15}.

The structural properties of quasicrystals are generally well understood\cite{ste18}. However, there is no counterpart to Bloch's theorem, and the structural description does not easily translate into a theory of elementary excitations. Major questions, such as the localization properties of the eigenstates or the measure of the density of states, remain unanswered even for the simplified models. The progress in one-dimensional models such as the Fibonacci chain \cite{koh83,ost83,koh87} have shown that the physics of elementary excitations in quasicrystals is extremely rich. The density of states has multifractal properties \cite{pie95}, and even the basic definitions of localization have to be reexamined due to the neither localized nor extended eigenstates \cite{sut86}. Most of the results obtained in one dimension rely on techniques such as transfer matrix methods which cannot be applied in higher dimensions. Two and three-dimensional results on elementary excitation spectra have been mostly restricted to numerical calculations. In some cases, particular eigenstates have been identified\cite{sut86,koh86,ara88,kka14}, but a general description and labeling of eigenstates has not been found. Understanding the properties of even a subset of the spectrum is valuable in the absence of a general description.    

Early numerical studies on the tight-binding model\cite{oda86,cho85} on the two-dimensional quasicrystal Penrose lattice have identified an abundance of eigenstates at the center of the spectrum.  Kohmoto and Sutherland \cite{koh86}  showed that there is a set of degenerate states exactly at zero energy, giving a delta function peak in the density of states. These states have non-zero density only on a finite number of lattice sites and are called strictly localized states. Six independent types of these states were identified for the Penrose lattice vertex model \cite{ara88}.  Similar states have been identified in the presence of gauge fields, called Aharonov-Bohm cages\cite{vid98}, and in some flat band models with non-trivial topology \cite{laf18}. These states lie strictly at zero energy for bipartite lattices \cite{sutbip86} and would be at the Fermi energy for the critically important case of half-filling \cite{kts17,kog20,kog21}. For a periodic crystal, such strictly localized states can exist only if confined to a unit cell.  It is not clear if strictly localized states are a general feature of quasicrystalline models or appear only for a specific subset. We refer to strictly localized states as localized states (LS) throughout, which should not be confused with the exponentially decaying localized states.

In a recent paper \cite{mok20}, we used the perpendicular space description of the Penrose lattice to investigate the LS identified in Ref.\cite{ara88}.  This method provides an easy way to label and count the LS once their real space structure is known. While the first identification of these states relied on the presence of `closed strings' of three edge vertices in the Penrose Lattice, subsequent work \cite{rsc95} has shown that these states exist even in local isomorphism classes without closed strings.  The presence of LS for another quasicrystal lattice, the octagonal Ammann-Beenker lattice (ABL), has also been reported \cite{jpi06}. In a recent paper \cite{kog20} Koga has identified 12 types of LS in the ABL and has conjectured that an infinite sequence of LS types with decreasing frequencies exist in this system. The total frequency of LS is calculated as $f_{Ex}=\frac{(\sqrt{2}-1)^2}{2}=3/2-\sqrt{2}$ Based on extrapolation from exact diagonalization of finite lattices obtained by deflation.   

We have two aims in this paper. First, we outline a numerical method to identify and count LS. Counting is based on identifying the null space of the Hamiltonian, which can be efficiently made through QR decomposition \cite{glo89}. We have been able to count LS frequencies for lattices containing up to 100 000 sites. Diagonalization of the position operator projected onto the null space of the  Hamiltonian provides an easy method to form compact LS. Second, we use this method to identify 20 independent types of LS for the ABL and count their frequencies through perpendicular space projections. This counting gives us an exact lower bound for the frequency of LS  $f_{LS}=30796-21776\sqrt{2} \simeq 8.547 \% .$ Our numerical results for the largest lattice sizes are close to this analytical result, and both results are in agreement with Ref.\cite{kog20} exact frequency.   

In the next section, we introduce the ABL, its perpendicular space image, and define the vertex tight-binding model on it. Section \ref{sec:Numerical} introduces the numerical methods used to generate the ABL and count the LS. Numerical calculations for both the Penrose lattice and ABL are given in this section, while the next Section \ref{sec:LS} contains the twenty independent types of LS for the ABL and their properties. A comparison of our results for the two lattices as well as general properties of the LS concludes the paper in Section \ref{sec:Conclusion}.

\section{Ammann-Beenker Lattice and the vertex model}
\label{sec:ABL}

Quasicrystals can be described as projections of sections of higher dimensional periodic lattices. While alternative methods exist, cut and project construction retains information regarding the projected out dimensions \cite{sos87}. The projected out dimensions form the so-called perpendicular space of the quasicrystal, which has been instrumental in understanding the x-ray spectra\cite{els85,kal85} as well as the elementary excitation properties \cite{mjp16,sja08,kka14,mjk17}. Penrose lattice, the most widely studied quasicrystal model, can be obtained as the projection of a five-dimensional cubic lattice to two dimensions \cite{bru81}. With this projection, the perpendicular space is three-dimensional; however, it reduces to four parallel two-dimensional pentagons rather than filling a volume \cite{hen86}. In a recent paper\cite{mok20},  we used the images of LS in these four pentagons to count and label the LS in the Penrose lattice.

In this paper, we concentrate on the Ammann-Beenker lattice (ABL), a quasicrystal with eight-fold rotational symmetry \cite{amm92,bee82}. Two tiles are used to construct the ABL, a square and a rhombus with a $\pi/4$ angle. Following Beenker, we define the ABL by projection from a four-dimensional cubic lattice. Let $\hat{u}_i$ with $i=0,1,2,3$ define mutually orthogonal unit vectors, $\hat{u}_i \cdot \hat{u}_j =\delta_{i,j}$ spanning the four-dimensional space
\begin{equation}
    \label{eq:r4}
    \vec{r}_4=x_0 \hat{u}_0 +x_1 \hat{u}_1 +x_2 \hat{u}_2 +x_3 \hat{u}_3,
\end{equation}
with $x_i \in \mathbb{R}$. This space can be partitioned into 4-cubes $k_i-1<x_i<k_i$ with $k_i \in \mathbb{Z}$, and each cube is associated with a point in the four dimensional cubic lattice
\begin{equation}
\label{eq:kvec}
    \vec{k}=k_0 \hat{u}_0 + k_1 \hat{u}_1 + k_2 \hat{u}_2+ k_3 \hat{u}_3.
\end{equation}

An alternative set of vectors which span the same space can be given.  Defining $\eta=e^{i \frac{\pi}{4}}$,  and
\begin{eqnarray}
\label{eq:CVectors}
\vec{c}_1 &=& \sum_{m=0}^{3} \Re \left(\eta^m \right) \hat{u}_m = \hat{u}_0 + \frac{1}{\sqrt{2}} \hat{u}_1 - \frac{1}{\sqrt{2}} \hat{u}_3, \\ \nonumber
\vec{c}_2 &=&\sum_{m=0}^{3} \Im \left(\eta^m \right) \hat{u}_m = \hat{u}_2 + \frac{1}{\sqrt{2}} \hat{u}_1 + \frac{1}{\sqrt{2}} \hat{u}_3, \\ \nonumber
\vec{c}_3 &=&\sum_{m=0}^{3} \Re \left((-\eta)^m \right) \hat{u}_m = \hat{u}_0 - \frac{1}{\sqrt{2}} \hat{u}_1 + \frac{1}{\sqrt{2}} \hat{u}_3, \\ \nonumber
\vec{c}_4 &=&\sum_{m=0}^{3} \Im \left((-\eta)^m \right) \hat{u}_m = \hat{u}_2 - \frac{1}{\sqrt{2}} \hat{u}_1 - \frac{1}{\sqrt{2}} \hat{u}_3. 
\end{eqnarray}
This set is mutually orthogonal yet not normalized, $\vec{c}_i \cdot \vec{c}_j=2 \delta_{i,j}.$ 

Only a subset of the points in the four dimensional cubic lattice are projected to two dimensions. We define a real vector $\vec{\gamma}$ and a plane with the equations
\begin{eqnarray}
\label{eq:CutPlane}
\left(\vec{r}_4-\vec{\gamma} \right) \cdot \vec{c}_3=0, \\ \nonumber
\left(\vec{r}_4-\vec{\gamma} \right) \cdot \vec{c}_4=0.
\end{eqnarray}
If the open cube corresponding to 4 dimensional lattice point $\vec{k}$ has an intersection with this plane, then we express it in the new basis:
\begin{equation}
    \vec{k}=\frac{x_R}{2} \vec{c}_1 + \frac{y_R}{2} \vec{c}_2 + \frac{x_\perp- \vec{\gamma}\cdot\vec{c_3}}{2} \vec{c}_3 +\frac{y_\perp- \vec{\gamma}\cdot\vec{c_4}}{2} \vec{c}_4.
\end{equation}
The expansion coefficients $x_R,y_R$ define the real space position of a point belonging to the ABL, forming the projection part of the cut-and-project construction. We can define two orthogonal unit vectors $\hat{i},\hat{j}$ spanning the real space, and the projected real space point is at the coordinates
\begin{equation}
    \vec{r}_R=x_R \hat{i} + y_R \hat{j} = \sum_{m=0}^{3} k_m \hat{e}_m,
\end{equation}
where the four star vectors of the ABL are:
\begin{equation}
\label{eq:StarVec}
    \hat{e}_0=\hat{i},\; \hat{e}_1=\frac{1}{\sqrt{2}} (\hat{i}+\hat{j}),\; \hat{e}_2=\hat{j},\; \hat{e}_3=\frac{1}{\sqrt{2}} (-\hat{i}+\hat{j}). 
\end{equation}
In real space two points of the ABL are connected by $\pm \hat{e}_i$, thus all bonds lie along one of the star vectors. Similarly the star vectors can be viewed as the projections of the four dimensional unit vectors $\hat{u}_i$ onto the real space.

It is also useful to consider the projected out coordinates of the point $\vec{k}$ which satisfies the condition \ref{eq:CutPlane}. This process defines the perpendicular space projection $\vec{R}_\perp= x_\perp \hat{i}_\perp+ y_\perp \hat{j}_\perp$ where
\begin{equation}
    x_\perp=\left(\vec{k}+\vec{\gamma}\right) \cdot \vec{c}_3, \;  y_\perp=\left(\vec{k}+\vec{\gamma}\right) \cdot \vec{c}_4,
\end{equation}
and
$\hat{i}_\perp,\hat{j}_\perp$ are two orthogonal unit vectors spanning the perpendicular space. Thus any point belonging to the ABL has both a real space position, and a perpendicular space position, corresponding to projection of the cut from the four dimensional lattice into two orthogonal planes. 

The condition that the open cube corresponding to $\vec{k}$ has an intersection with the plane defined by Eq.\ref{eq:CutPlane} constrains the perpendicular space position to lie in an octagon of side length one. As the projection mapping is linear and $\sqrt{2}$ is irrational, the octagon is densely and uniformly filled with projected points. While all points belonging to ABL have a perpendicular space image inside this octagon, not all points inside the octagon correspond to a point in the ABL. We also require $\vec{\gamma}$ to be chosen so that the intersection with the open cubes and the plane is unambiguous {\it, i.e.}, the lattice is not singular \cite{bee82}.

\begin{figure}[!htb]
    \centering
    \includegraphics[trim=8mm 8mm 8mm 8mm,clip,width=0.48\textwidth]{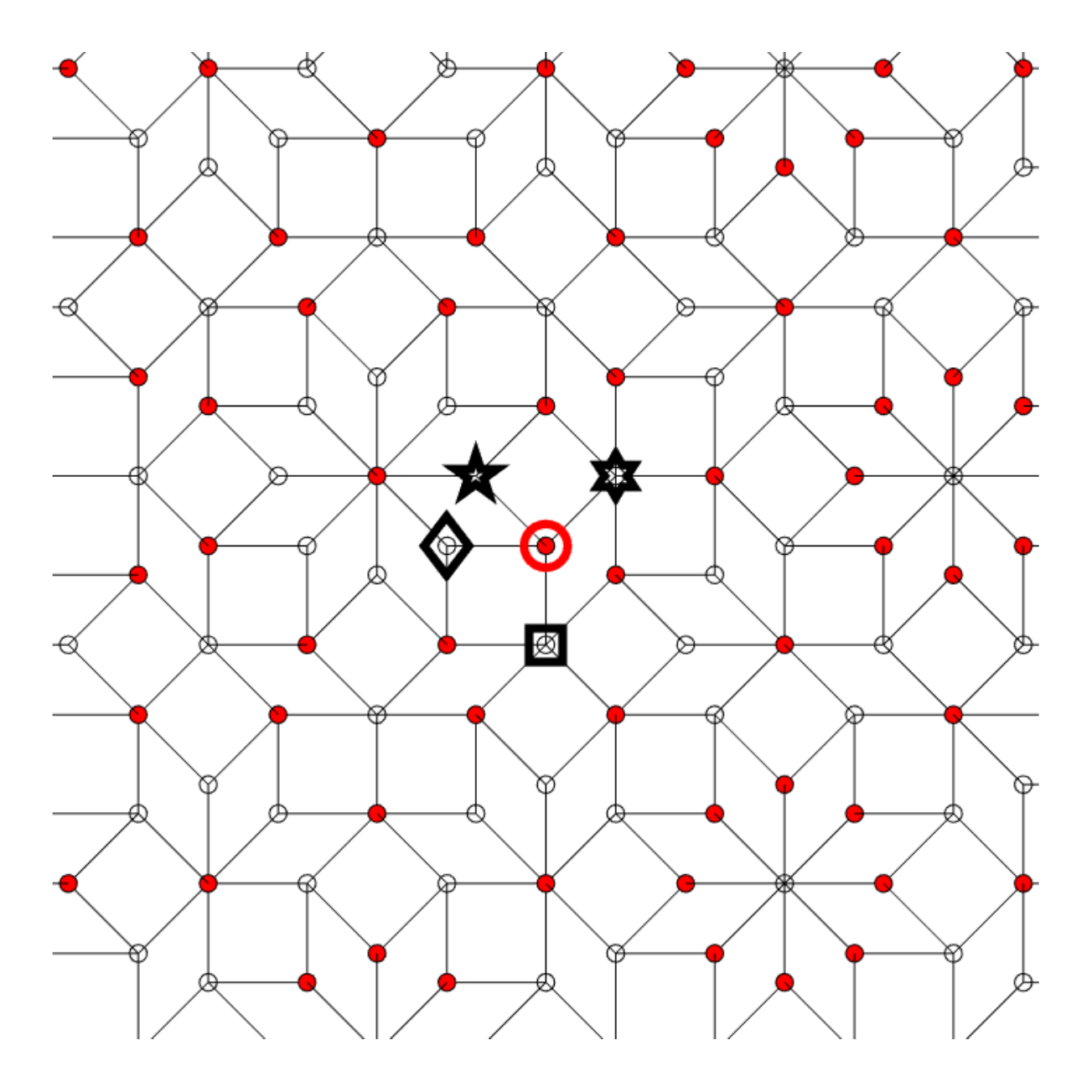}
    \caption{A finite section of the ABL. All bonds are parallel to the four star vectors in Eq.\ref{eq:StarVec}. The perpendicular space images of the five marked points are given in Fig. \ref{fig:ABL_PerpSpace}}
    \label{fig:ABL_RealSpace}
\end{figure}

\begin{figure}[!htb]
    \centering
    \includegraphics[clip,width=0.48\textwidth]{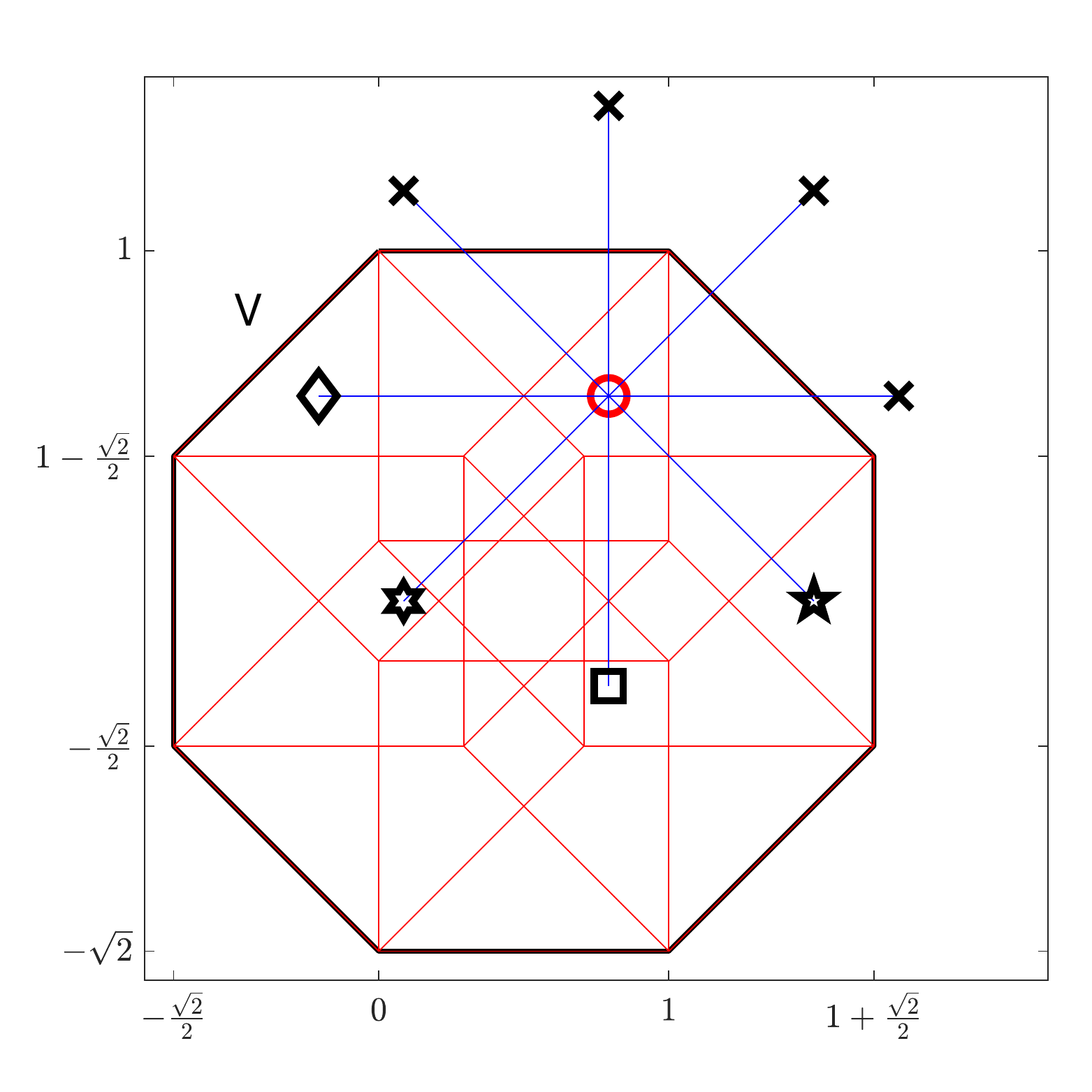}
    \caption{ Perpendicular space images of all the points in ABL lie within the octagon V shown in the figure. The regions inside $V$ separate perpendicular space images of vertices with a different number of edges.   The perpendicular space images of the four points marked in Fig.\ref{fig:ABL_RealSpace} are shown. Notice that vertex marked with a circle has only four nearest neighbors in real space because only four out of the eight points reached by star vectors from its perpendicular space image lie in $V$.   }
    \label{fig:ABL_PerpSpace}
\end{figure}

A finite section of the ABL is displayed in Fig.\ref{fig:ABL_RealSpace}, and the corresponding perpendicular space image is in Fig.\ref{fig:ABL_PerpSpace}. A point in the ABL can have between 3 and 8 nearest neighbors, all of them connected along the star vector directions. Consider the point denoted by the (red) circle in both figures. This point can have at most eight neighbors in the directions $\pm \hat{e}_m$. A translation by $\hat{e}_m$ in real space corresponds to a translation by $\hat{u}_m$ in the four dimensional cubic lattice. The definitions of the $\vec{c}$ basis Eq(\ref{eq:CVectors}) show that a translation by $\hat{u}_m$ corresponds to translation by $(-1)^m \hat{e}_m$ in perpendicular space. When we consider the eight possible points reached from the perpendicular space position of the point denoted by the (red) circle perpendicular, we see that only four of them lie in the octagon $V$. Hence this point has only four nearest neighbors. The smaller sections within $V$ identify regions for vertices with a different number of edges, and following Beenker \cite{bee82} we call a point with $n$ edges a $T(9-n)$ vertex. For example, $T1$ vertices with eight neighbors have perpendicular space images that lie within the small octagon at the center of $V$.   

The mapping between the four-dimensional cubic lattice and the perpendicular space is linear, and the fact that $\sqrt{2}$ is irrational guarantees that no two points in the ABL have the same perpendicular space image. Hence, areas in perpendicular space can be used to measure frequencies of vertices in the ABL. For example, $T1$ vertices with 8 edges have perpendicular space images inside an octagon of width $\sqrt{2}-1$ at the center of $V$, which itself is an octagon of width $\sqrt{2}+1$. The frequency of $T1$ vertices is then
\begin{equation}
    f_{T1}=\frac{\left(\sqrt{2}-1\right)^2}{\left(\sqrt{2}+1\right)^2}=17-12\sqrt{2} \simeq 0.02943.
\end{equation}

We now define a tight binding hopping Hamiltonian on the ABL, using a single state at each vertex and uniform hopping to nearest neighbor sites,
\begin{equation}
\label{eq:Hamiltonian}
    {\cal H}=-\sum_{<i,j>} |\vec{R}_i\rangle \langle \vec{R}_j |,
\end{equation}
with $|\vec{R}_i\rangle$ denoting the Wannier wavefunction which is localized at the ABL point $\vec{R}_{i}$ and the sum is carried out over all edges in the ABL.  

All the tiles forming the ABL are quadrilaterals; hence, the ABL vertices can be separated into two sublattices so that a vertex in one sublattice has nearest neighbors only in the other sublattice. Due to this bipartite symmetry, the spectrum of Eq.\ref{eq:Hamiltonian} is symmetric around zero energy. If $|\Psi_1\rangle=\sum_i \Psi_i |\vec{R}_i\rangle$ is an eigenstate with energy $E$, then reversing the sign of the wavefunction in one of the sublattices creates $|\Psi_2\rangle=\sum_i (-1)^{\sigma_i} \Psi_i |\vec{R}_i\rangle$ with $\sigma_i=\{0,1\}$ the sublattice index of site $i$. By applying ${\cal H} |\Psi_2\rangle$, we see that energy $-E$ also belongs to the spectrum.  

In this paper, we are interested in LS which have zero energy. Still, bipartite symmetry can be used to require the LS to have support only on one of the sublattices. If a state $|\Psi_1\rangle$ with zero energy has non-zero density in both sublattices $|\Psi_2\rangle$ as constructed above, will be independent of $|\Psi_1\rangle$. In this case two new LS $|\Psi_\pm \rangle=|\Psi_1\rangle \pm |\Psi_2\rangle$ can be constructed which are localized to only one sublattice.  In the next section, we detail the numerical method we use to obtain and investigate the LS. 

\section{Numerical Method}
\label{sec:Numerical}

The spectrum of the Hamiltonian Eq.\ref{eq:Hamiltonian} and similar tight-binding models on quasicrystals have been investigated by several methods  \cite{tsu91,jpi06,sbe90,jag94,zja00,zij04,bsi91,psb92,jsc97,jag04,kts17}. Generation of finite lattices through inflation of a small set of tiles or their construction using the dual grid method are used to form Hamiltonians consisting of tens of thousands of sites. However, diagonalization of a finite system with open boundaries yields eigenstates that may depend on edges' presence. An alternative route to observe bulk behavior uses quasicrystal approximants, periodic lattices with large unit cells approximating quasicrystal symmetries \cite{dun89}. In this case, both the eigenstates' dependence on quasimomentum and their scaling behavior with changing unit cell size give useful information. 

As we aim to investigate LS, which have zero density outside a certain radius, we opt to use finite lattices with open boundary conditions. We generate these lattices starting from a single point defined to be the origin in real space, but we also specify its perpendicular space coordinate. The first neighbors of this lattice point can be generated following the algorithm outlined in Fig.\ref{fig:ABL_PerpSpace}, {\it i.e.} by deciding if the perpendicular space images of possible neighbors lie in the octagon $V$. Repeating the same process for the first neighbors generates the second neighbors, and the process can be extended up to the desired depth. We call such a finite section of the lattice the D-deep neighborhood of the central point. 

\begin{figure}[!htb]
    \centering
    \includegraphics[trim=8mm 8mm 8mm 8mm,clip,width=0.48\textwidth]{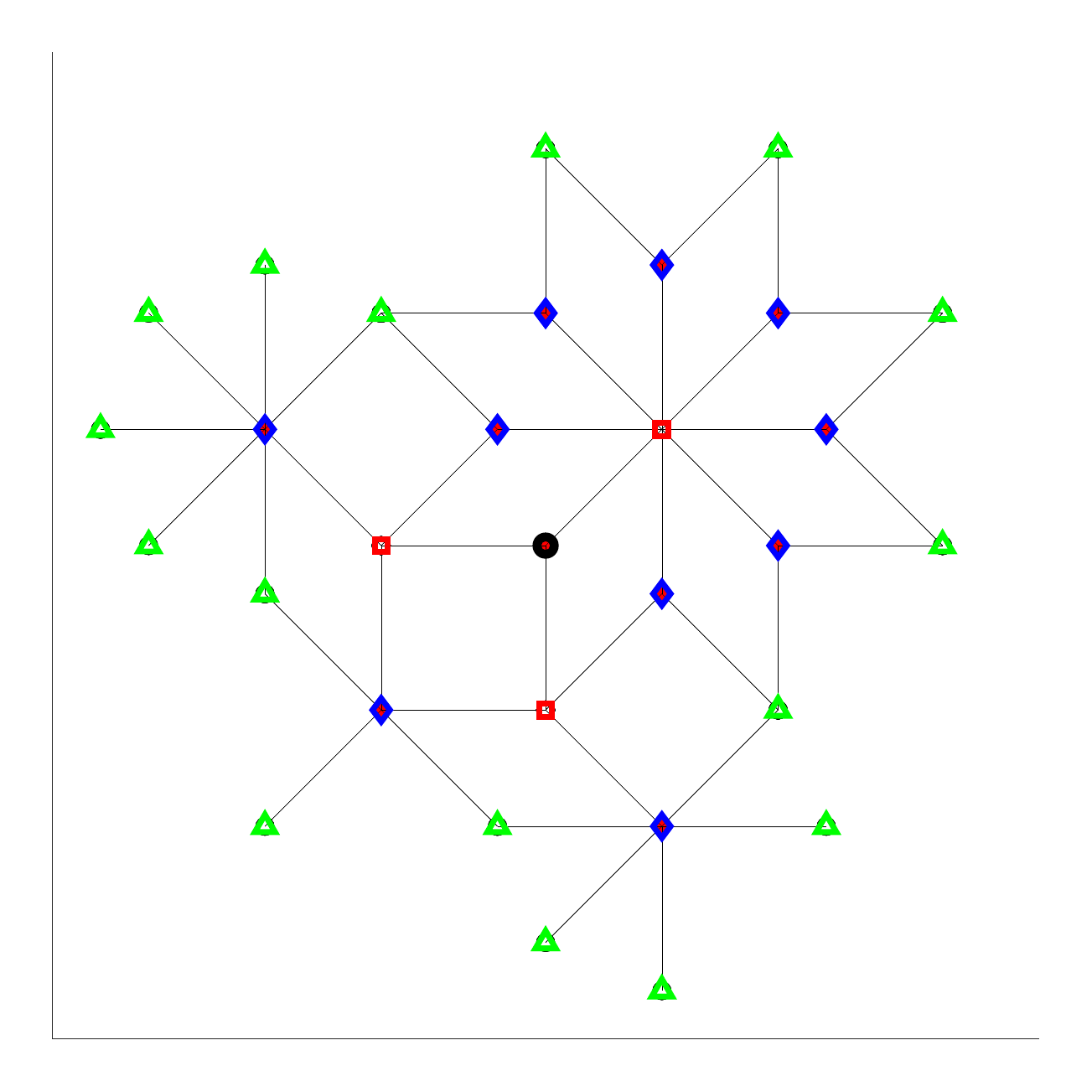}
    \caption{The 3-deep neighborhood of the point marked by the black circle. First neighbors (red squares) and third neighbors (green triangles) form one sublattice, while the second neighbors (blue diamonds) and the central point form the other sublattice. The perpendicular space images of all the points are marked with the same symbols in Fig.\ref{fig:Neighborhood_PerpSpace}.}
    \label{fig:Neighborhood_RealSpace}
\end{figure}

\begin{figure}[!htb]
    \centering
    \includegraphics[trim=8mm 8mm 8mm 8mm,clip,width=0.48\textwidth]{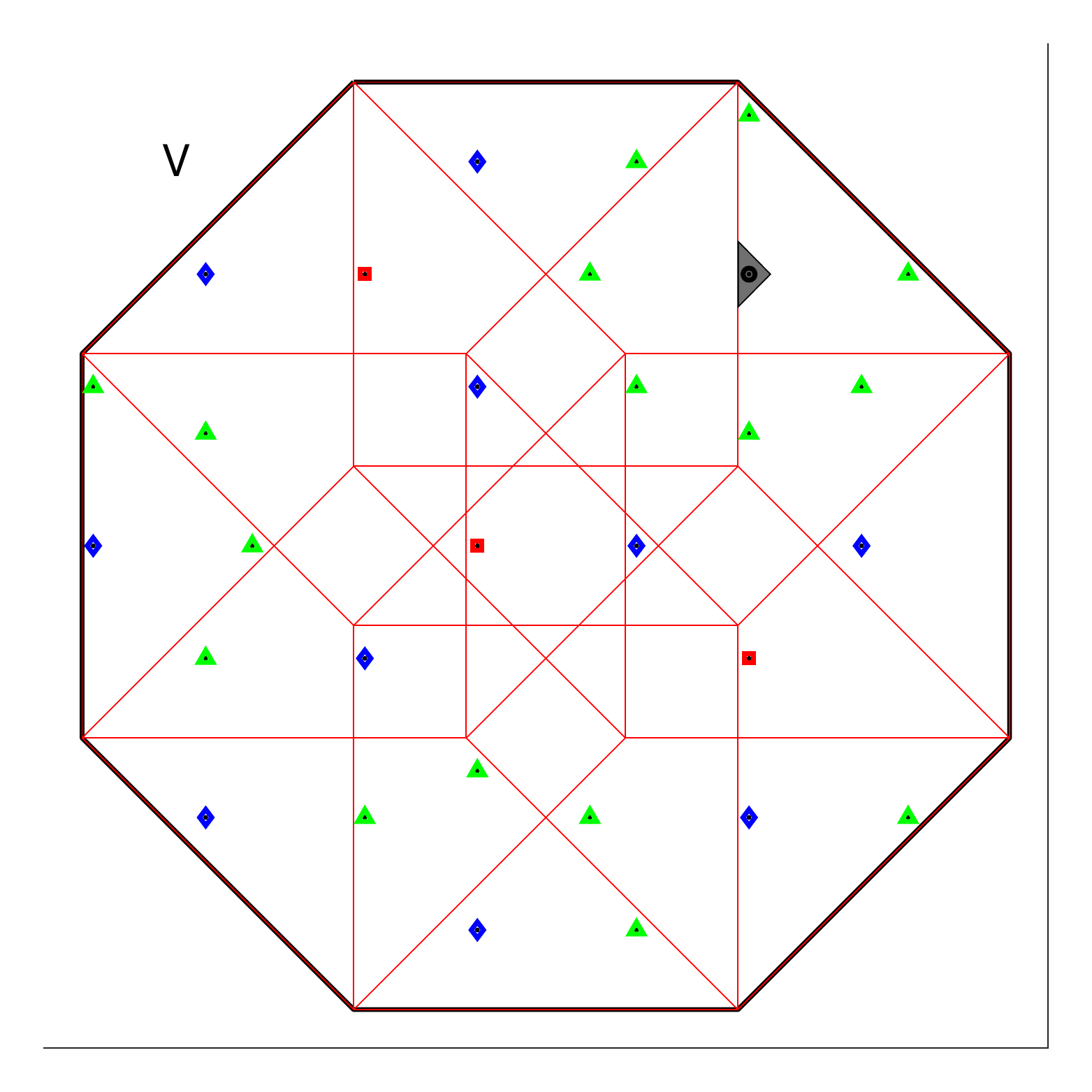}
    \caption{Perpendicular space images of all the points shown in Fig.\ref{fig:Neighborhood_RealSpace}. If the perpendicular space position of the initial point is moved within the marked triangle, all the points in the 3-deep neighborhood move without changing their type. The ratio of this triangle area to the area of the octagon $V$ gives the frequency of this exact neighborhood appearing in the infinite ABL.}
    \label{fig:Neighborhood_PerpSpace}
\end{figure}

As an example we display the 3-deep neighborhood of the point with perpendicular space coordinates $x_\perp=1,y_\perp=\frac{1}{2}$ in Fig.\ref{fig:Neighborhood_RealSpace}. The perpendicular space images of all the vertices in this neighborhood are displayed in Fig.\ref{fig:Neighborhood_PerpSpace}. The perpendicular space image  offers a simple way to count the frequency of occurrence of this finite configuration of sites in the infinite ABL. As long as the initial point's perpendicular space coordinate lies in the shaded triangle shown in   Fig.\ref{fig:Neighborhood_PerpSpace}, all the points in the 3-deep neighborhood will have perpendicular space images inside the octagon $V$. As soon as the initial point's perpendicular space coordinates go outside the triangle, at least one point will either change its vertex type or cease to exist. Thus, the frequency of this particular neighborhood can be found by the ratio of the triangle area to the area of $V$.


\begin{figure}[!htb]
    \centering
    \includegraphics[trim=8mm 8mm 8mm 8mm,clip,width=0.48\textwidth]{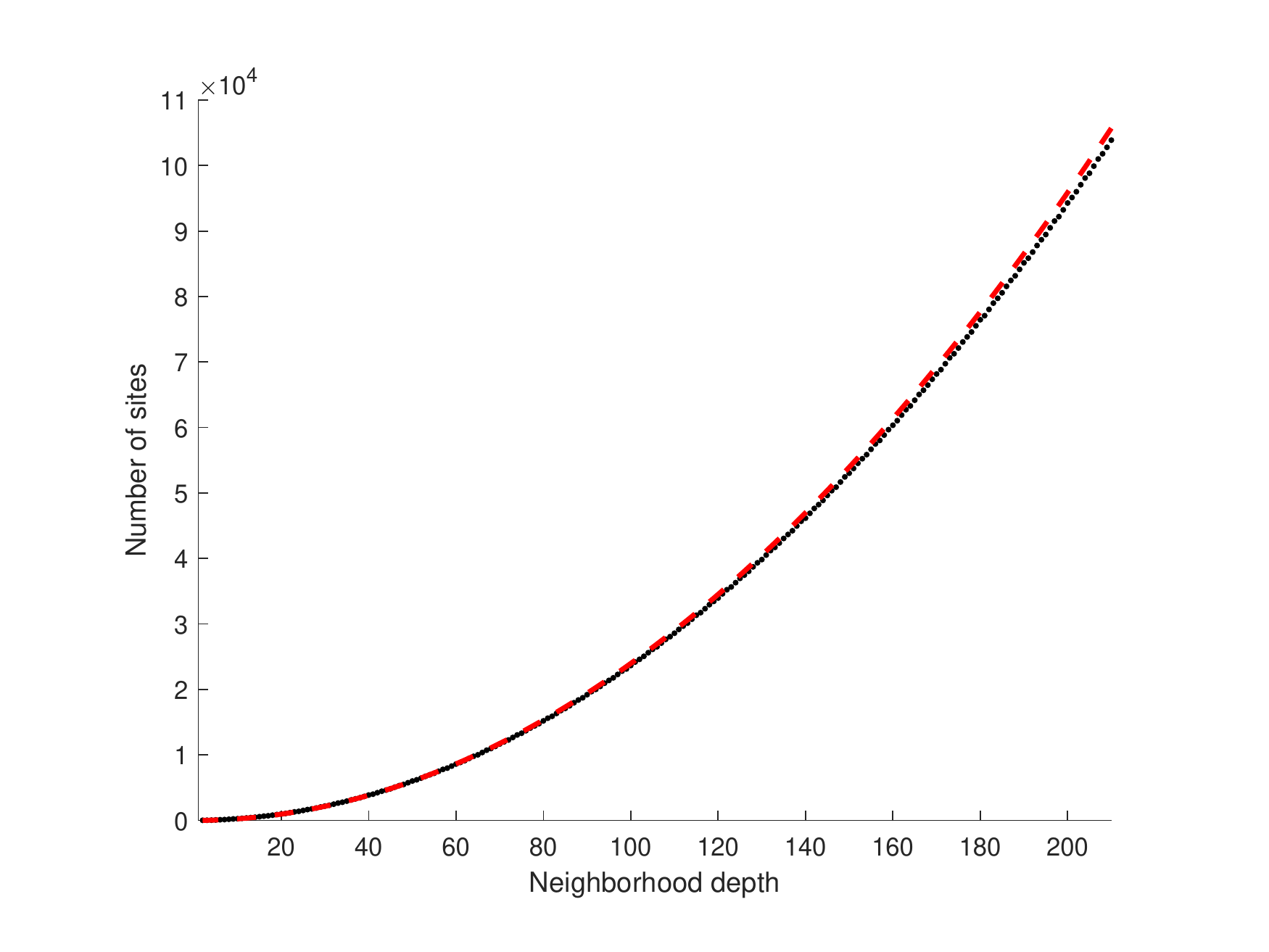}
    \caption{The number of sites in the D-deep neighborhood of an ABL point as a function of neighborhood depth. Numerical calculation (black dots) is in good agreement with the estimate Eq.\ref{eq:N_Sites_Estimate} (red dashed line). While the exact number of neighbors depends on the initial point's perpendicular space position, the variation is too small to resolve at this scale.} 
    \label{fig:Site_Count}
\end{figure}

The first step in judging the computational burden of finding the LS in a neighborhood is estimating the number of sites in the D-deep neighborhood. While the exact number of sites will depend on the perpendicular space coordinates of the starting point, the overall uniformity of the ABL ensures that the variation remains small for large $D$. Consider the largest distance one can cover in the x-direction by hopping across $D$ bonds in the ABL. While choosing all $D$ bonds to be oriented along $\hat{e}_0$ would give a distance of D, more than two subsequent hops in the same direction would take the perpendicular space image of the point outside of $V$. Thus such a series of connected same direction bonds do not appear in the ABL. One has to maximize the function $R =\hat{i} \cdot (n_0 \hat{e}_0 + n_1 \hat{e}_1 + n_3 \hat{e}_3)$, not only with the constraint $D=n_0+n_1+n_3$ but also with $\hat{i} \cdot (n_0 \hat{e}_0 - n_1 \hat{e}_1 - n_3 \hat{e}_3) \simeq 0$ so that the perpendicular space image remains close to the initial position. An estimate can be obtained by regarding the last condition as an equality, giving $n_1=n_3=\frac{n_0}{\sqrt{2}}$ thus $n_0=(\sqrt{2}-1) D$. The estimated radius of the neighborhood is then $R=2(\sqrt{2}-1)D$. The approximate area covered by the neighborhood $A=\pi R^2$ will contain $A/2$ squares and $A/\sqrt{2}$ rhombuses. Except for tiles on the boundary, every tile contributes two edges, and each edge connects two vertices. Combining these with the average number of edges for a vertex, which is $10-4\sqrt{2}$ for the ABL, we obtain
\begin{equation}
\label{eq:N_Sites_Estimate}
    N_{\mathrm{sites}}\simeq\frac{4 \left(3\sqrt{2}-1\right) \pi}{17} D^2,
\end{equation}
for the number of sites in a D-deep neighborhood. In Fig.\ref{fig:Site_Count}, we compare this estimate with the number of sites obtained by the numerical generation of the lattice. The difference in the number of sites for different starting points is too small to observe on this scale, and we can say that generally, 200-deep neighborhoods would contain about 100 000 sites.  

Although it is a finite section of the ABL, a D-deep neighborhood is uniquely suited to the LS calculation. The only sites that have missing bonds because of the edge are $D^{th}$ neighbors of the initial point. Thus, all of the edge sites belong to the same sublattice.  The diagonalization of the Hamiltonian on this finite lattice will result in a manifold of zero energy states, some of which exist due to the edge's presence. However, bipartite symmetry allows us to constrain the LS to only one sublattice. Once this operation is carried out, the zero energy states localized on the sublattice including the $(D-1)^{th}$ neighbors are guaranteed to be LS states of the infinite ABL. As an example, consider a 9-deep neighborhood of a given initial point. Take a state with zero energy found by numerical diagonalization of the Hamiltonian. In general, it will have an amplitude on both the odd neighbor sublattice (first-third-..-ninth) and the even neighbor sublattice (initial point-second neighbors...-eighth neighbors). Due to the bipartite symmetry, if we take the amplitudes on the odd sublattice to be zero, we still have an eigenstate that is defined only on the even sublattice. Such a state will also be a LS of the infinite ABL, as all the nearest neighbors of the $8^{th}$ neighbor sites are already present in the finite Hamiltonian. While this finite-size calculation would miss the LS of the infinite ABL that cross the edge, it is guaranteed to find all the LS which lie entirely within the defined neighborhood. Thus, we expect that large neighborhoods to provide numerical lower bounds for the number of LS in the ABL.   

The Hamiltonian for a D-deep neighborhood can be expressed as a sparse matrix of size $N_{\mathrm{sites}}$. Finding all the eigenvalues is necessary to be able to describe the properties of the full spectrum. However, such a calculation suffers from three shortcomings for our purpose. First is the presence of unwanted edge states, unavoidable for a finite lattice. The second problem is that eigenvalues' numerical accuracy may make it hard to decide if an eigenstate has exactly zero energy. Finally, finding the full spectrum creates unnecessary computational cost if one is only interested in zero energy eigenvalues. Typically matrix diagonalization scales with $O(N_{\mathrm{sites}}^3)$ which constrains system size the tight-binding calculations for quasicrystals.

Instead of finding the spectrum, we speed up the calculation of the LS by focusing on finding the null space of the Hamiltonian and by taking advantage of the bipartite symmetry. Let's consider a D-deep neighborhood of an initial point which has $N_{\mathrm{even}}$ sites in the sublattice containing the initial point and $N_{\mathrm{odd}}$ sites in the other sublattice. The Hamiltonian will be $N_{\mathrm{sites}} \times N_{\mathrm{sites}} $ matrix when expressed in the site $|\vec{R}\rangle$  basis with $N_{\mathrm{sites}}=N_{\mathrm{even}}+N_{\mathrm{odd}}$. By reorganizing the rows and columns so that odd and even sublattice sites form blocks, the Hamiltonian matrix takes the form
\begin{equation}
\label{eq:MatrixC}
    {\cal H}=\begin{bmatrix}
0 & {\cal C} \\
{\cal C}^T & 0
\end{bmatrix}.
\end{equation}
Here ${\cal C}$ is a sparse $N_{\mathrm{even}} \times N_{\mathrm{odd}}$ matrix and ${\cal C}^T$ is its transpose. LS in the odd (even) sublattice form the null space of the ${\cal C}$ (${\cal C}^T$).  If the neighborhood depth D is even, we are interested in the LS in the odd sublattice; thus, the null space of ${\cal C}$.  

Calculation of the null space size of ${\cal C}$ can be accomplished by a number of numerical methods. We choose QR decomposition in which the matrix ${\cal C}$ is expressed as the multiplication of two matrices
\begin{equation}
    {\cal C}= {\cal Q} {\cal R} 
\end{equation}
where ${\cal Q}$ is a unitary matrix and ${\cal R}$ is right triangular \cite{glo89}. The number of zeros on the diagonal of ${\cal R}$ gives the size of the null space. We use sparse QR decomposition as implemented in MATLAB to calculate the null space size. The sparsity of ${\cal C}$ speeds up the calculation immensely, for a depth of $D=200$ with $N_{even}=47339$ and $N_{odd}=46942$ the vectors forming the null space is calculated in $t\simeq 30 s$ on a laptop computer with an Intel i5 processor. 

We estimate the LS fraction from the result of the D-deep neighborhood calculation as follows. First, assume that $D$ is even. We construct the Hamiltonian in the form of Eq.\ref{eq:MatrixC}, and count the number of odd sublattice sites  $N_{odd}$. We calculate the size of the null space of ${\cal C}$ call the number of LS defined on the odd sublattice $N^{LS}_{odd}$. We repeat the same process for depth $D-1$, for which the Hamiltonian can be obtained trivially by truncating the Hamiltonian of neighborhood $D$. Now we are interested in the LS on the even sublattice, which are obtained by QR decomposition of the truncated ${\cal C}^T$. This calculation gives us $N_{even}$ and $N^{LS}_{even}$. Then our estimate for the LS fraction is
\begin{equation}
\label{eq:LS_estimation}
    f_{LS}\simeq \frac{N^{LS}_{even}+N^{LS}_{odd}}{N_{even}+N_{odd}}.
\end{equation}

\begin{figure}[!htb]
    \centering
    \includegraphics[trim=2mm 2mm 2mm 2mm,clip,width=0.48\textwidth]{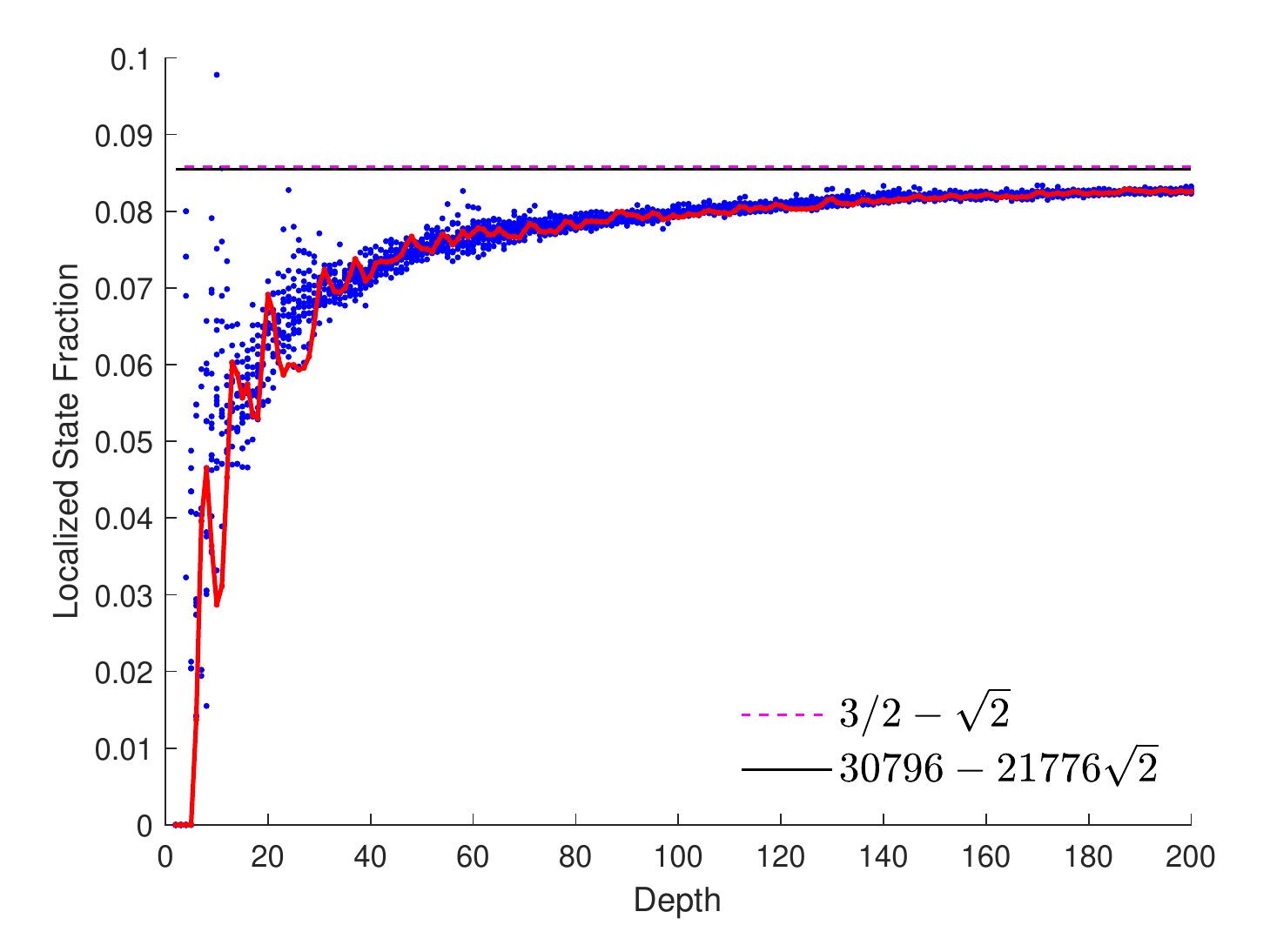}
    \caption{The numerically calculated LS frequency compared with the analytical lower bound Eq\ref{eq:AnalyticLowerBound} and the conjectured exact frequency \cite{kog20} which are almost indistinguishable at this scale.  The LS frequency is estimated with Eq.\ref{eq:LS_estimation} for a finite neighborhood. The perpendicular space coordinates for the initial point of the neighborhood is randomly selected. Data for 15 different initial points are shown with (blue) dots. The red line shows the variation with depth for a single starting point.  The LS frequency is expected to approach the exact frequency from below as the edge effects become less pronounced in larger neighborhoods. }
    \label{fig:Frequency_ABL}
\end{figure}

In Fig. \ref{fig:Frequency_ABL} we show the LS fraction for fifteen  different starting points as a function of neighborhood depth. The initial point's perpendicular space position does not cause large variations once the depth becomes larger than 50. The LS fraction increases as the boundary effects get smaller. As our method finds all the LS that stay within the calculated neighborhood, it is natural to expect that the fraction to increase as the boundary region becomes smaller compared to the bulk region.   
The average for fifteen different starting points gives $f_{LS}=0.0838$ at depth 200.  In the next section, we calculate an analytical lower bound for the LS fraction as 
\begin{equation}
\label{eq:AnalyticLowerBound}
f_{LS}=30796-21776\sqrt{2} \simeq 0.08547.
\end{equation}
The agreement between the two lower bounds shows that we are unlikely to miss an LS type with more than $0.1\%$ frequency. Both of these values are quite close to the recent\cite{kog20} conjectured exact frequency  $f_{Ex}=3/2-\sqrt{2}\simeq 0.08579$. 

\begin{figure}[!htb]
    \centering
    \includegraphics[trim=2mm 2mm 2mm 2mm,clip,width=0.48\textwidth]{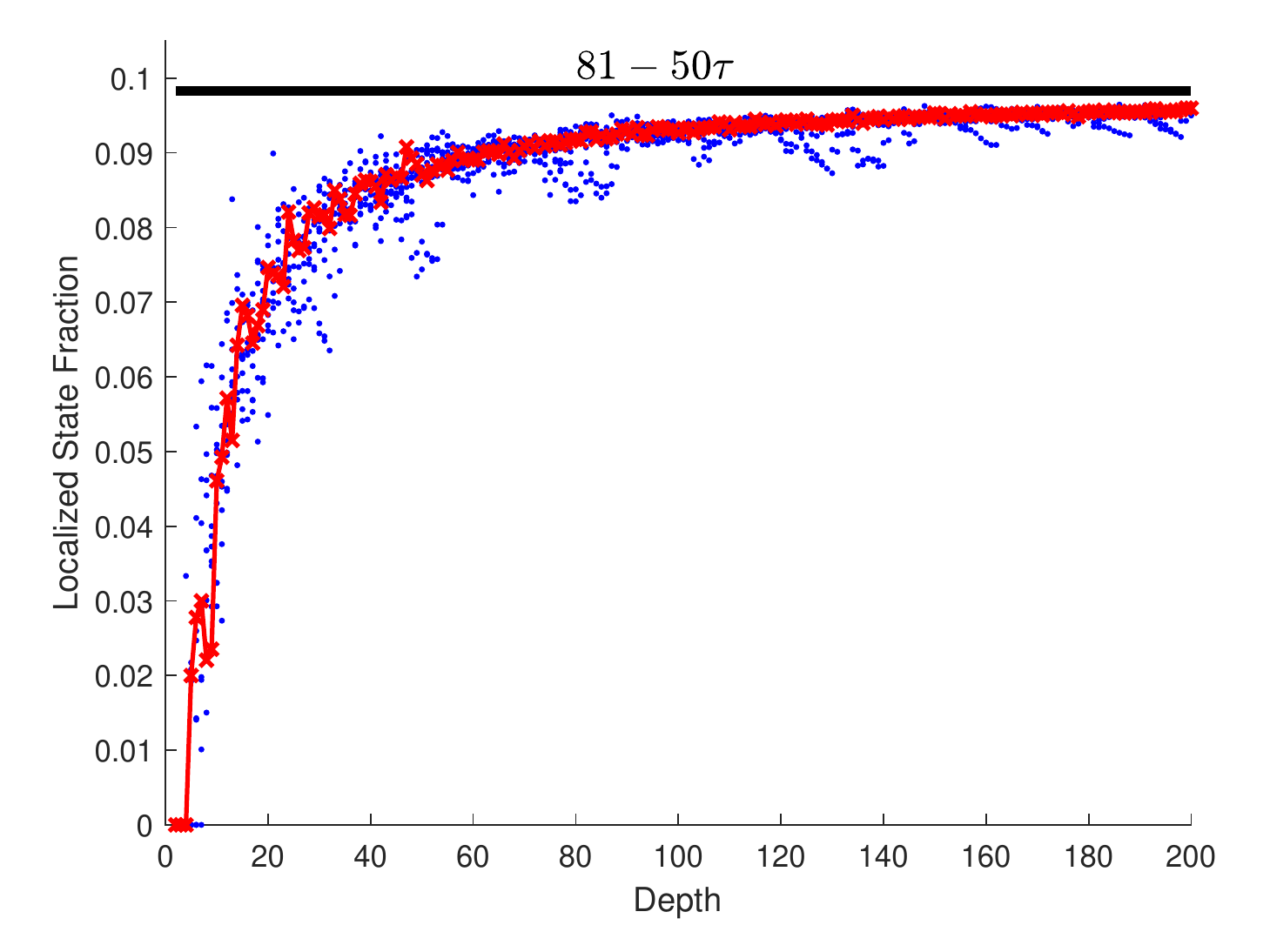}
    \caption{ The LS fraction for the Penrose Lattice calculated for ten different starting points (blue dots), compared with the analytical lower bound from Ref.\cite{ara88}. The calculated fractions for one starting point are connected by red lines to display the variation for a single starting point. There are 100735 sites in a neighborhood of depth 205.}
    \label{fig:Frequency_PL}
\end{figure}

As a test of our numerical method, we repeat the same calculation for the Penrose lattice. The analytical lower bound for LS fraction for the Penrose lattice is $81-50 \tau$, with $\tau=(1+\sqrt{5})/2$ \cite{ara88,kts17}. We show the numerically calculated fraction in Fig.\ref{fig:Frequency_PL} for ten distinct starting points. The agreement between the analytical limit and our numerical calculation is similar for both the ABL and the Penrose lattice.

The QR decomposition we use to count the LS also provides states that span the null space. As these states are degenerate, any linear combination of them is still a LS. The typical result of the numerical calculation is an eigenstate distributed over a large section of the neighborhood. However, a better understanding of the zero-energy manifold can be obtained by defining types of LS. By definition, any LS has finite support. Thus, Conway's theorem ensures that there are infinitely many copies of the same LS related by translations. Following Ref.\cite{ara88}, we define all these copies as belonging to the same LS type.  While any LS can be defined as a new LS type, the aim is to find a finite set of LS types independent from each other, spanning the whole manifold of zero energy states. For the Penrose Lattice, six LS types were identified with these properties \cite{ara88,kts17,mok20}. 

In general, a more compact LS type will have a higher frequency and will span a larger portion of the LS manifold. Thus, it is desirable to identify linear combinations of LS  which have a small number of sites in their support. Previously, an optimization algorithm has been used to find states with high inverse participation ratios \cite{rsc95}. Instead of focusing on a single LS, we use a method that was first proposed for the calculation of maximally localized Wannier functions in a lattice \cite{kiv82}. The position operators $\hat{X}, \hat{Y}$ are trivially defined by their action on the states $|\vec{R}_i \rangle$ defined at each site.  We first define a projection operator 
\begin{equation}
    \hat{P}_{LS}=\sum | \Psi_{LS} \rangle \langle \Psi_{LS} |,
\end{equation}
where the sum is over all the LS found by our numerical calculation. Now we can define a projected position operator 
\begin{equation}
    \hat{X}_{LS}=\hat{P}_{LS} \hat{X} \hat{P}_{LS}.
\end{equation}
The eigenstates of the projected position operator are as compact as possible in the $\hat{x}$ direction given the constraint that they are built up of functions within the LS manifold. This diagonalization is efficiently calculated as the projected position operator for any D-deep neighborhood can be represented by a matrix with the same dimension as the Hamiltonian's null space. In our calculation, the diagonalization of $\hat{X}_{LS}$ results in a block diagonal form, as there are various LS that are related by a translation along the $y$ direction. Thus, we follow by defining the $\hat{Y}_{LS}$ operator in a similar manner and diagonalize it within degenerate the blocks to localize the LS in both directions. As a result of this process, we find a set of twenty LS types which are investigated in the next section.

\section{Types and Frequencies of Localized states}
\label{sec:LS}
\begin{table*}[!ht]

\caption{Properties of the 20 LS types for the ABL. The LS support is either on the sublattice $\sigma$ containing the central T1 vertex (0) or the other (1) sublattice. The LS requires a neighborhood depth, shown in the Depth column, around the central vertex to exist. Each LS has a density that is eight-fold symmetric around the central vertex. However, the wavefunction either stays the same or is multiplied by -1 under a $\pi/4$ rotation as listed in the $C_8$ column. The width of the allowed perpendicular space octagon, as well as resulting frequency, is also listed.  }
\label{tb:LStable}
\begin{tabular}{|c|c|c|c|c|c|c|}
\hline
Type & $\sigma$ & Depth & $C_8$ & $W_\perp$       & $f_{LS}$        & $f_{LS}\simeq $   \\ \hline \hline
A    &  1      & 2     & -1    & \multirow{2}{*}{$\sqrt{2}-1$}  & $(\sqrt{2}-1)^4=$ &\multirow{2}{*}{$2.9437 \; 10^{-2}$}  \\  
B    &  0      & 3     & -1    &   &  $17-12\sqrt{2}$& \\  \hline 
C    &  1      & 6     & +1    &   &  & \multirow{4}{*}{$5.0506 \; 10^{-3}$}  \\  
D    &  1      & 6     & -1    & $(\sqrt{2}-1)^2=$  &$(\sqrt{2}-1)^6=$ &  \\  
E    &  0      & 7     & -1    &  $3-2\sqrt{2}$  & $99-70\sqrt{2}$ & \\  
F    &  0      & 7     & +1    &   &  &  \\  \hline
G    &  1      & 12    & -1    &   &  &\multirow{6}{*}{$8.6655 \; 10^{-4}$} \\  
H    &  0      & 13    & -1    &   &  &  \\  
I    &  1      & 14    & +1    &  $(\sqrt{2}-1)^3=$ & $(\sqrt{2}-1)^8=$ &  \\  
J    &  1      & 14    & -1    &  $-7+5\sqrt{2}$ & $577-408\sqrt{2}$ &  \\  
K    &  0      & 17    & -1    &   &  &  \\  
L    &  0      & 17    & +1    &  &  &  \\  \hline 
M    &  1      & 30    & +1    &  &  &\multirow{8}{*}{$1.4868 \; 10^{-4}$}  \\   
N    &  1      & 30    & -1    &  &  &  \\   
O    &  0      & 31    & +1    &   &  &  \\   
P    &  0      & 31    & -1    &   $(\sqrt{2}-1)^4=$ &  $(\sqrt{2}-1)^{10}=$ & \\   
Q    &  1      & 34    & +1    &  $17-12\sqrt{2}$ & $3363-2378\sqrt{2}$ &  \\ 
R    &  1      & 34    & -1    &   &  &  \\  
S    &  0      & 41    & -1    &   &  &  \\  
T    &  0      & 41    & +1    &  &  &  \\  \hline 

\end{tabular}
\end{table*}

\begin{figure}[!htb]
    \centering
    \includegraphics[trim=8mm 8mm 8mm 8mm,clip,width=0.4\textwidth]{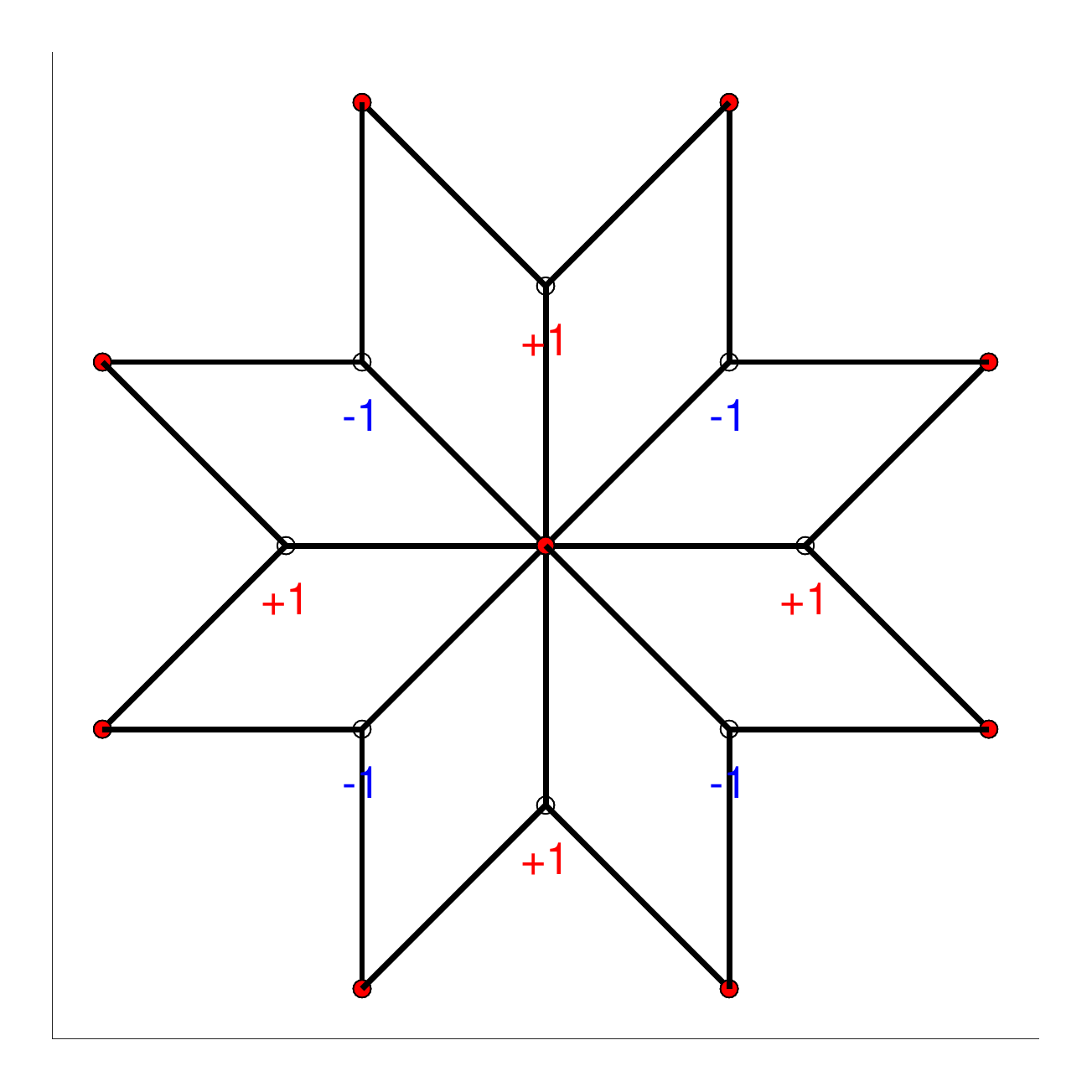}\\
    \includegraphics[trim=8mm 8mm 8mm 8mm,clip,width=0.4\textwidth]{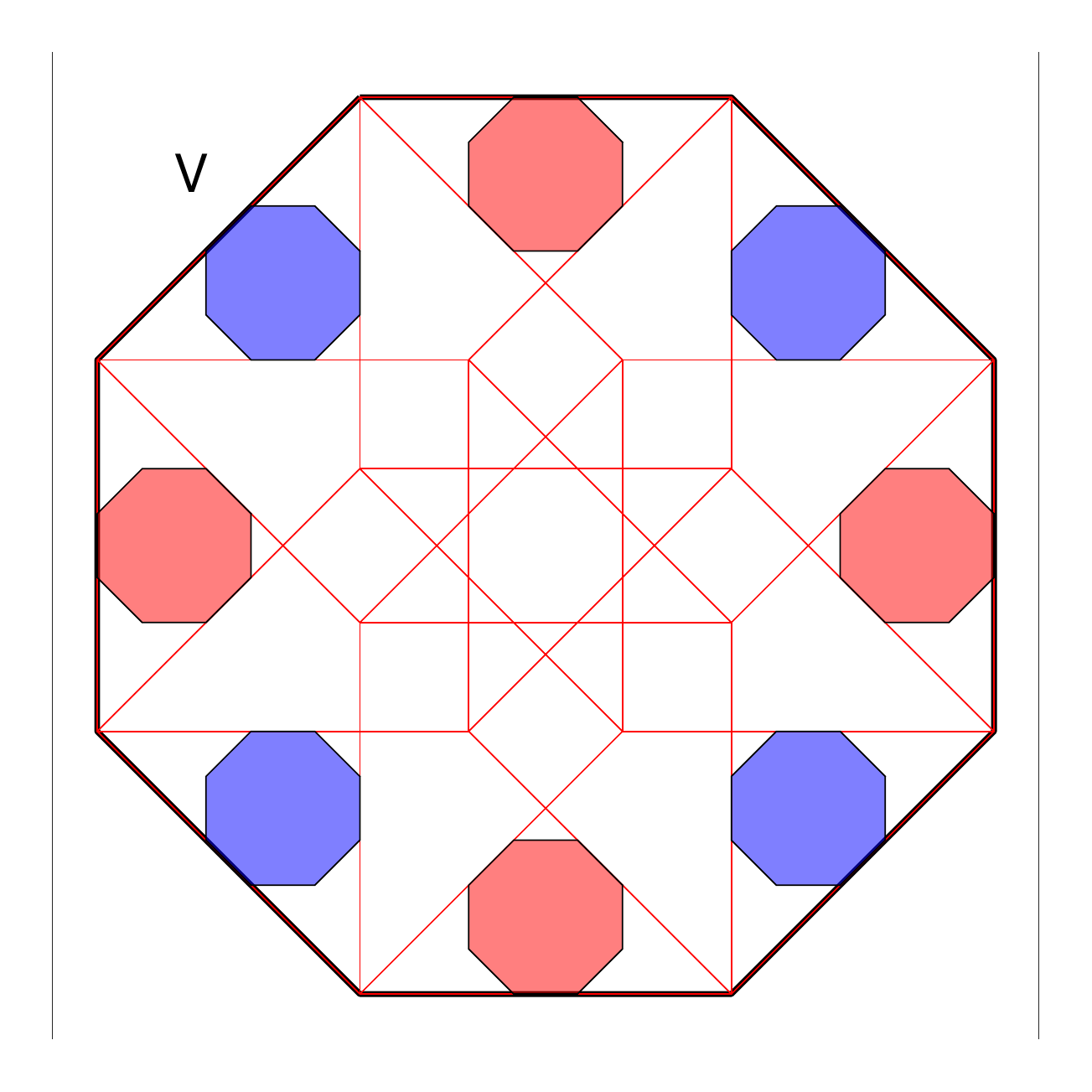}
    \caption{Type-A LS wavefunction and corresponding perpendicular space allowed regions for eight points in the support. The perpendicular space allowed regions are colored based on the sign of the wavefunction (red for positive, blue for negative) here and in all subsequent figures. All type-A LS support consists of 3 edge vertices. The outermost vertices are the nearest neighbors of the support of Type-A wavefunction. The perpendicular space position of the central point determines what other bonds are connected to these outermost vertices. }
    \label{fig:TA_RealSpace}
\end{figure}
\begin{figure}[!htb]
    \centering
    \includegraphics[trim=8mm 8mm 8mm 8mm,clip,width=0.4\textwidth]{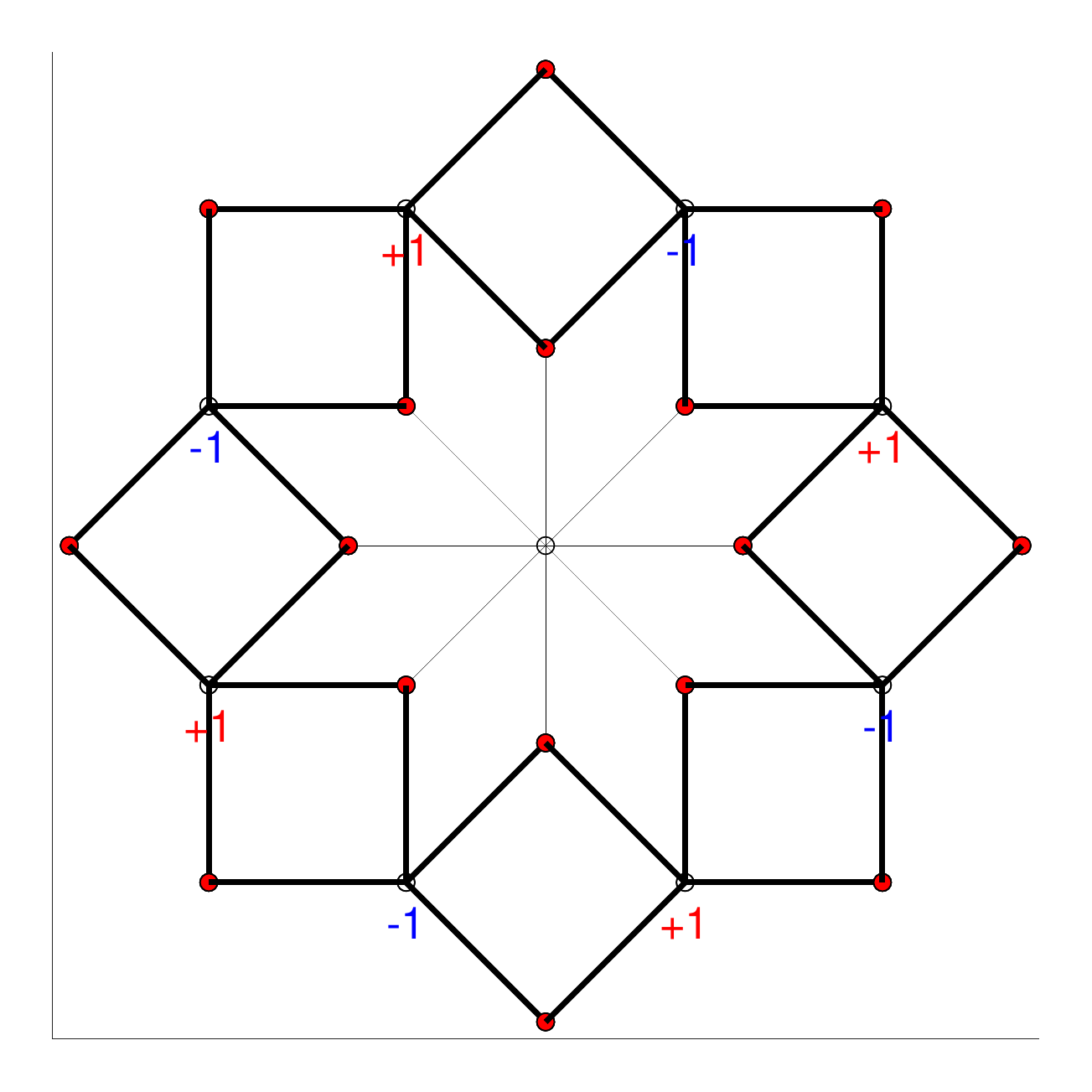}
    \includegraphics[trim=8mm 8mm 8mm 8mm,clip,width=0.4\textwidth]{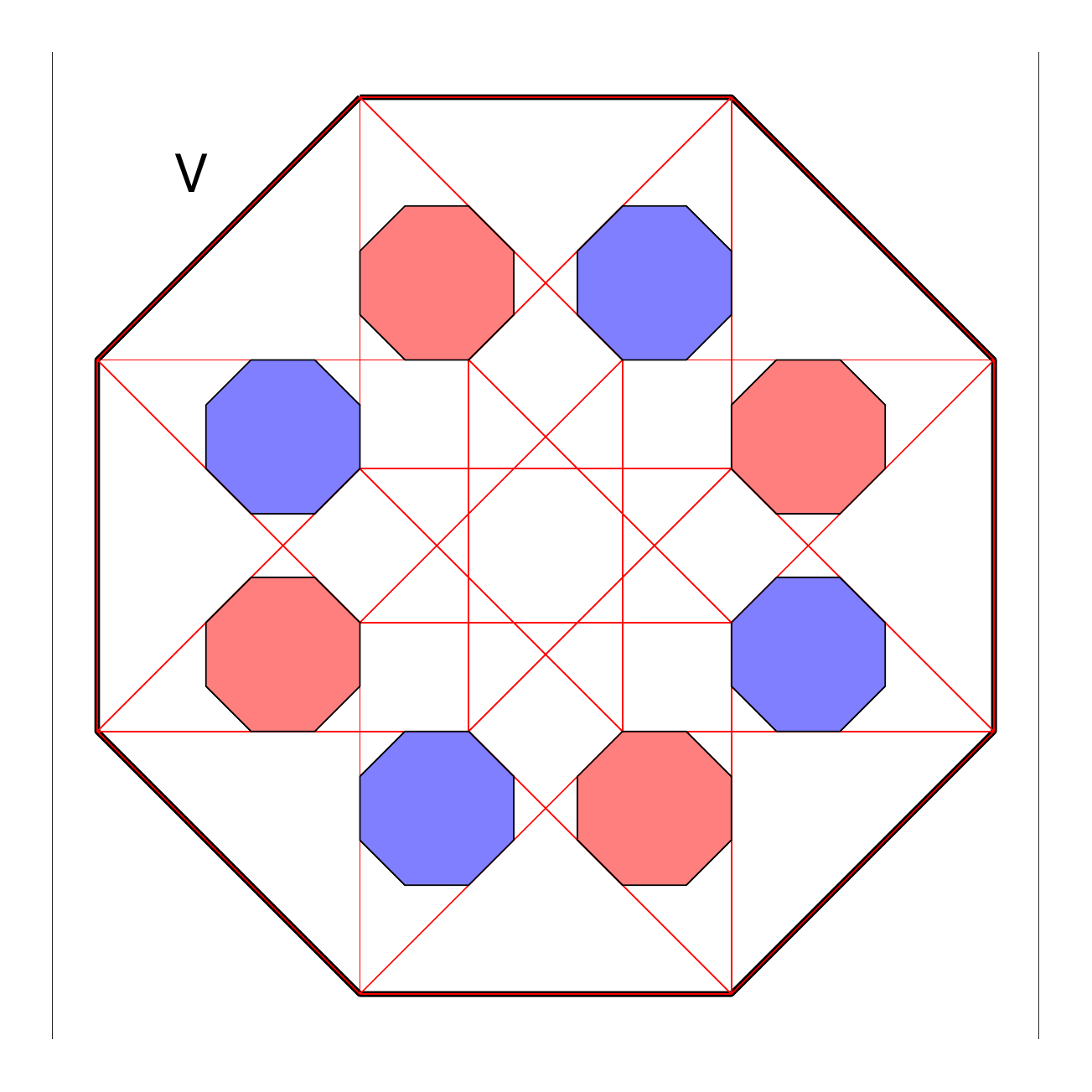}
    \caption{Type-B LS wavefunction and corresponding allowed regions in perpendicular space. Type-B wavefunction is non-zero only on T5 vertices; there is never overlap with a type-A state. }
    \label{fig:TB_RealSpace}
\end{figure}

\begin{figure}[h]
    \centering
    \includegraphics[trim=8mm 8mm 8mm 8mm,clip,width=0.4\textwidth]{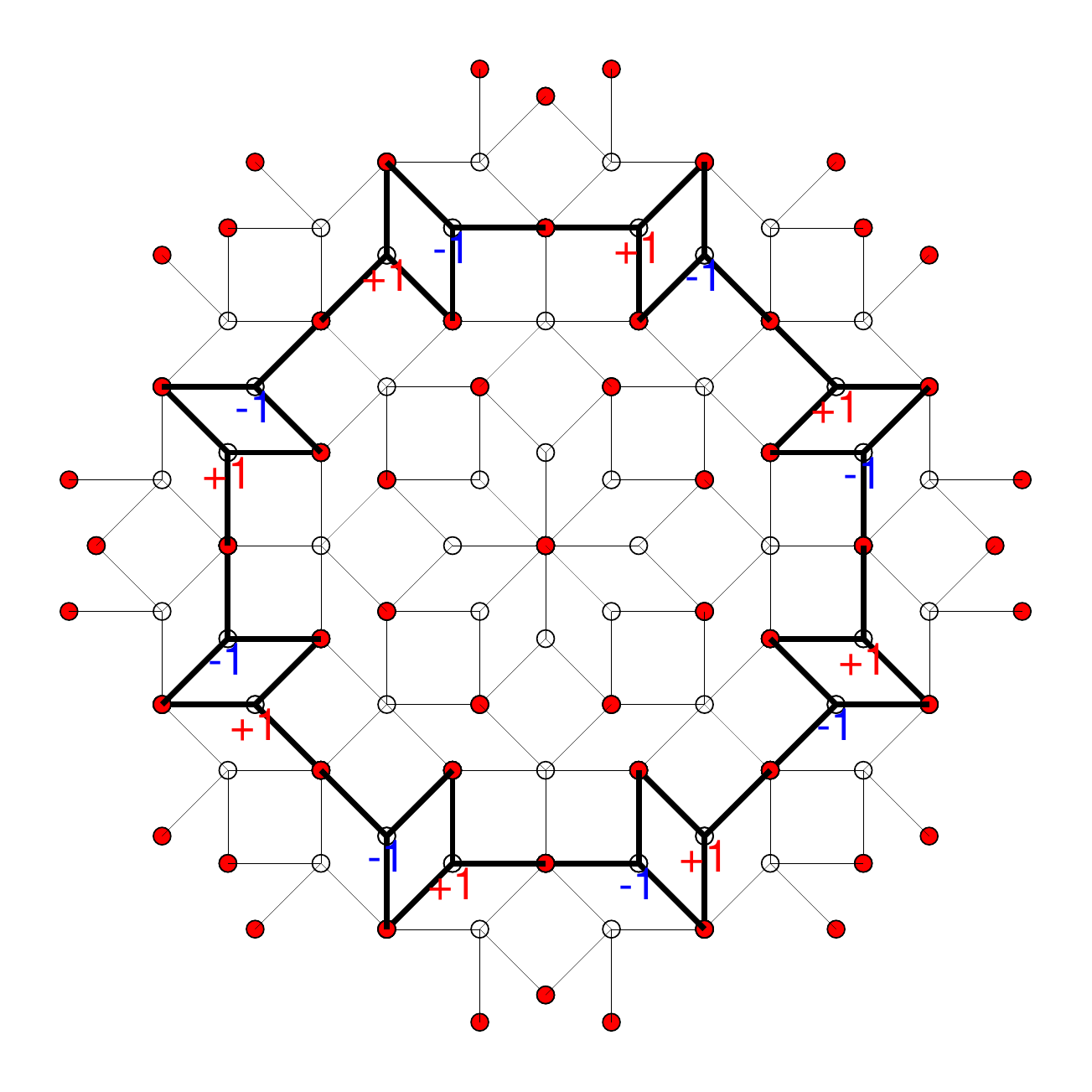}\\
    \includegraphics[trim=8mm 8mm 8mm 8mm,clip,width=0.4\textwidth]{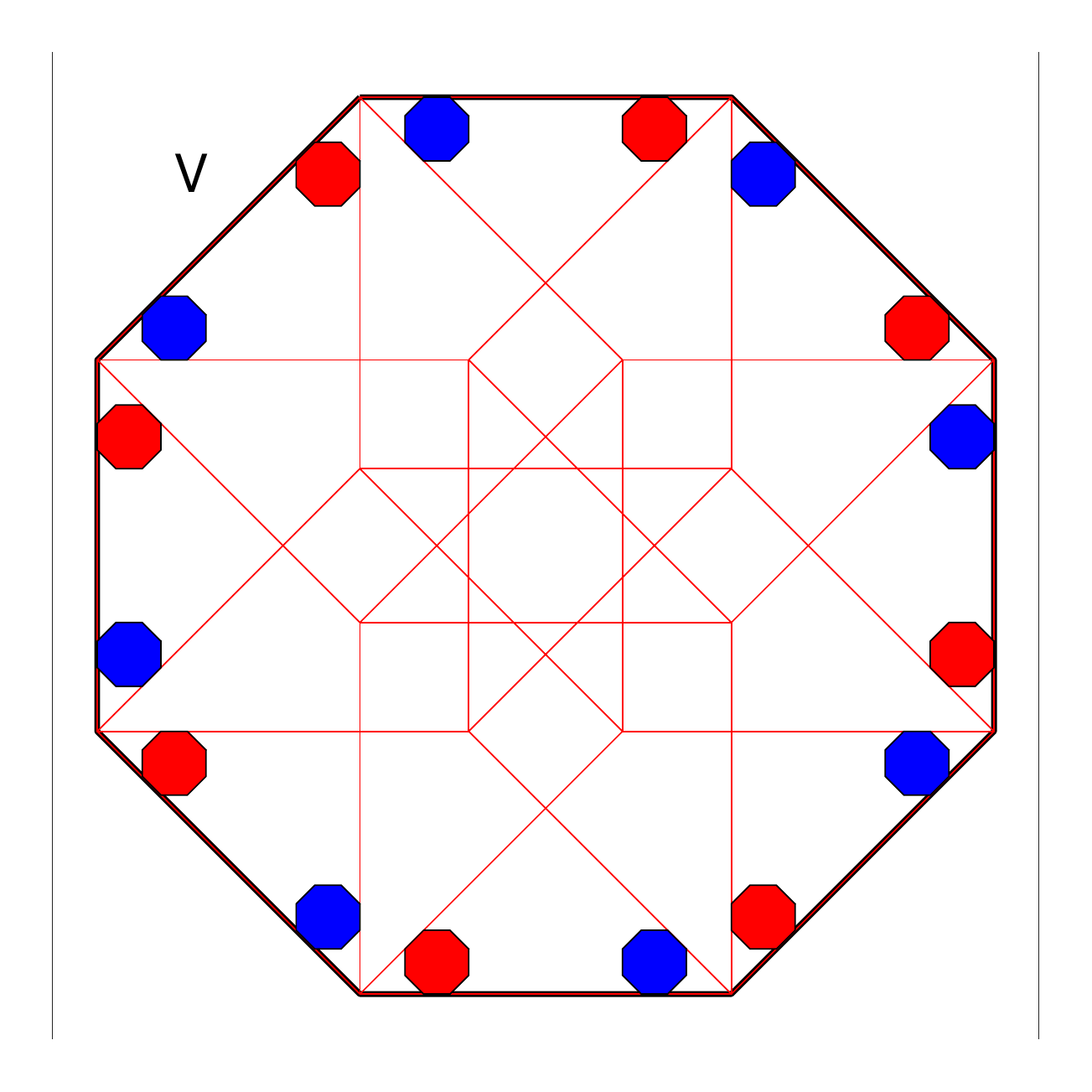}
    \caption{Type-C LS wavefunction and perpendicular space allowed regions. Although type-C support also consists of T6 vertices like type-A, no vertex appears in the support of both type-A and type-C, as can be seen by comparing the two types' perpendicular space images. Regions previously uncovered by other types in perpendicular space can be used to establish independence. }
    \label{fig:TC_RealSpace}
\end{figure}

\begin{figure}[h]
    \centering
    \includegraphics[trim=8mm 8mm 8mm 8mm,clip,width=0.4\textwidth]{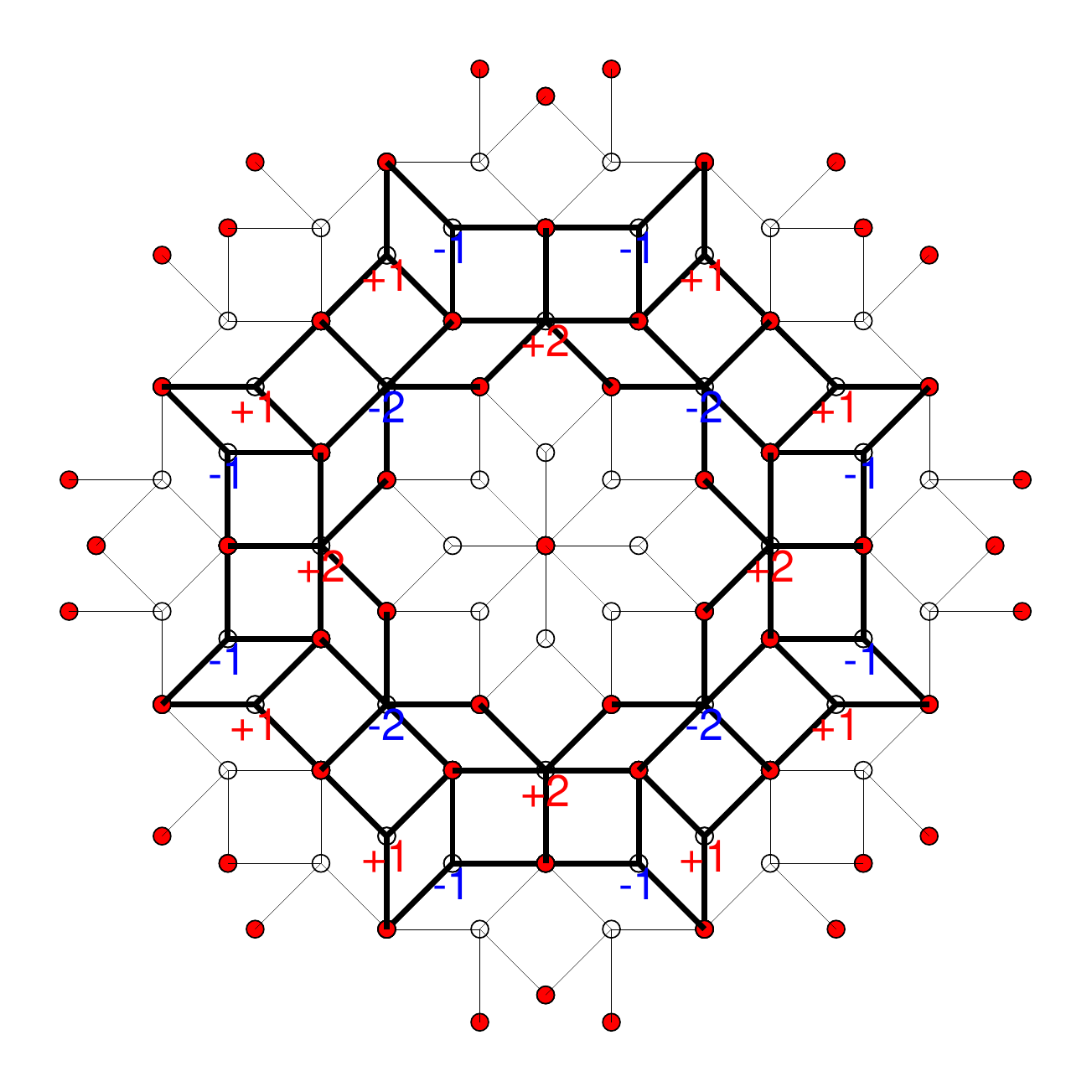}\\
    \includegraphics[trim=8mm 8mm 8mm 8mm,clip,width=0.4\textwidth]{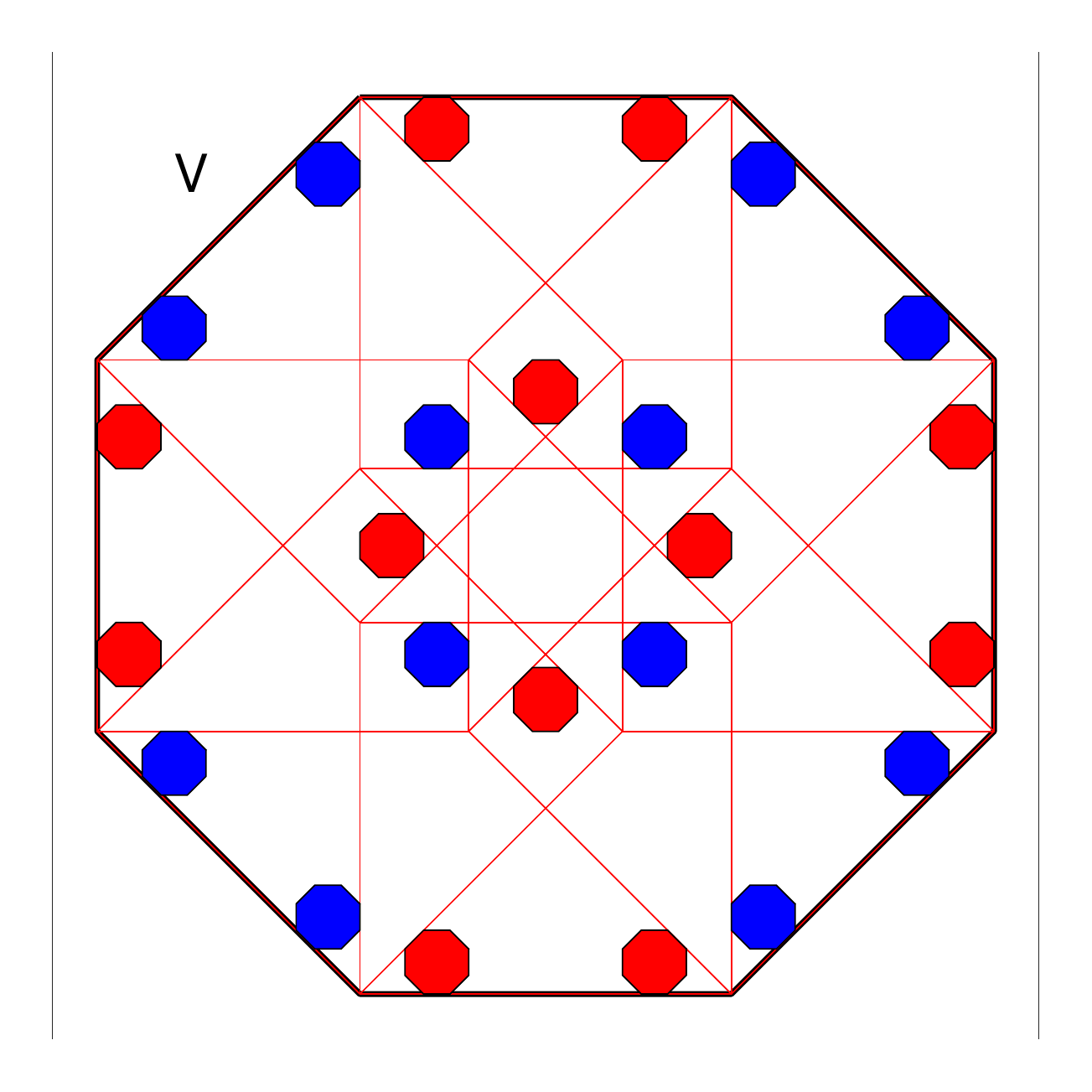}
    \caption{Type-D LS wavefunction and perpendicular space allowed regions. Type-D is the first LS type that has support on 5-edge vertices, thus independent from previous types. Although type-C and type-D share 16 overlapping sites, their overlap is zero as one is even and the other is odd under $\pi/4$ rotation.}
    \label{fig:TD_RealSpace}
\end{figure}

\begin{figure}[!htb]
    \centering
    \includegraphics[trim=8mm 8mm 8mm 8mm,clip,width=0.4\textwidth]{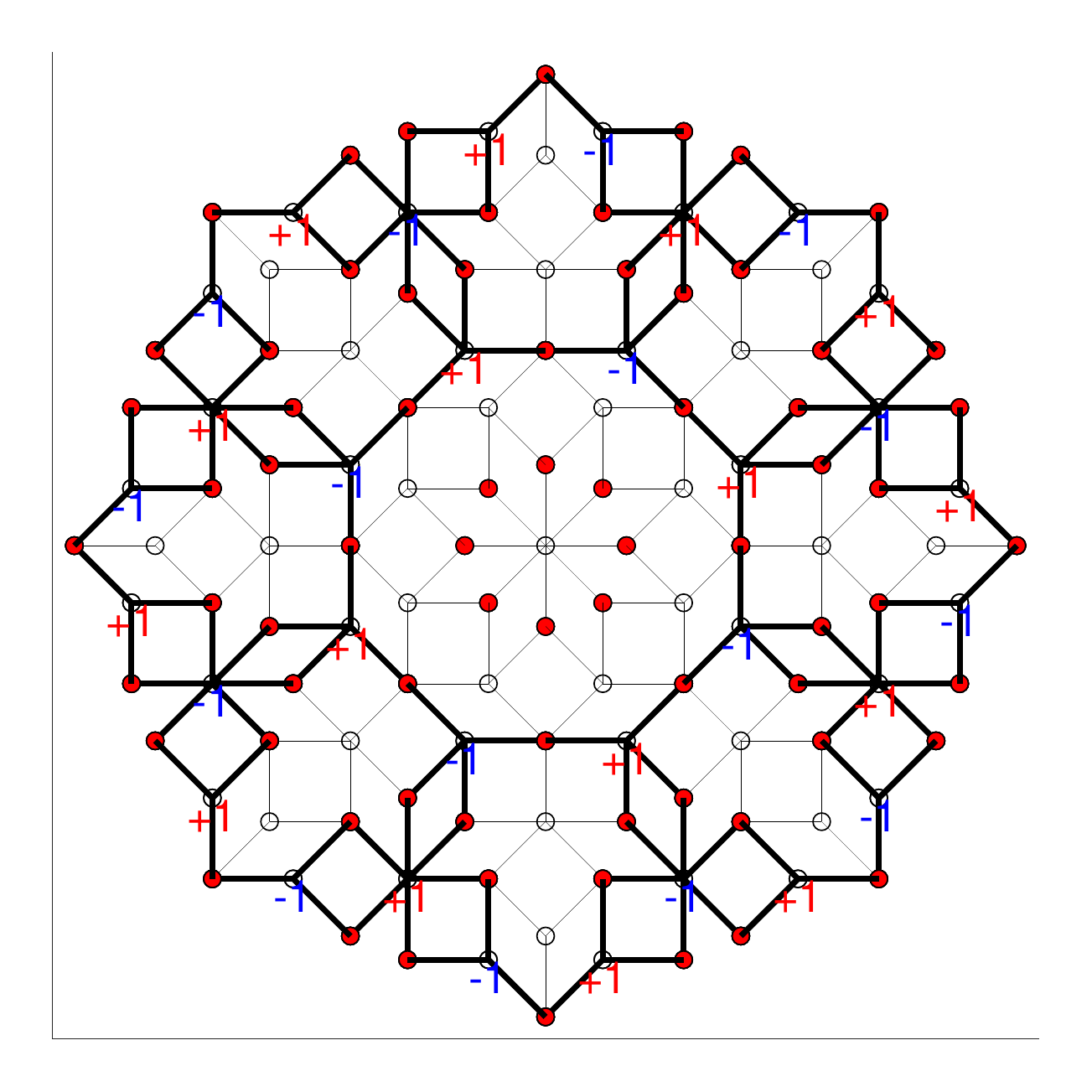}\\
    \includegraphics[trim=8mm 8mm 8mm 8mm,clip,width=0.4\textwidth]{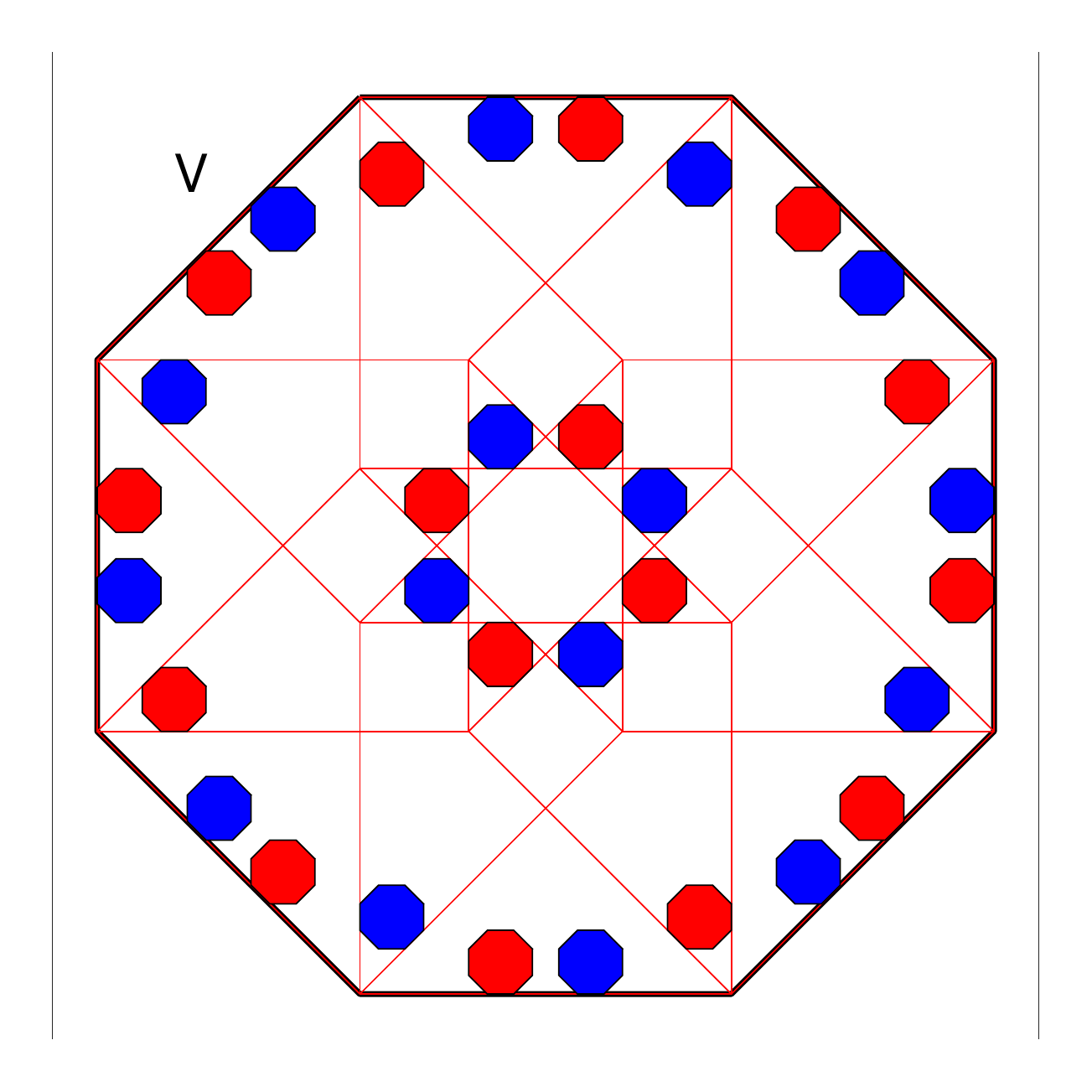}
    \caption{Type-E LS wavefunction and perpendicular space allowed regions. Independence of Type-E from the previous LS types is established by noticing that its support contains six edge vertices.}
    \label{fig:TE_RealSpace}
\end{figure}

\begin{figure}[!htb]
    \centering
    \includegraphics[trim=8mm 8mm 8mm 8mm,clip,width=0.4\textwidth]{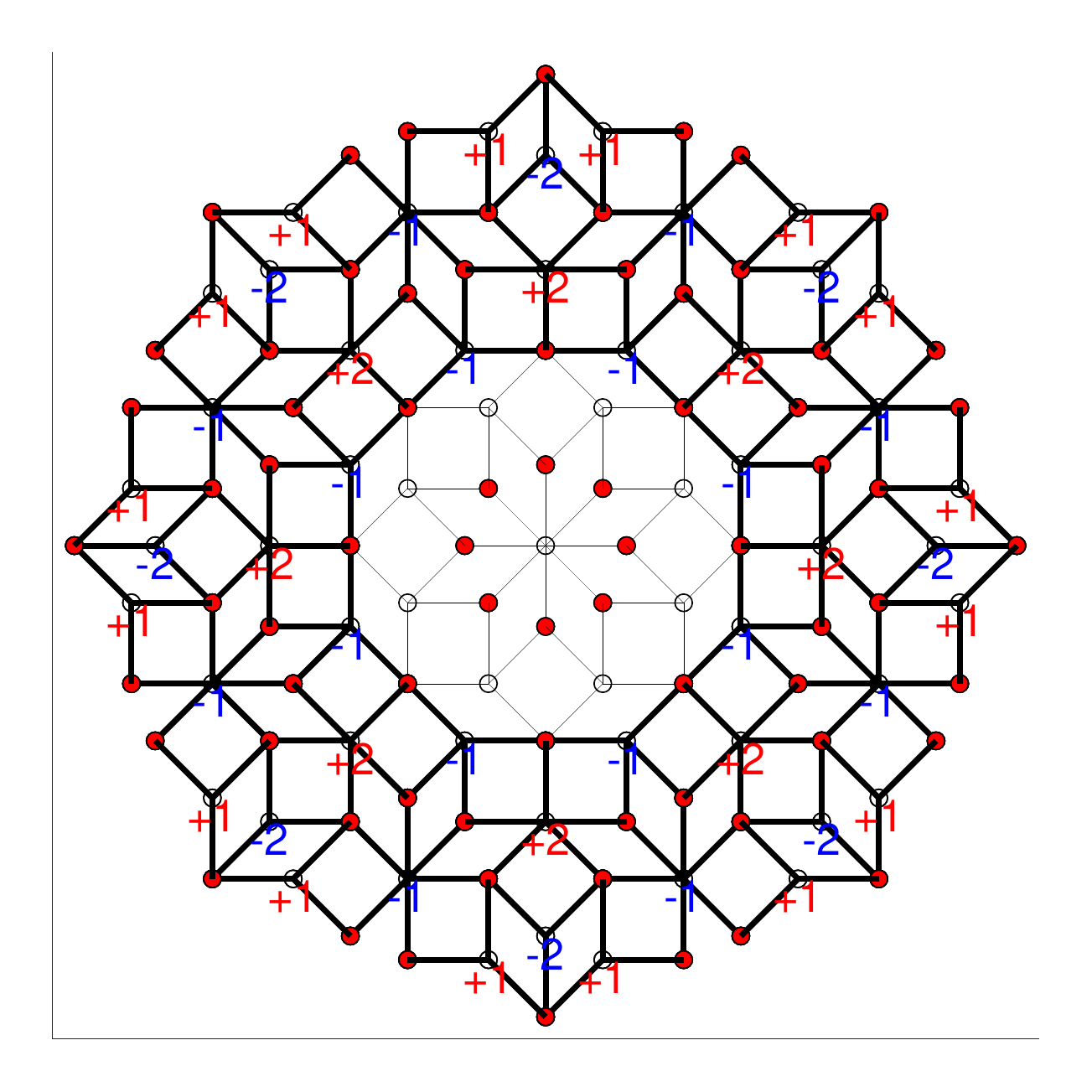}\\
    \includegraphics[trim=8mm 8mm 8mm 8mm,clip,width=0.4\textwidth]{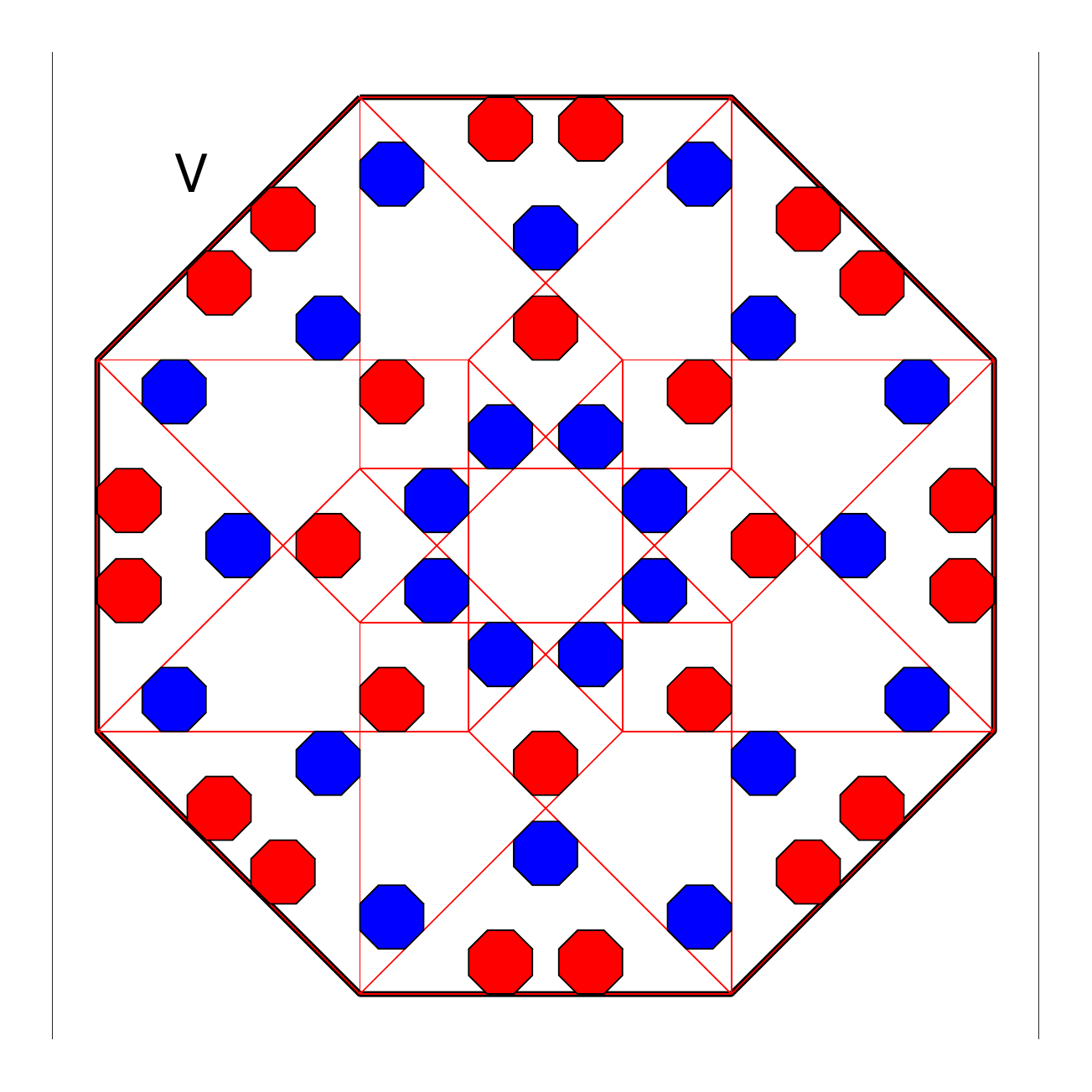}
    \caption{Type-F LS wavefunction and perpendicular space allowed regions. Type-F has 5-edge vertices which are not in the support of any previous types. Although it shares a significant amount of allowed regions with type-E, rotational symmetry proves these types' orthogonality. While they are independent, type-F and type-A need not be orthogonal, as can be seen by overlapping allowed areas of three edge vertices. }
    \label{fig:TF_RealSpace}
\end{figure}

The previous section's numerical method allows us to identify twenty LS types. In each case, the numerically obtained wavefunction can easily be verified without worrying about the accuracy of the energy eigenvalue, as the amplitudes of an LS can be taken as integers. We display the wavefunctions of types A to F  in Figs.\ref{fig:TA_RealSpace}-\ref{fig:TF_RealSpace} and the remaining fourteen types in the appendix. Once the wavefunction is determined in real space, we use the perpendicular space image of its support to calculate its exact frequency. We display the perpendicular space allowed regions for each LS next to its real space form.

The existence of LS in the ABL was apparent in numerical studies of the spectrum.\cite{jpi06}. The most abundant LS, which we call type-A and type-B, are immediately obvious by comparison with the LS of the Penrose lattice. However, we find significant differences between the LS of the Penrose lattice and ABL when all of the zero-energy manifold is investigated.

For the Penrose lattice, four out of the six types of LS break the five-fold rotational symmetry. While five-fold symmetry can be restored with a redefinition of the LS types, breaking of the symmetry results in more compact LS, which yields LS types with higher frequency. All twenty LS types of the ABL we identify below have eight-fold symmetric densities around a T1 eight edge vertex. The LS found from our numerical calculations always had support which encircles at least one eight edge vertex. While the eight-fold symmetry can be reduced by linear combinations of the LS types we give below, this does not decrease but rather increases the number of sites in the support. 

The support of LS for the Penrose lattice excluded some vertex types \cite{mok20}. Local connectivity ruled out specific regions of the perpendicular space, resulting in forbidden sites. For the ABL, we find that every type of vertex, from T6 with three neighbors to T1 with eight neighbors, can support an LS. While the vertices with a smaller number of edges are more likely to harbor LS, even eight-edge and seven-edge vertices have LS, as shown in types G and L. We have not been able to exclude any regions in perpendicular space as forbidden sites using the methods developed for the Penrose lattice. 


We count the frequency of any type of LS by identifying the allowed region for one of the vertices in the LS's support. Due to the eight-fold symmetry of all our LS types, the allowed areas are always octagons. The ratio of the area of an allowed octagon to the area of $V$ gives the LS frequency. In table \ref{tb:LStable} we give a summary of the properties of the LS types, together with their frequencies. 

The most abundant LS are the type-A and type-B which are found around each T1 vertex as given in Fig.\ref{fig:TA_RealSpace} and Fig.\ref{fig:TB_RealSpace}. Both have eight next-nearest neighbor sites arranged on a ring around a T1 vertex with the wavefunction alternating in sign. Types A and B need 2 and 3 deep neighborhoods to exist around the central T1 vertex, which does not put any further constraints on that vertex's perpendicular space position. Thus a Type-A LS can be uniquely labeled by a perpendicular space vector $|TA,\vec{r}_\perp\rangle$ where $\vec{r}_\perp$ lies in a octagon of width $\sqrt{2}-1$. The frequency of both types is equal to the frequency of T1 vertices
\begin{equation}
    f_{1}=(\sqrt{2}-1)^4=17-12\sqrt{2} \simeq 2.9437 \; 10^{-2} .
\end{equation}
Types A and B are independent of each other as one is defined only on three edge vertices while the other's support consists of four edge vertices. 

The next set of LS are formed by types C,D,E, and F. All of which have the same size perpendicular space octagons as allowed regions, and corresponding frequencies
\begin{equation}
    f_{2}=(\sqrt{2}-1)^6= 99-70\sqrt{2}\simeq 5.0506 \; 10^{-3}.
\end{equation}
The real space configuration and their allowed areas in perpendicular space are given in Figs.\ref{fig:TC_RealSpace},\ref{fig:TD_RealSpace},\ref{fig:TE_RealSpace},\ref{fig:TF_RealSpace}.These four types have allowed regions outside what was covered by types A and B, consequently, they are independent of these types. Among each other, types C to F are not only independent but orthogonal. Types C and D do not overlap with E and F in perpendicular space. Types C and F have wavefunctions that are unchanged under a $\pi/4$ rotation, while types D and E acquire a $-1$ sign under the same rotation, which establishes mutual orthogonality. These states' support contains 3, 4, 5, and 6 edge sites, but not 7 or 8 edge sites.   

The next set of LS has 6 types all of which have the frequency
\begin{equation}
    f_{3}=(\sqrt{2}-1)^8=577-408\sqrt{2}\simeq 8.6655 10^{-4},
\end{equation}
and their real and perpendicular configurations are given in Figs. \ref{fig:TG_RealSpace},\ref{fig:TH_RealSpace},\ref{fig:TI_RealSpace},\ref{fig:TJ_RealSpace},\ref{fig:TK_RealSpace},\ref{fig:TL_RealSpace}. Independence of each type from others can be established by investigating the unique perpendicular space regions covered by each type, as well as their symmetry under rotation. For example type-G has 8-edge vertices in its support, and type-L has 7-edge vertices making them independent from all other types. 

Finally we identify 8 more LS types with the frequency
 \begin{equation}
    f_{4}=(\sqrt{2}-1)^{10}=3363-2378\sqrt{2}\simeq 1.4868 10^{-4},
\end{equation}
which are displayed in the appendix. The sum of the frequencies of these 20 LS give us a lower bound
\begin{equation}
    f_{LS}\geq 30796-21776 \sqrt{2} \simeq    0.08547.
\end{equation}

In a recent paper\cite{kog20} Koga has used deflation of the neighborhood around a T1 vertex to obtain larger structures with up to $\sim 10^9$ vertices. By systematically counting the appearance of smaller structures within the larger lattices and exact diagonalization of the Hamiltonian he was able to identify new LS appearing at each deflation. The numerical structure for the first 7 deflations fit a pattern of $m(m+1)$ independent LS types for the $m^{th}$ deflation of the lattice. Our results which are obtained for general neighborhoods rather than a symmetric lattice around a T1 vertex display a similar pattern. 

Based on the patterns observed in table \ref{tb:LStable}, we can conjecture that there are infinitely many independent LS types in the ABL which are organized into generations. At the $m^{th}$ generation there are $2m$ LS types. The allowed perpendicular space region for the each one of these $2m$ LS types is an octagon of width $(\sqrt{2}-1)^m$, corresponding to a frequency of $f_m=(\sqrt{2}-1)^{2m+2}$. Each generation has equal number of LS in both sublattices. If the generation index $m$ is even, LS in each sublattice have $m/2$ types with wavefunctions which are symmetric under $\pi/8$ rotation and $m/2$ types which are anti-symmetric. For odd $m$, $(m+1)/2$ of the LS in a sublattice have eigenvalue $-1$ under eight-fold rotation, and $(m-1)/2$ have eigenvalue $+1$. If this pattern holds for every generation then the total frequency of LS is 
\begin{equation}
    f_{Ex}=\sum_{m=1}^{\infty} 2 m f_m =  3/2 -\sqrt{2},
\end{equation}
as found in \cite{kog20}.

Another interesting property of the LS of the ABL are their almost uniform density. Unlike the Penrose lattice LS types, all the twenty LS types have wavefunctions made from only $\pm 1$ and $\pm 2$. Furthermore if we relax the 8 fold rotation symmetry we can find a basis of independent LS which have entirely uniform density on their support. 

We can compare this value with the results of the numerical calculation as given in Fig.\ref{fig:Frequency_ABL}, and see that numerical calculation approaches the analytical values from below as the neighborhood depth is increased. Our numerical method cannot count the LS which cross the boundary of the finite lattice we are using. Thus the deviation from the exact result should be essentially a boundary effect. We expect the numerically calculated frequency to approach the exact result with $1/R$ decay as a function of the radius $R$ of the neighborhood. More precisely we can estimate the number of a specific LS type which cross the boundary of the neighborhood as follows. If a LS type has frequency  $f_{x}$ and a support with radius $r_{x}$ the approximate number of LS of this type crossing the boundary would be $N_{miss}\simeq f_x 2 \pi R r_x$. Hence the numerically estimated frequency would be lower than the exact frequency by $f_{est}\simeq f_{exact}- f_x \frac{r_x}{R}$. This estimate assumes that the LS type is uniformly distributed over the lattice, this assumption is violated for the Penrose lattice which has natural boundaries for LS in the form of strings. We observe that there are large deviations in the numerical estimate for the Penrose lattice in Fig.\ref{fig:Frequency_PL} which is a consequence of strings crossing the boundary. 
For the ABL, a $1/R$ fit is consistent with the exact result being equal to the values given above. However, the fits are not restrictive enough to rule out a missing frequency of the order of $ 0.1 \% $.

\section{Conclusion}
\label{sec:Conclusion}

\begin{figure}[!htb]
    \centering
    \includegraphics[trim=8mm 8mm 8mm 8mm,clip,width=0.46\textwidth]{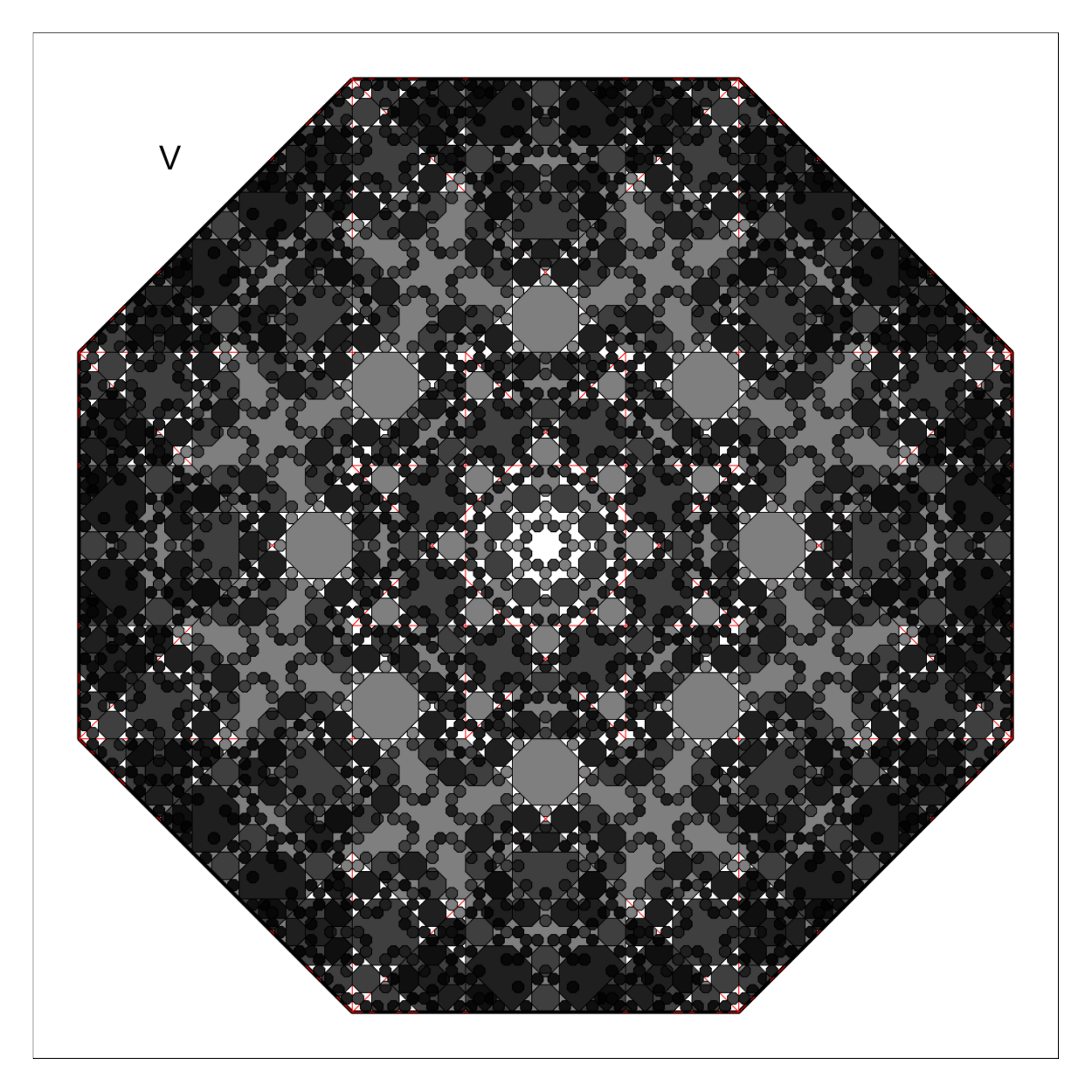}
    \caption{Allowed regions for all the LS superimposed. Most of the area inside V is covered with at least one LS allowed region. All vertex types can appear in the LS support. However, none of the regions corresponding to a particular vertex type is wholly covered. Especially the eight edge vertices close to the center of $V$ form the largest uncovered area. }
    \label{fig:AllT_PerpSpace}
\end{figure}

Motivated by the recent interest in the realization of quasicrystalline symmetry in engineered quantum systems, such as the eight-fold quasicrystal potential for cold atoms \cite{vie19}, we investigated the strictly localized states of the Ammann-Beenker lattice in the tight-binding limit. 

 We constructed finite lattices made up of all sites that can be reached from a given lattice point with a fixed number of hops to the nearest neighbors. This construction was efficiently carried out by considering the perpendicular space images of lattice points. Combining this neighborhood with bipartite symmetry enables the identification of LS without the inclusion of any edge states.  We calculated the null space of the Hamiltonian using a numerical method based on a sparse matrix QR decomposition algorithm. This method allowed us to count the number of LS in lattices of up to 100 000 sites.  Our numerical result indicates $f_{LS}\simeq 0.08338$ for the frequency of LS.
 
 We investigated the LS's general structure by calculating the eigenstates of the position operator projected into the zero energy manifold. As a result, we identified twenty independent LS types, all of which have eight-fold rotational symmetry around an eight-edge vortex. We calculated the frequency of each LS by identifying allowed perpendicular space areas for vertices in their support. When those areas are superimposed for all LS as in Fig.\ref{fig:AllT_PerpSpace}, one can see that most of the vertices in the ABL support at least one LS and all vertex types can host LS. There are areas in perpendicular space for which neither found an LS nor could out rule the existence of an LS.  The total frequency of LS gives $f_{LS}=30796-21776\sqrt{2} \simeq 0.08547$.
 
 This value is very close to  the recent conjecture for the exact fraction $f_{Ex}=3/2-\sqrt{2}$, and also the numerical results. The general organization of the LS types are parallel with Ref.\cite{kog20}, and our results for the fourth generation of LS as well as the patterns observed in the first four generations can be considered as further evidence for this conjecture. Both the uniformity of the LS for the ABL, and symmetry around an eightfold vertex are in marked contrast with the LS of the Penrose lattice. The reason for this contrast is not clear, indicating that the connection between the definition of a vertex model on a quasicrystalline lattice and its spectrum requires further investigation.      
 
We demonstrated the use of the perpendicular space method in Ref\cite{mok20} in another setting. It would be interesting to see if a similar labeling method can be developed for quasicrystal eigenstates that are not strictly localized. 

\appendix*
\section{LS types in the third and fourth generation }

The fourteen types of LS in the third and fourth generation are displayed in real space below. The figures also show the allowed regions for their vertices in perpendicular space and  the caption of each figure points out the reason for the independence of the LS type. 

\begin{figure*}[h!]
    \centering
    \includegraphics[trim=8mm 8mm 8mm 8mm,clip,width=0.45\textwidth]{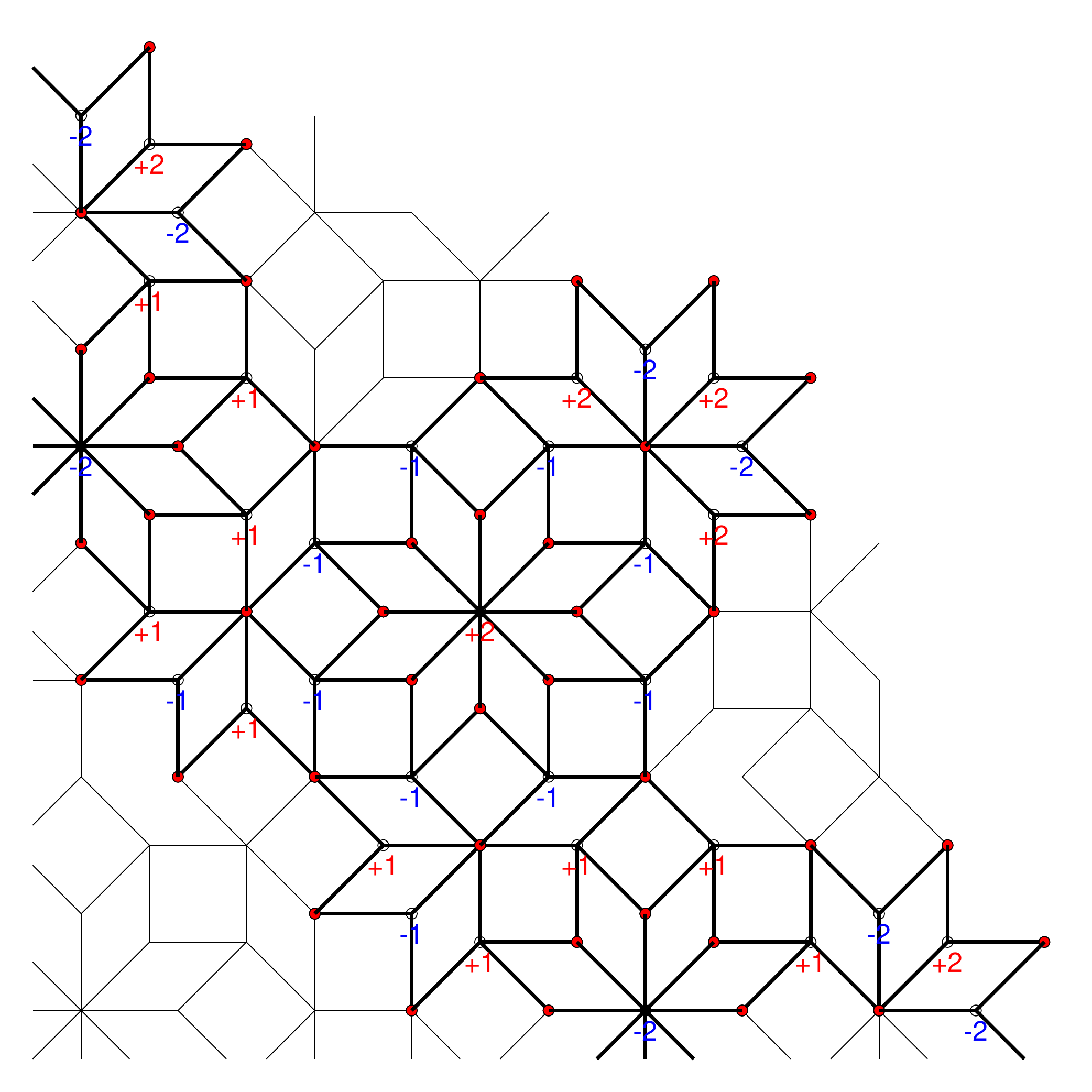}
    \includegraphics[trim=8mm 8mm 8mm 8mm,clip,width=0.45\textwidth]{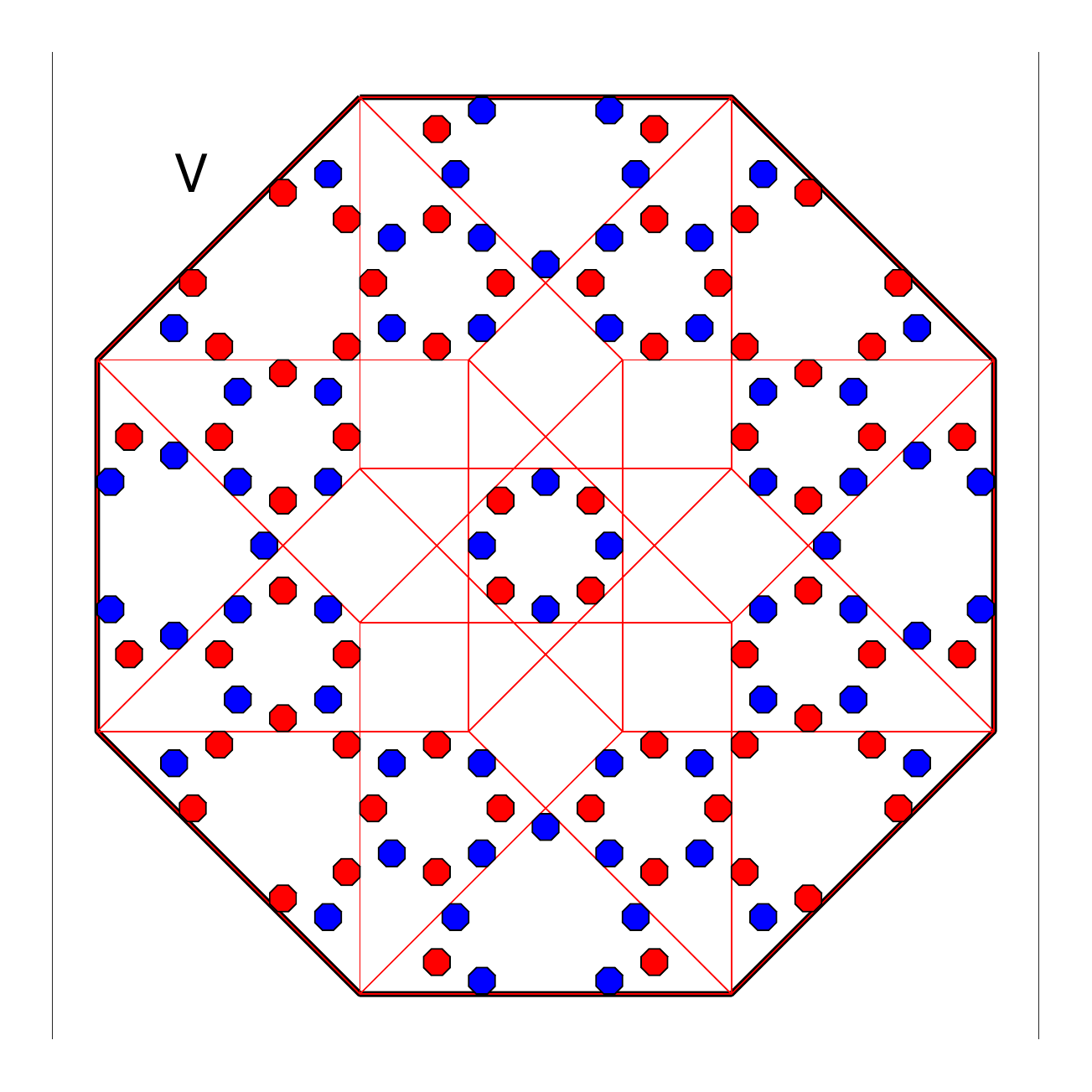}
    \caption{Type-G LS wavefunction and perpendicular space allowed regions. Only a quarter of the region in real space is shown in this and subsequent figures for visual clarity. The lattice and the state are four-fold symmetric around the eight-edge vertex in the lower-left corner. Type-G is the first LS to have support on eight-edge vertices. }
    \label{fig:TG_RealSpace}
\end{figure*}

\begin{figure*}[h!]
    \centering
    \includegraphics[trim=8mm 8mm 8mm 8mm,clip,width=0.45\textwidth]{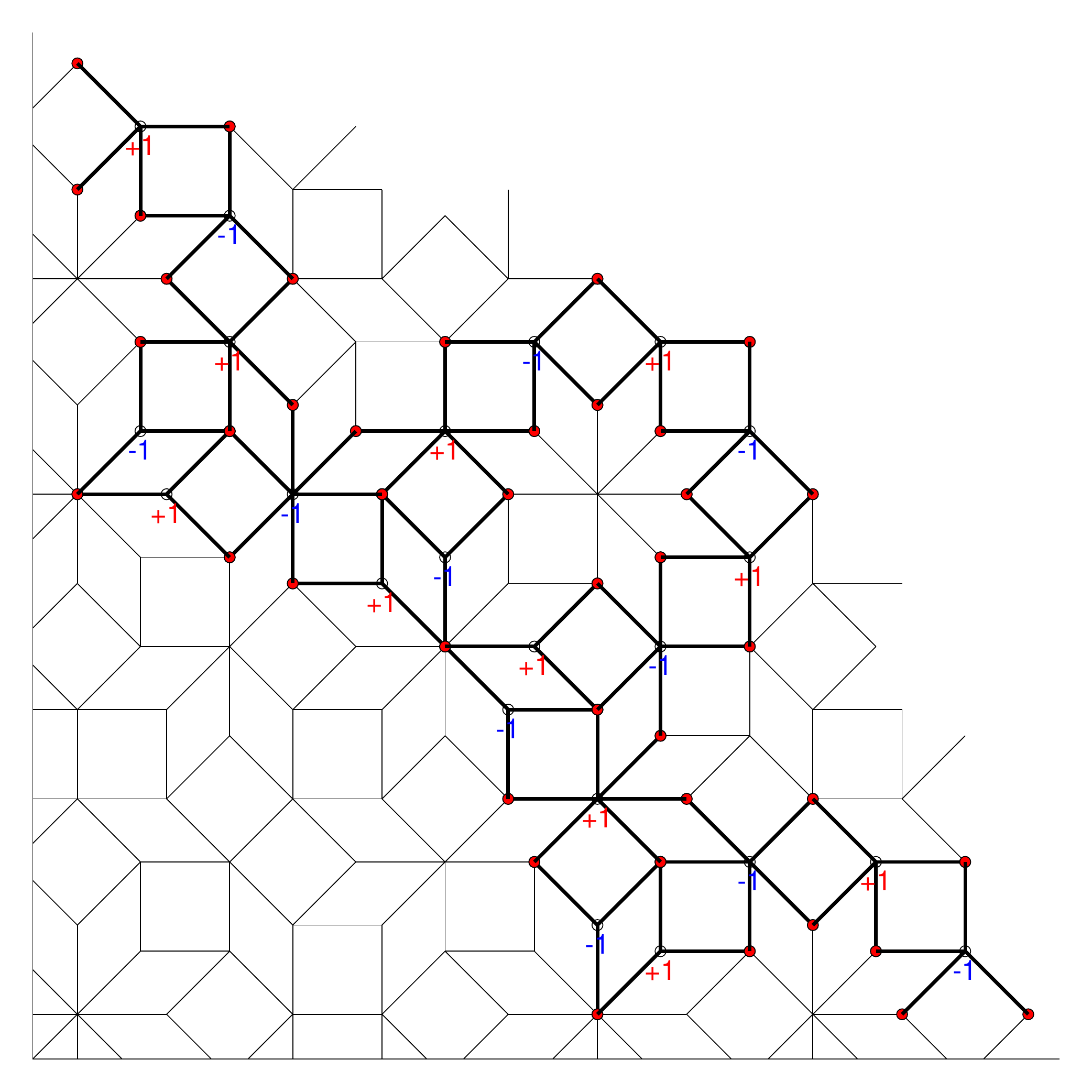}
    \includegraphics[trim=8mm 8mm 8mm 8mm,clip,width=0.45\textwidth]{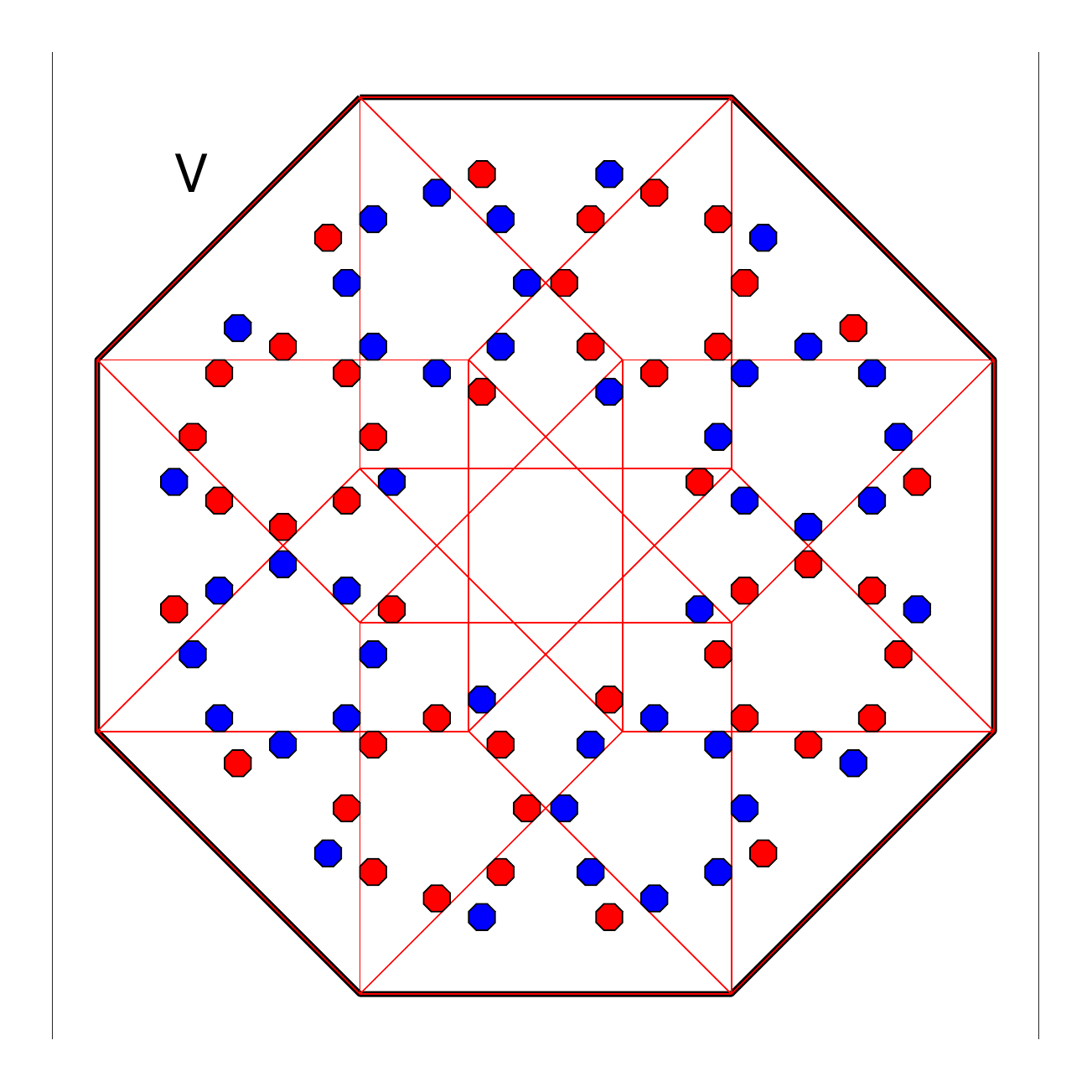}
    \caption{Type-H LS wavefunction and perpendicular space allowed regions. No previous type covers the allowed regions for both six-edge and five-edge vertices in the support.}
    \label{fig:TH_RealSpace}
\end{figure*}

\begin{figure}[!h]
    \centering
    \includegraphics[trim=8mm 8mm 8mm 8mm,clip,width=0.45\textwidth]{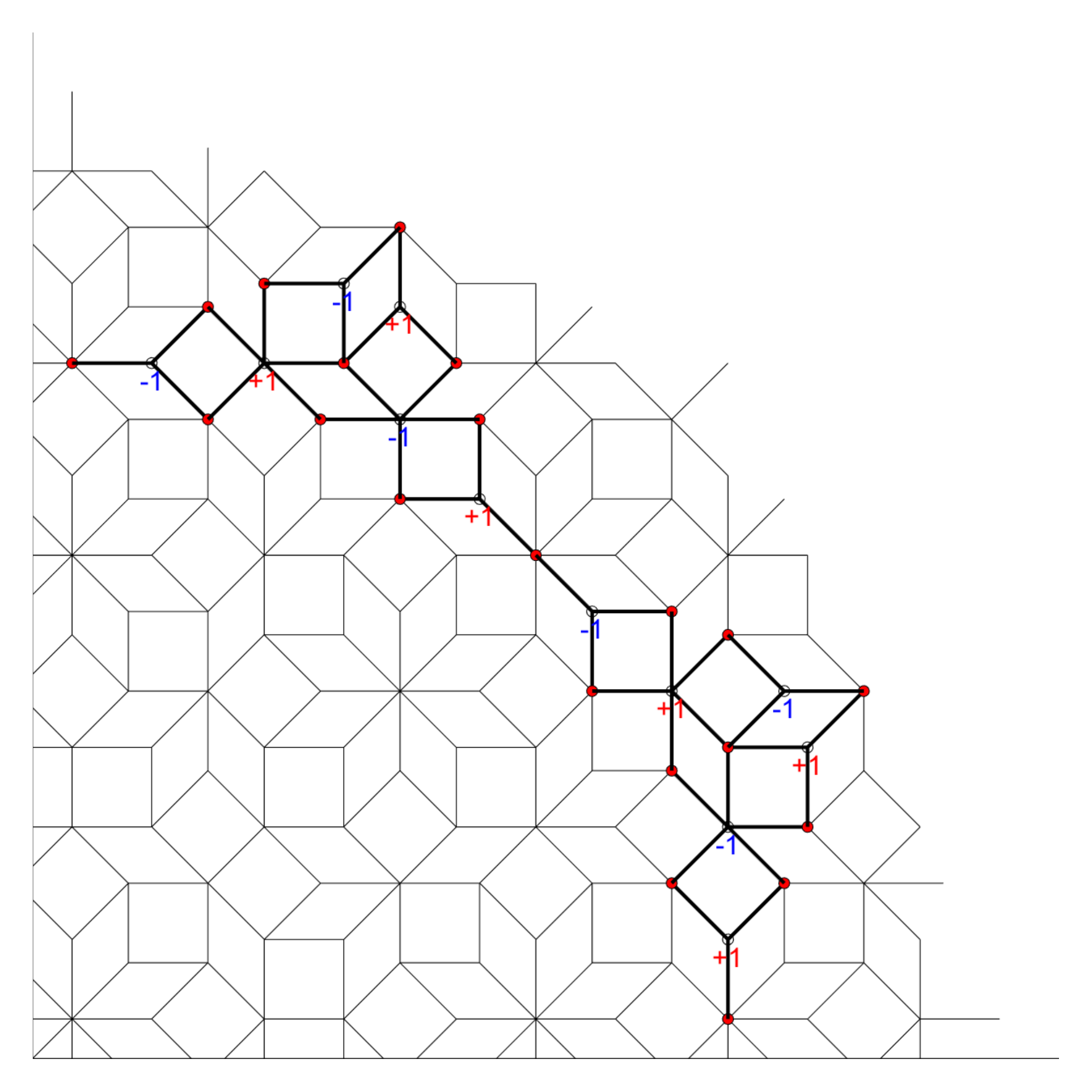}
    \includegraphics[trim=8mm 8mm 8mm 8mm,clip,width=0.45\textwidth]{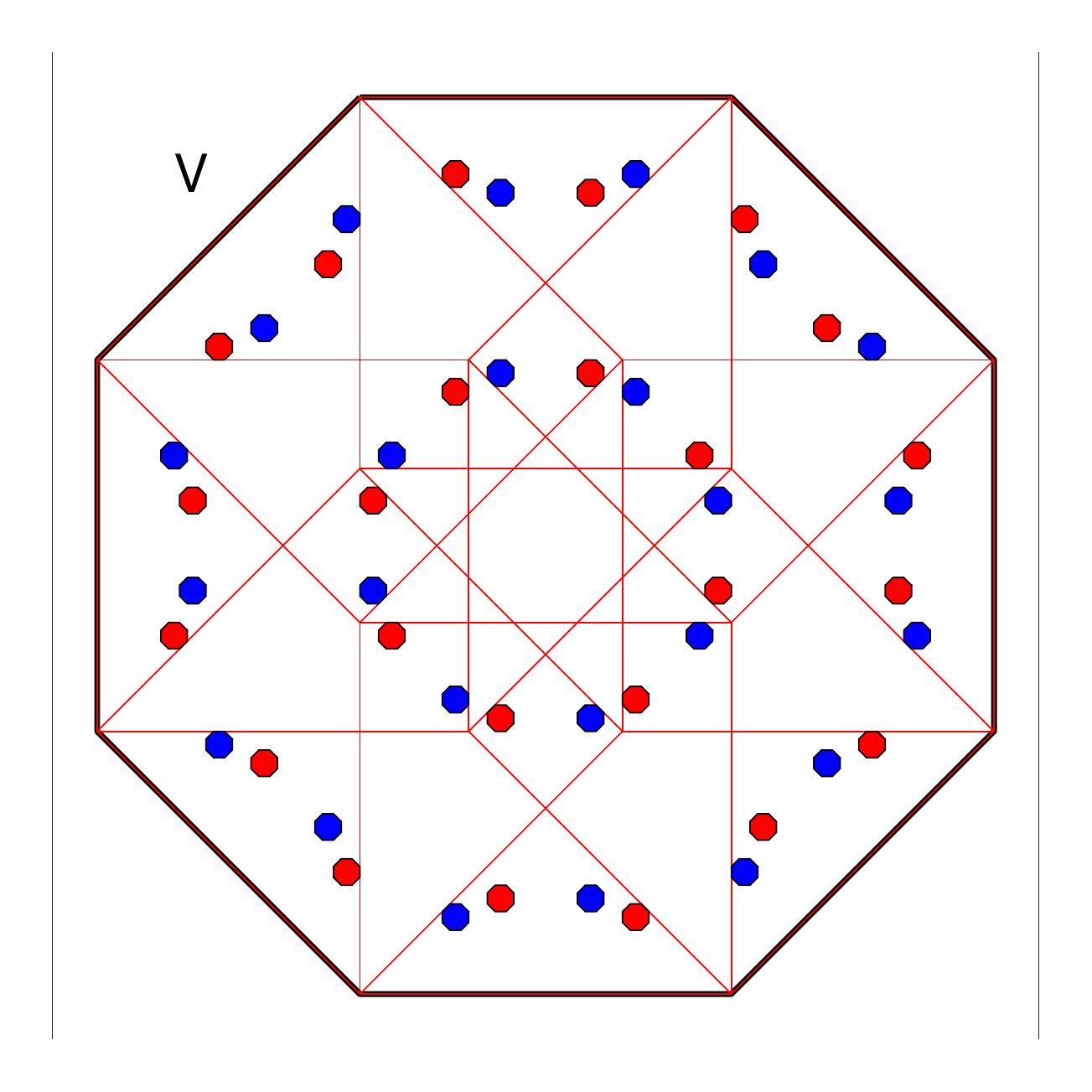}
    \caption{Type-I LS wavefunction and perpendicular space allowed regions. Independence is established by noticing that the five-edge regions in perpendicular space are not covered by previous types.}
    \label{fig:TI_RealSpace}
\end{figure}

\begin{figure}[!h]
    \centering
    \includegraphics[trim=8mm 8mm 8mm 8mm,clip,width=0.45\textwidth]{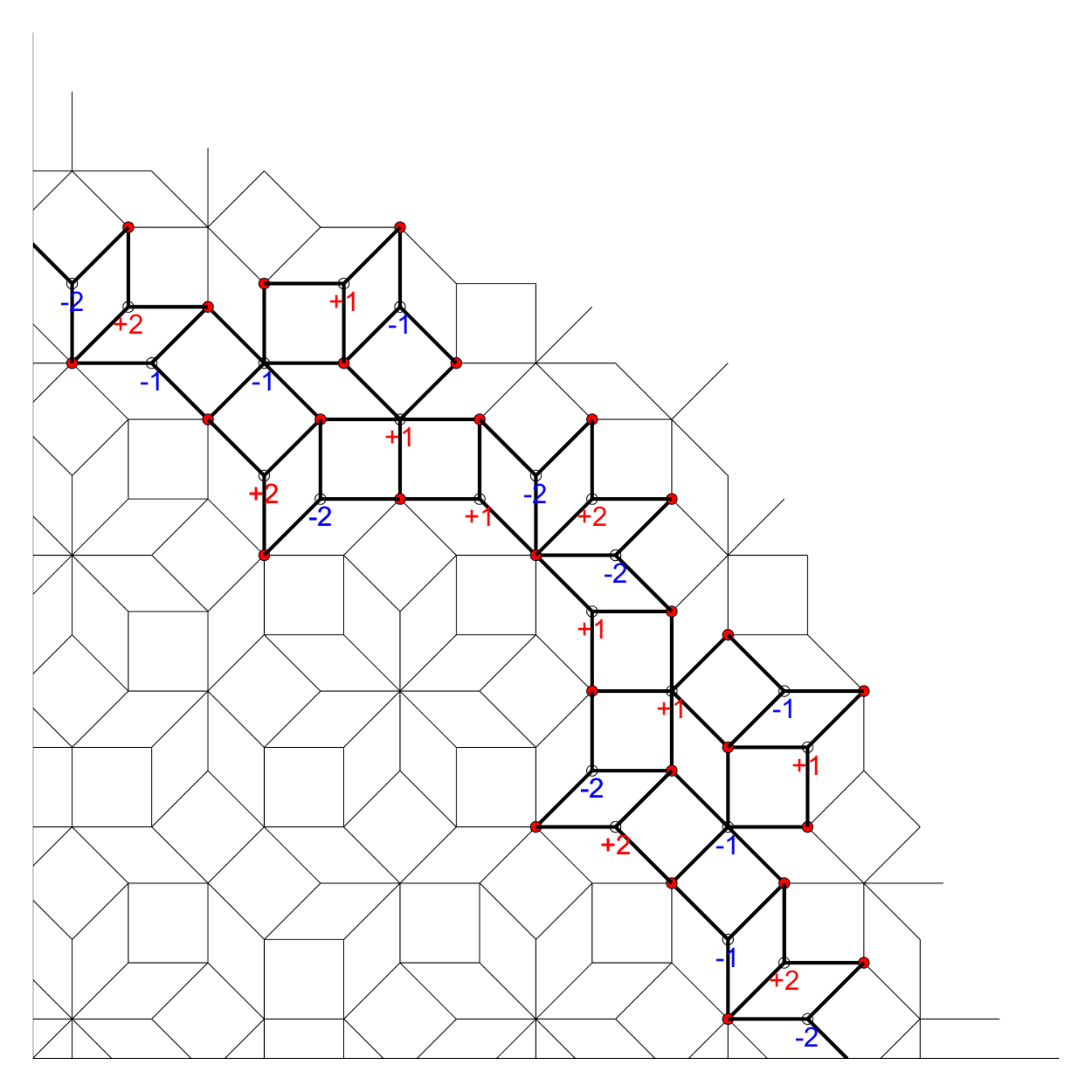}
    \includegraphics[trim=8mm 8mm 8mm 8mm,clip,width=0.45\textwidth]{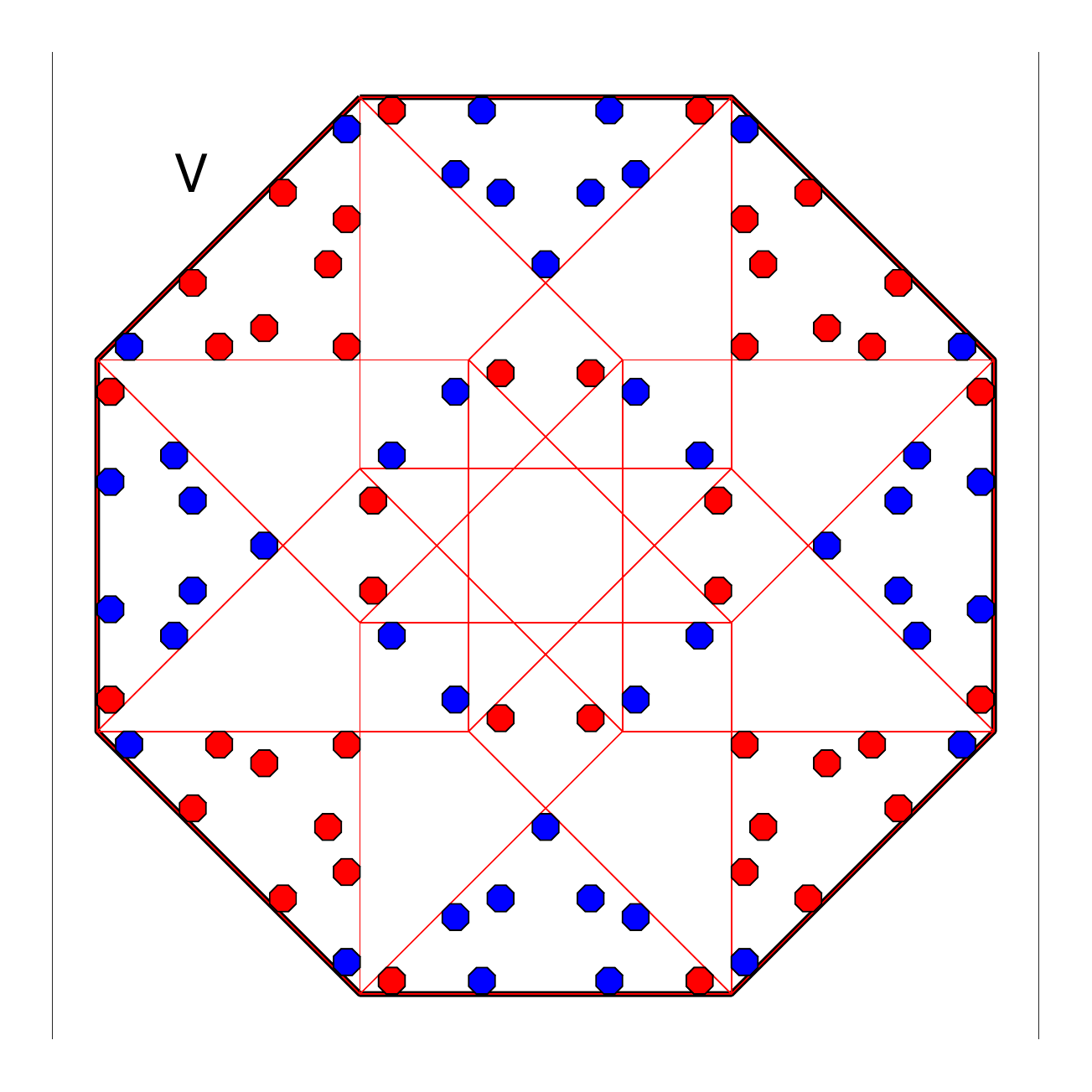}
    \caption{Type-J LS wavefunction and perpendicular space allowed regions. Type-J allowed regions cover all type-I allowed regions, including the five-edge vertices used to establish type-I's independence. These two types are orthogonal due to rotational symmetry; hence type-J is independent of all previous types. }
    \label{fig:TJ_RealSpace}
\end{figure}

\begin{figure}[!h]
    \centering
    \includegraphics[trim=8mm 8mm 8mm 8mm,clip,width=0.45\textwidth]{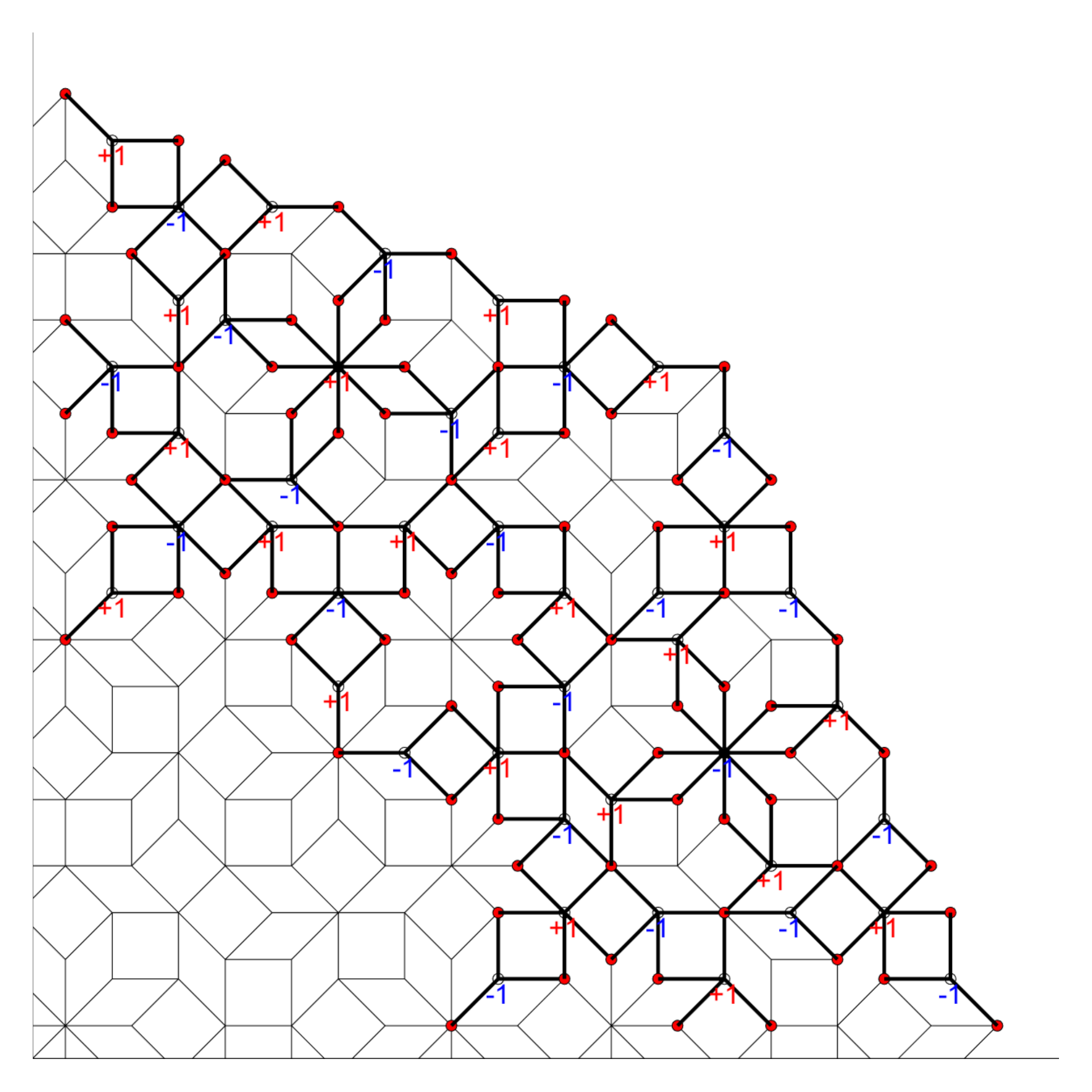}
    \includegraphics[trim=8mm 8mm 8mm 8mm,clip,width=0.45\textwidth]{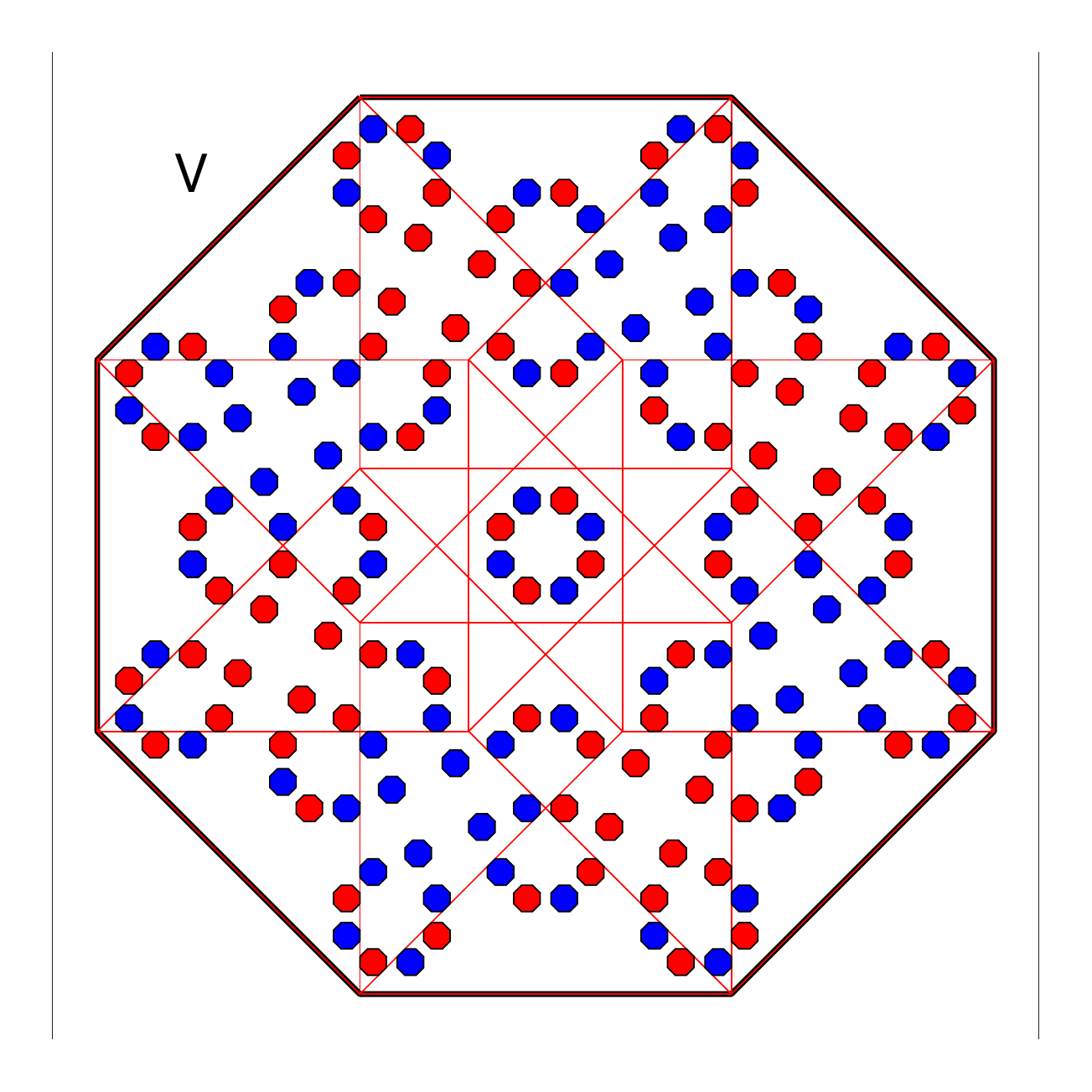}
    \caption{Type-K LS wavefunction and perpendicular space allowed regions. Any previous LS type does not cover the eight-edge vertex regions. }
    \label{fig:TK_RealSpace}
\end{figure}

\begin{figure}[!h]
    \centering
    \includegraphics[trim=8mm 8mm 8mm 8mm,clip,width=0.45\textwidth]{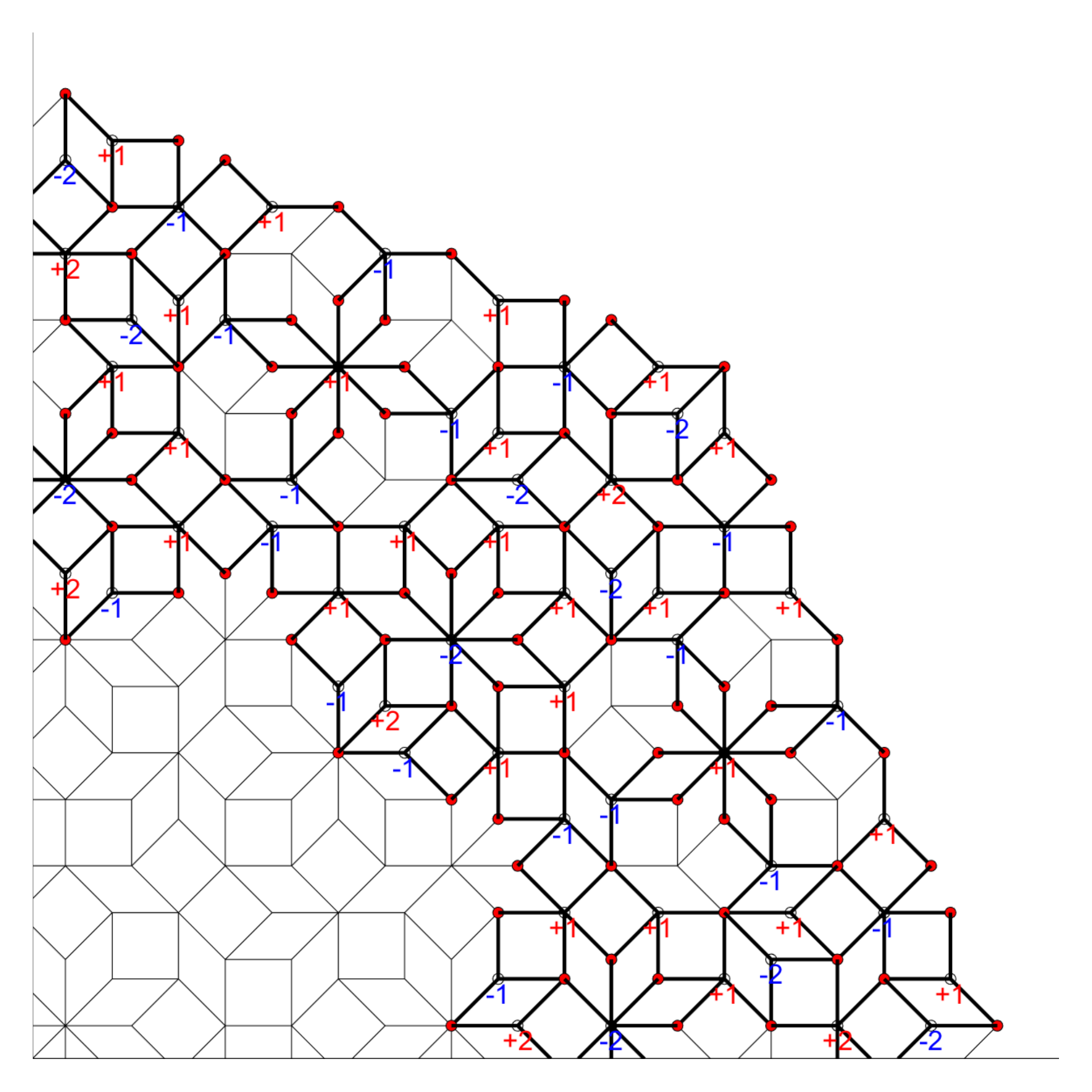}
    \includegraphics[trim=8mm 8mm 8mm 8mm,clip,width=0.45\textwidth]{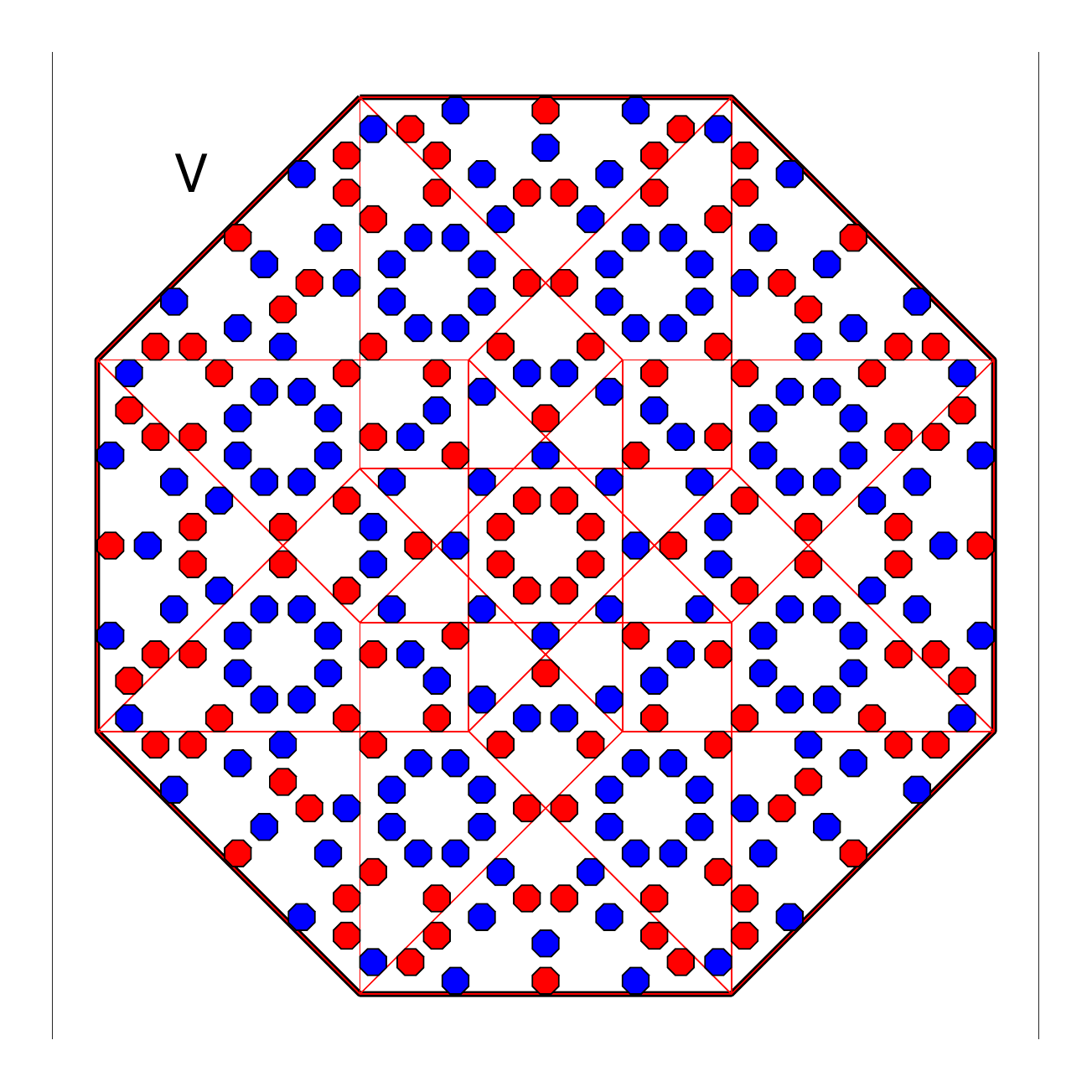}
    \caption{Type-L LS wavefunction and perpendicular space allowed regions.Type-L is the first LS type we identified which has seven-edge vertices in its support.}
    \label{fig:TL_RealSpace}
\end{figure}

%
%
%

\begin{figure*}[!h]
    \centering
    \includegraphics[trim=8mm 8mm 8mm 8mm,clip,width=0.40\textwidth]{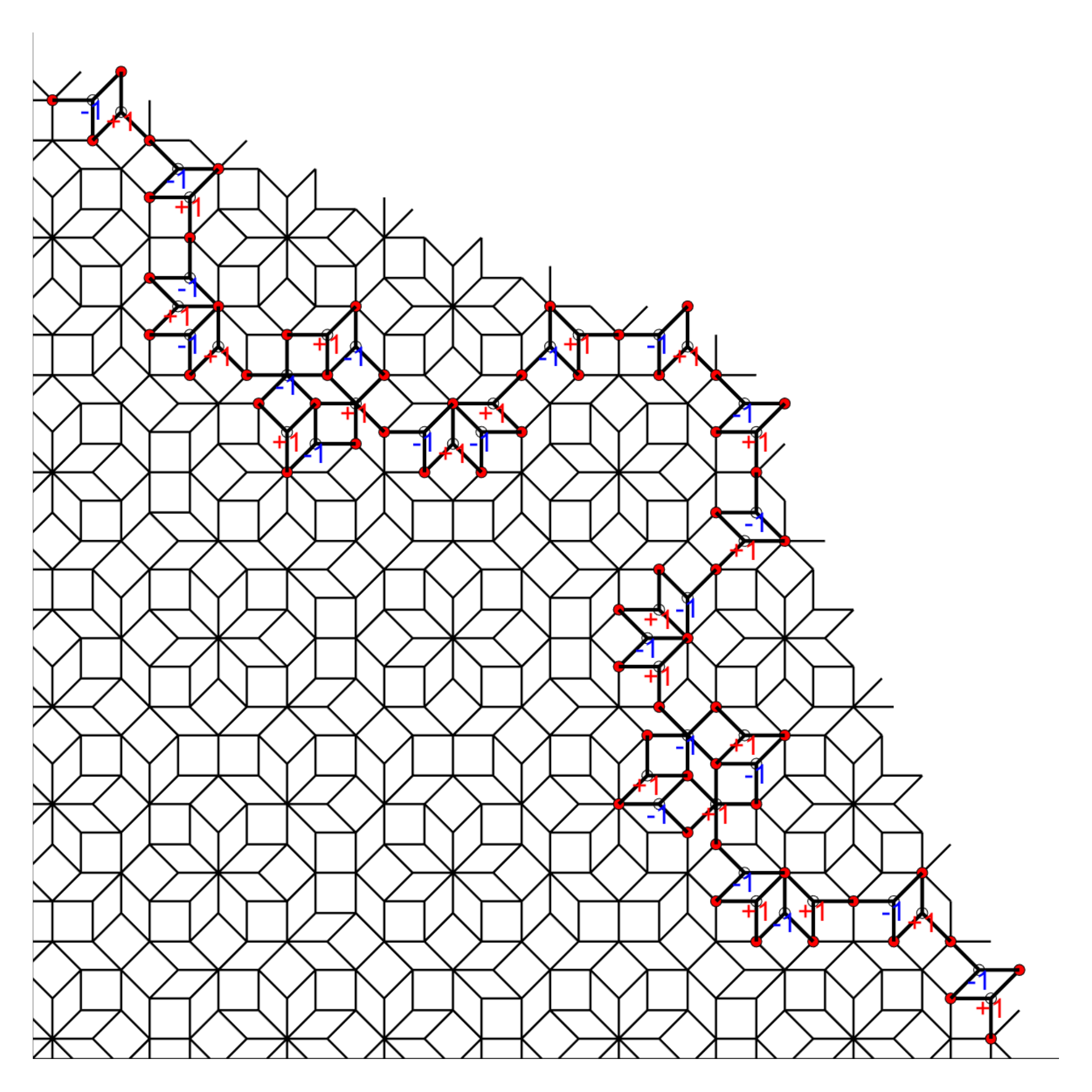}
    \includegraphics[trim=8mm 8mm 8mm 8mm,clip,width=0.40\textwidth]{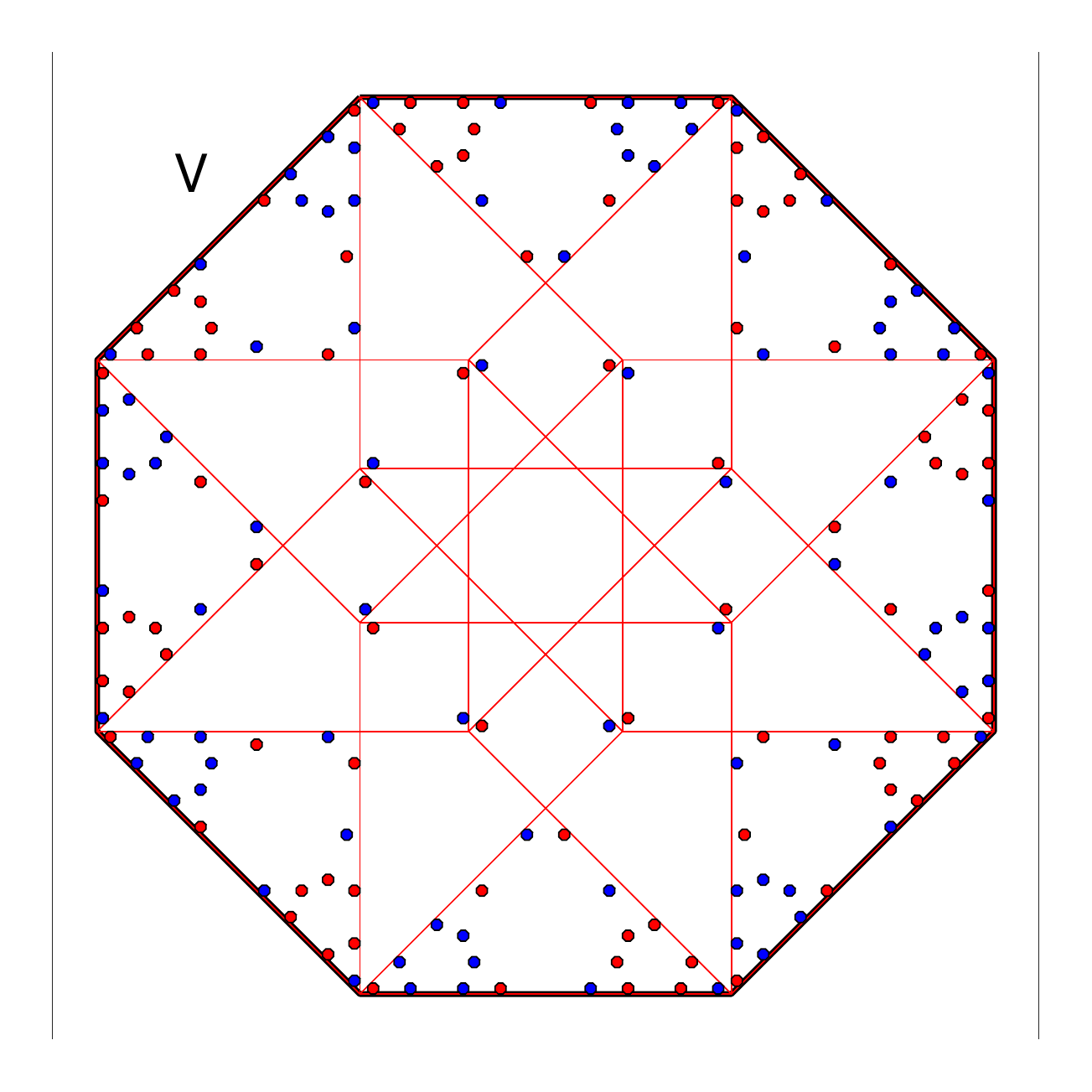}
    \caption{Type-M LS wavefunction and perpendicular space allowed regions. Independence is established through the 3-edge vertex regions farthest away from the origin. Type-N covers the same regions but is orthogonal to type-N through rotational symmetry.}
    \label{fig:TM_RealSpace}
\end{figure*}

\begin{figure*}[!h]
    \centering
    \includegraphics[trim=8mm 8mm 8mm 8mm,clip,width=0.45\textwidth]{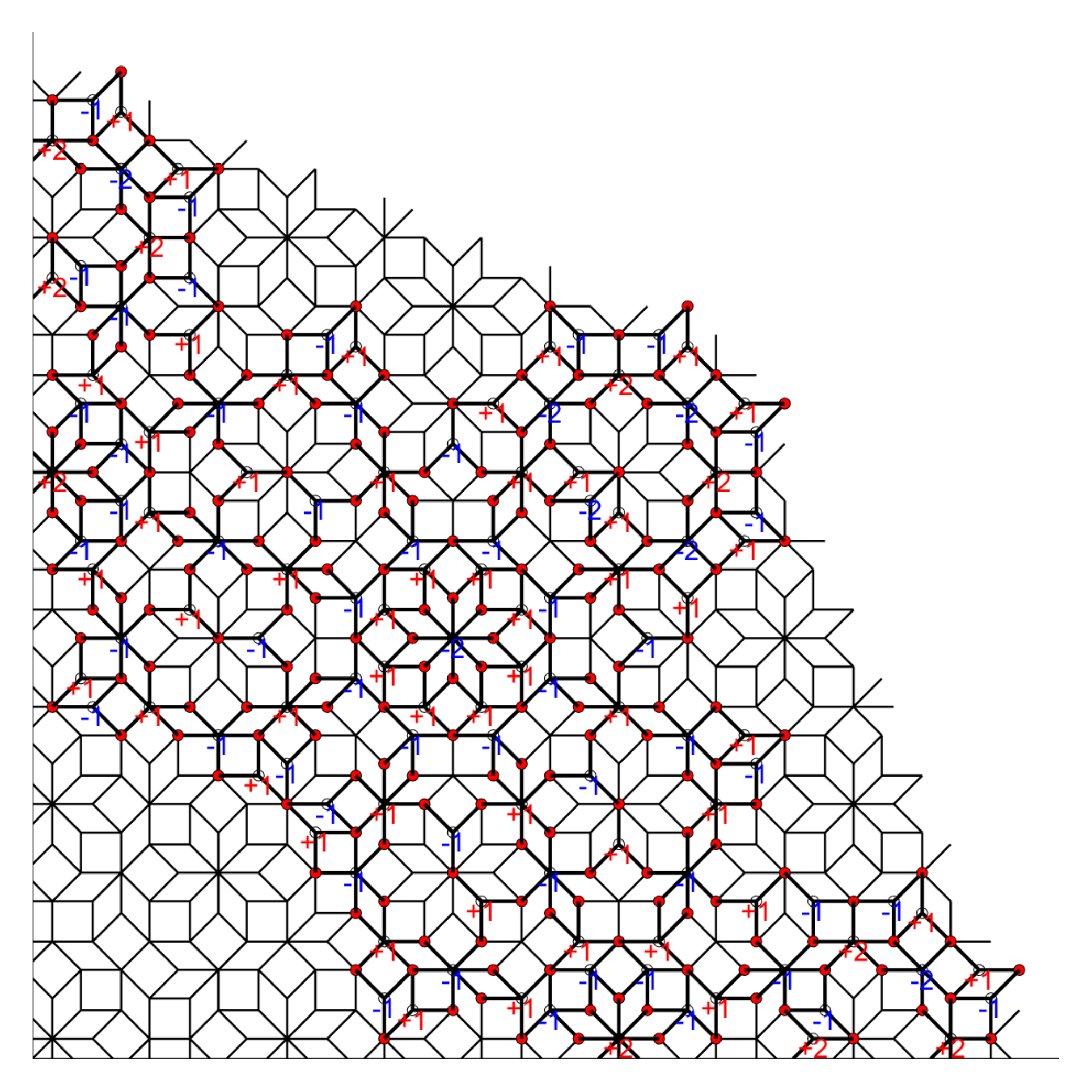}
    \includegraphics[trim=8mm 8mm 8mm 8mm,clip,width=0.45\textwidth]{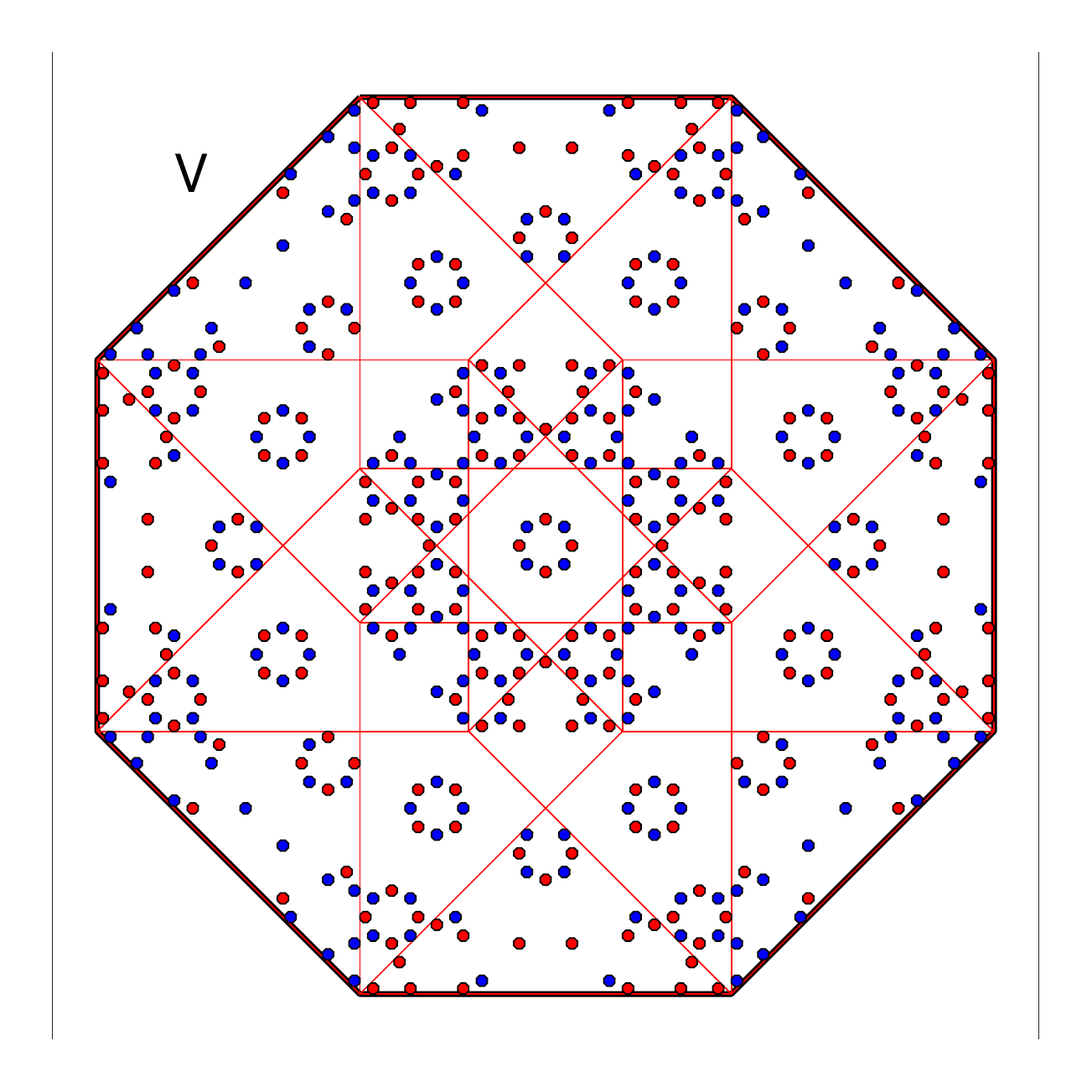}
    \caption{Type-N LS wavefunction and perpendicular space allowed regions.Type-N covers the same 3-edge regions which were first covered by type-M, but the two states are orthogonal by rotational symmetry.}
    \label{fig:TN_RealSpace}
\end{figure*}

\begin{figure*}[!h]
    \centering
    \includegraphics[trim=8mm 8mm 8mm 8mm,clip,width=0.45\textwidth]{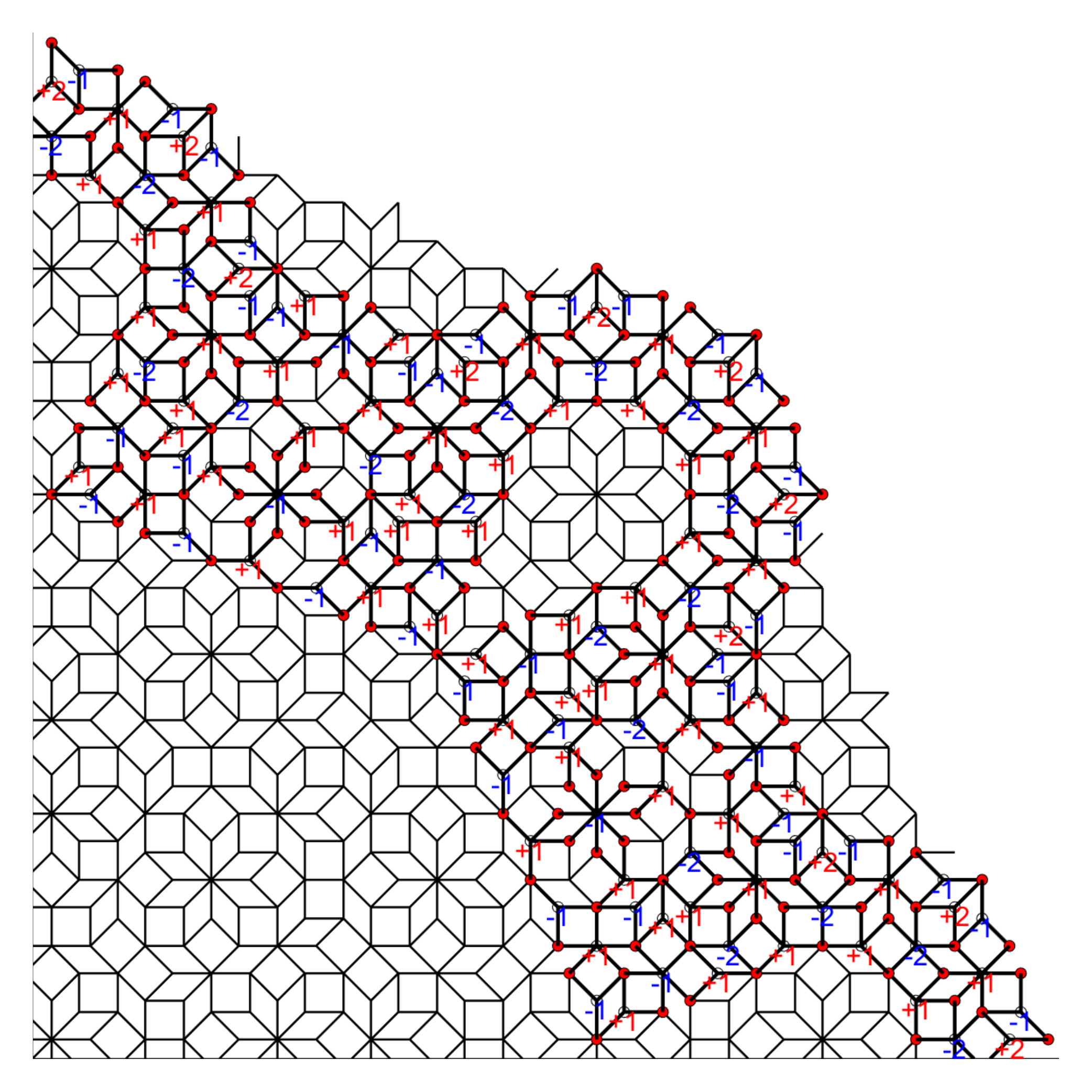}
    \includegraphics[trim=8mm 8mm 8mm 8mm,clip,width=0.45\textwidth]{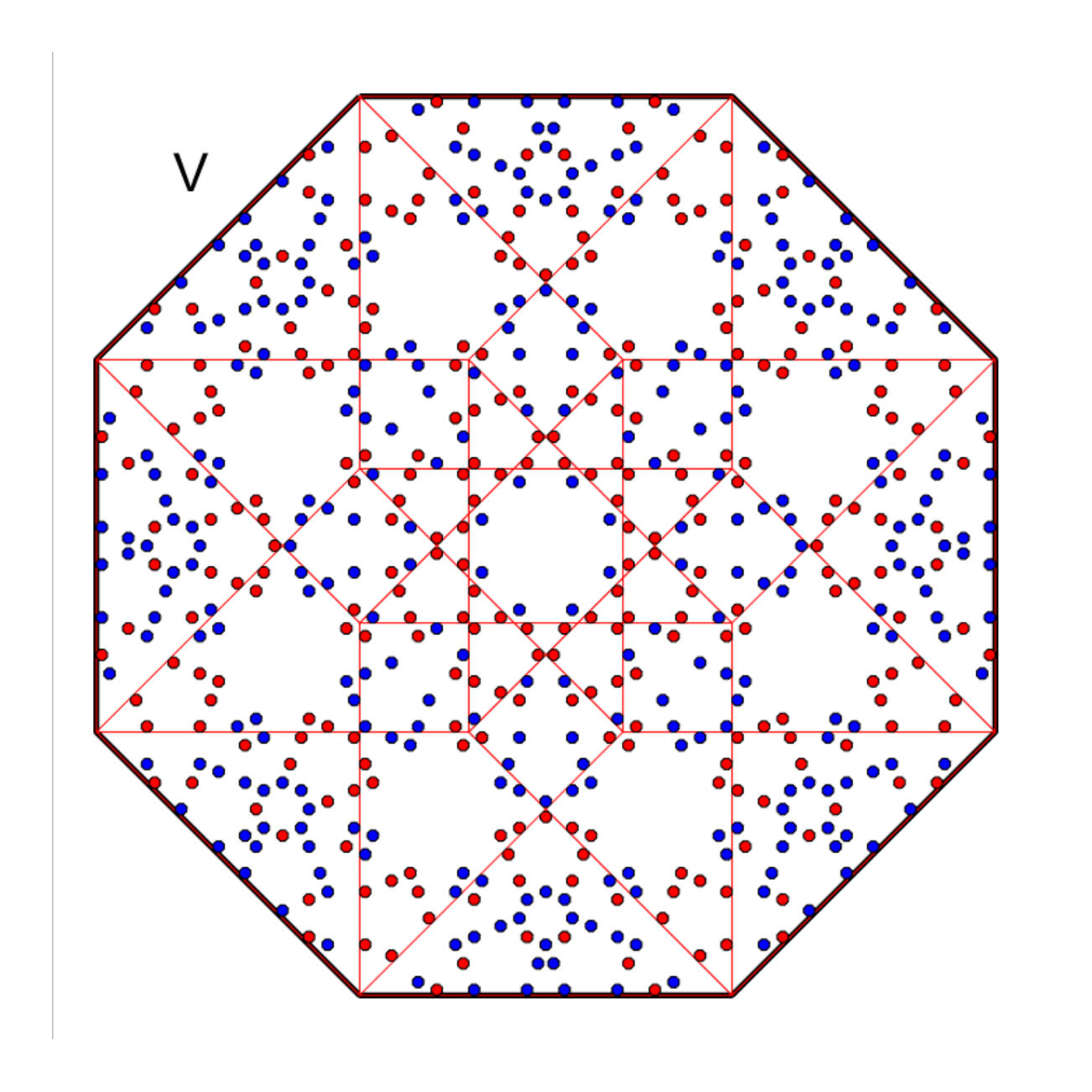}
    \caption{Type-O LS wavefunction and perpendicular space allowed regions. Type-O and Type-P cover new 7-edge regions and are both independent as they are orthogonal.}
    \label{fig:TO_RealSpace}
\end{figure*}

\begin{figure*}[!h]
    \centering
    \includegraphics[trim=8mm 8mm 8mm 8mm,clip,width=0.45\textwidth]{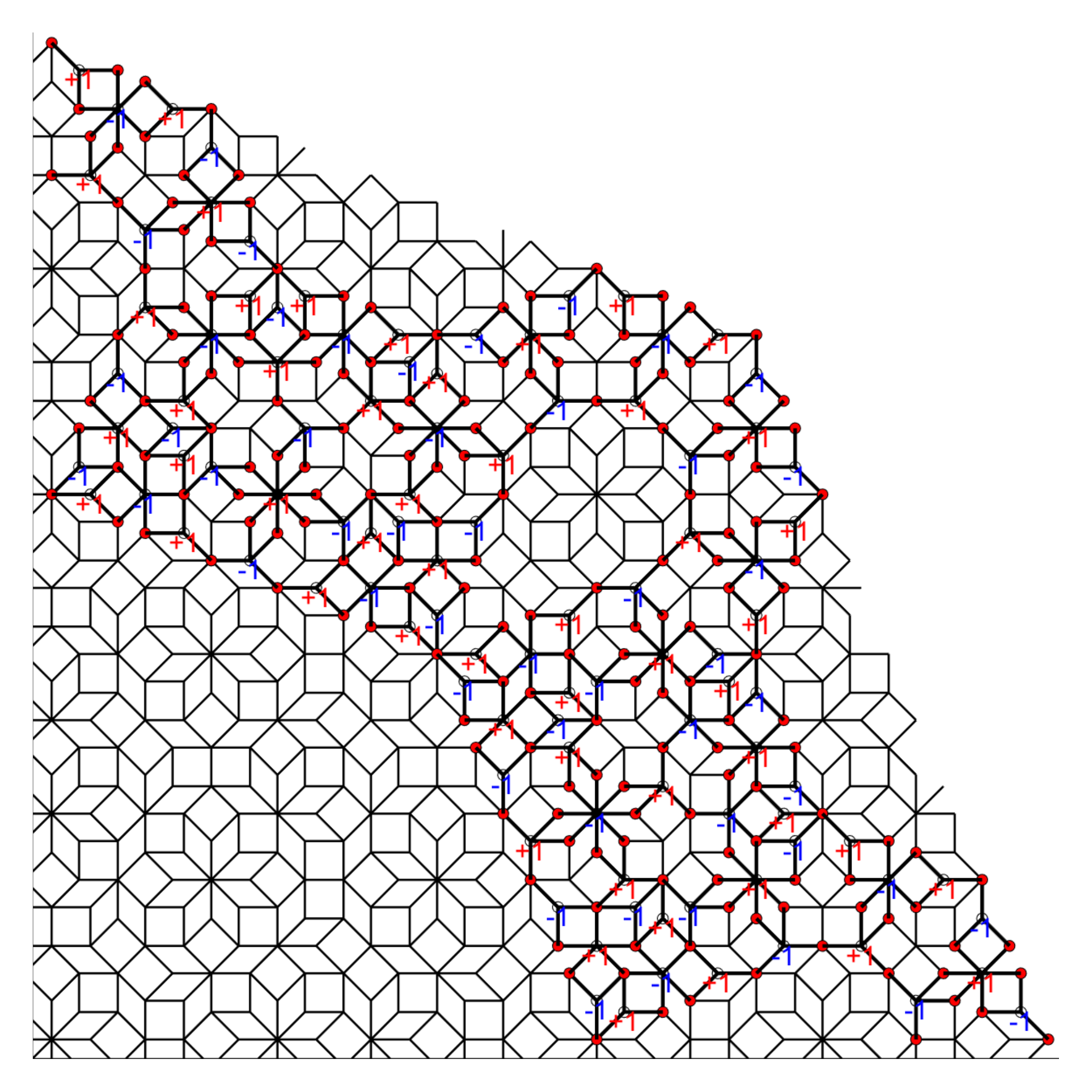}
    \includegraphics[trim=8mm 8mm 8mm 8mm,clip,width=0.45\textwidth]{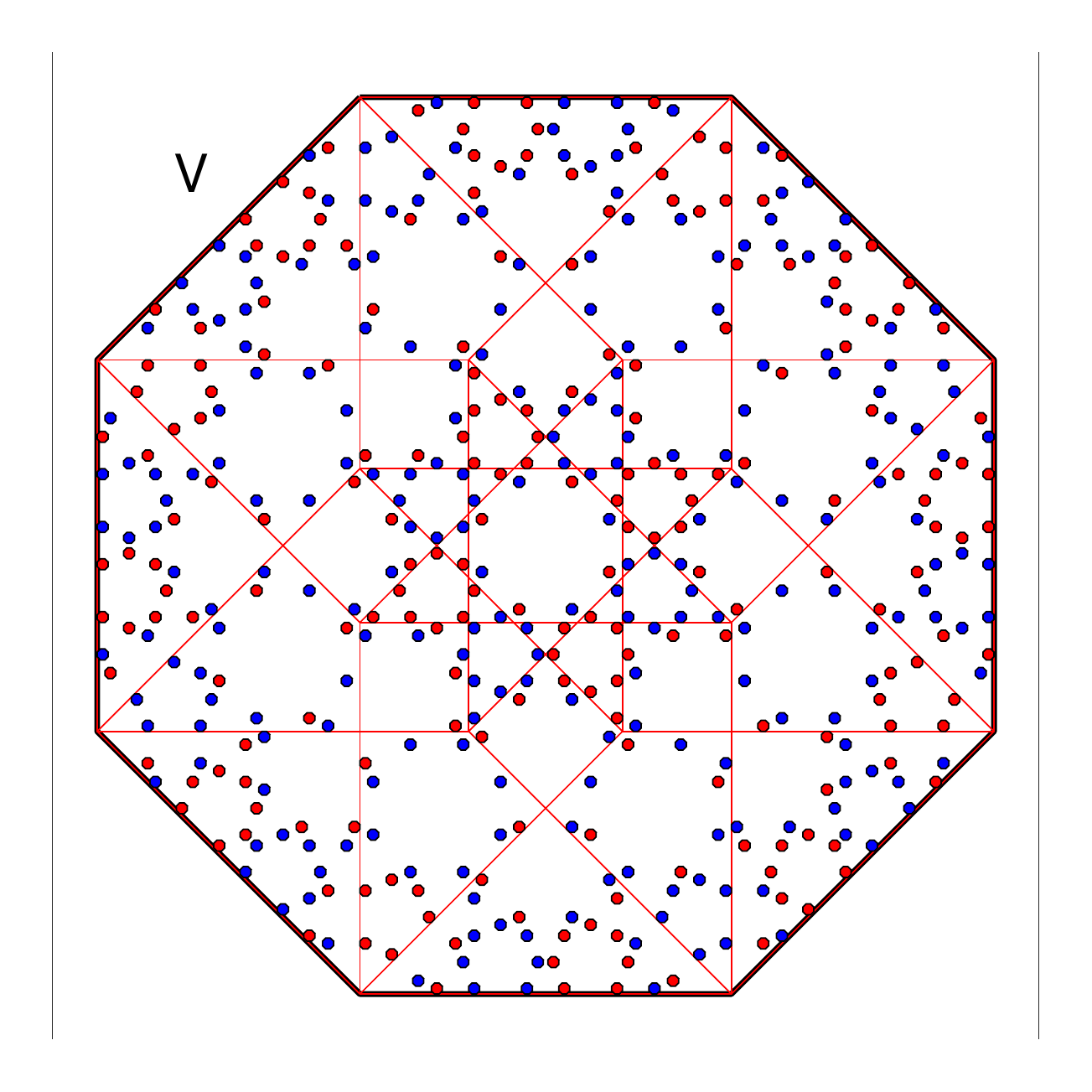}
    \caption{Type-P LS wavefunction and perpendicular space allowed regions.Type-O and Type-P cover new 7-edge regions and are both independent as they are orthogonal.}
    \label{fig:TP_RealSpace}
\end{figure*}

\begin{figure*}[!h]
    \centering
    \includegraphics[trim=8mm 8mm 8mm 8mm,clip,width=0.45\textwidth]{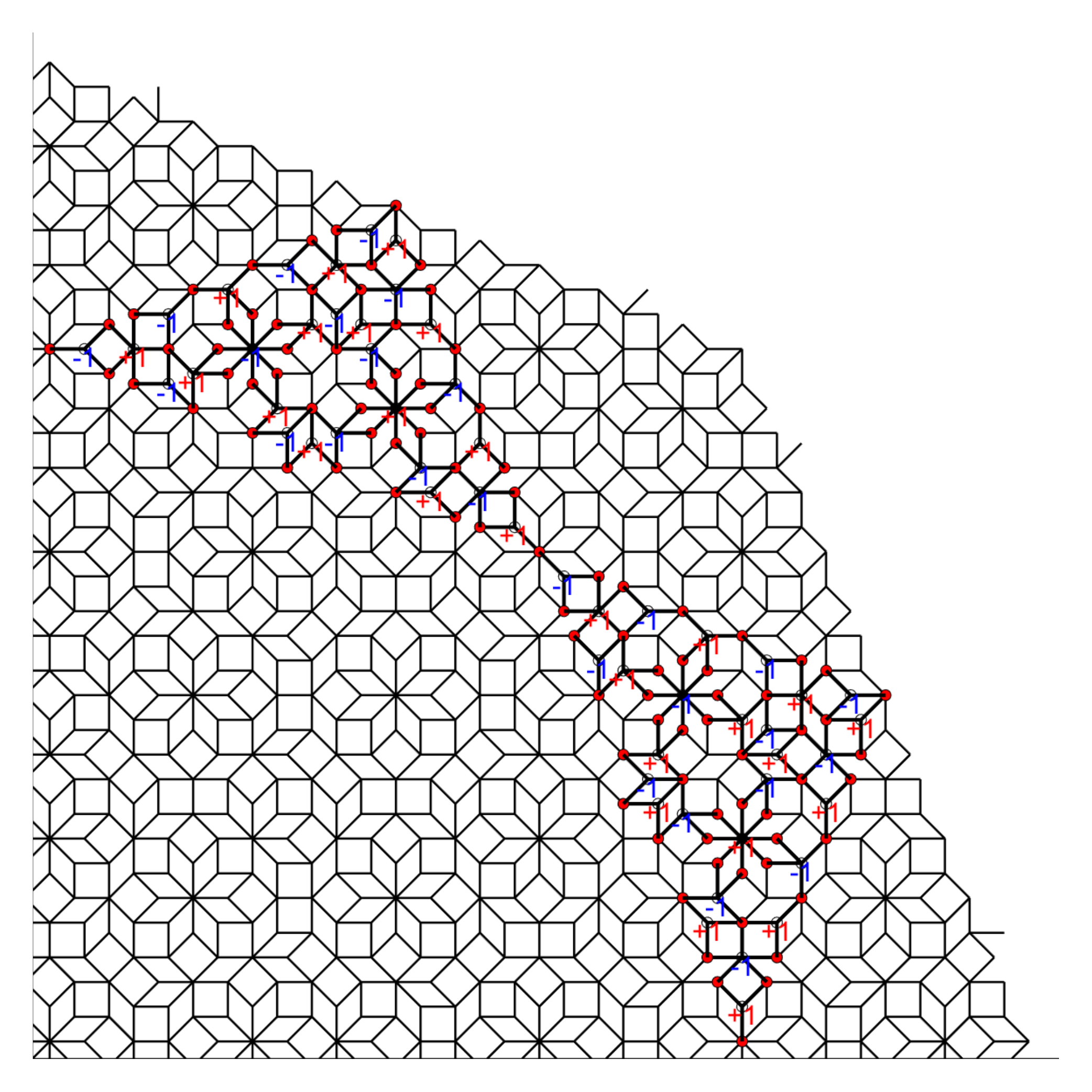}
    \includegraphics[trim=8mm 8mm 8mm 8mm,clip,width=0.45\textwidth]{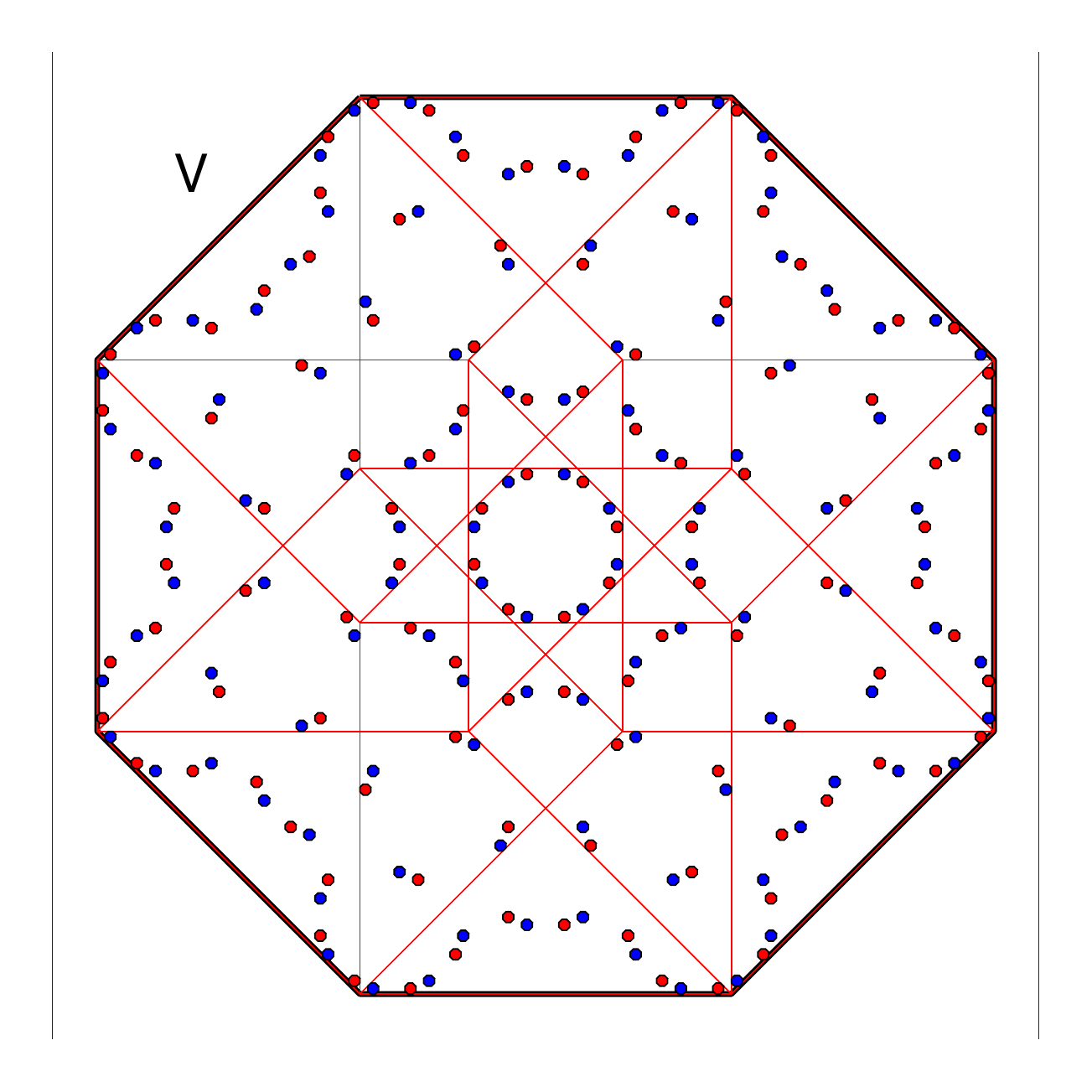}
    \caption{Type-Q LS wavefunction and perpendicular space allowed regions.Type-Q and Type-R cover new 8-edge regions and are both independent as they are orthogonal.}
    \label{fig:TQ_RealSpace}
\end{figure*}

\begin{figure*}[!h]
    \centering
    \includegraphics[trim=8mm 8mm 8mm 8mm,clip,width=0.45\textwidth]{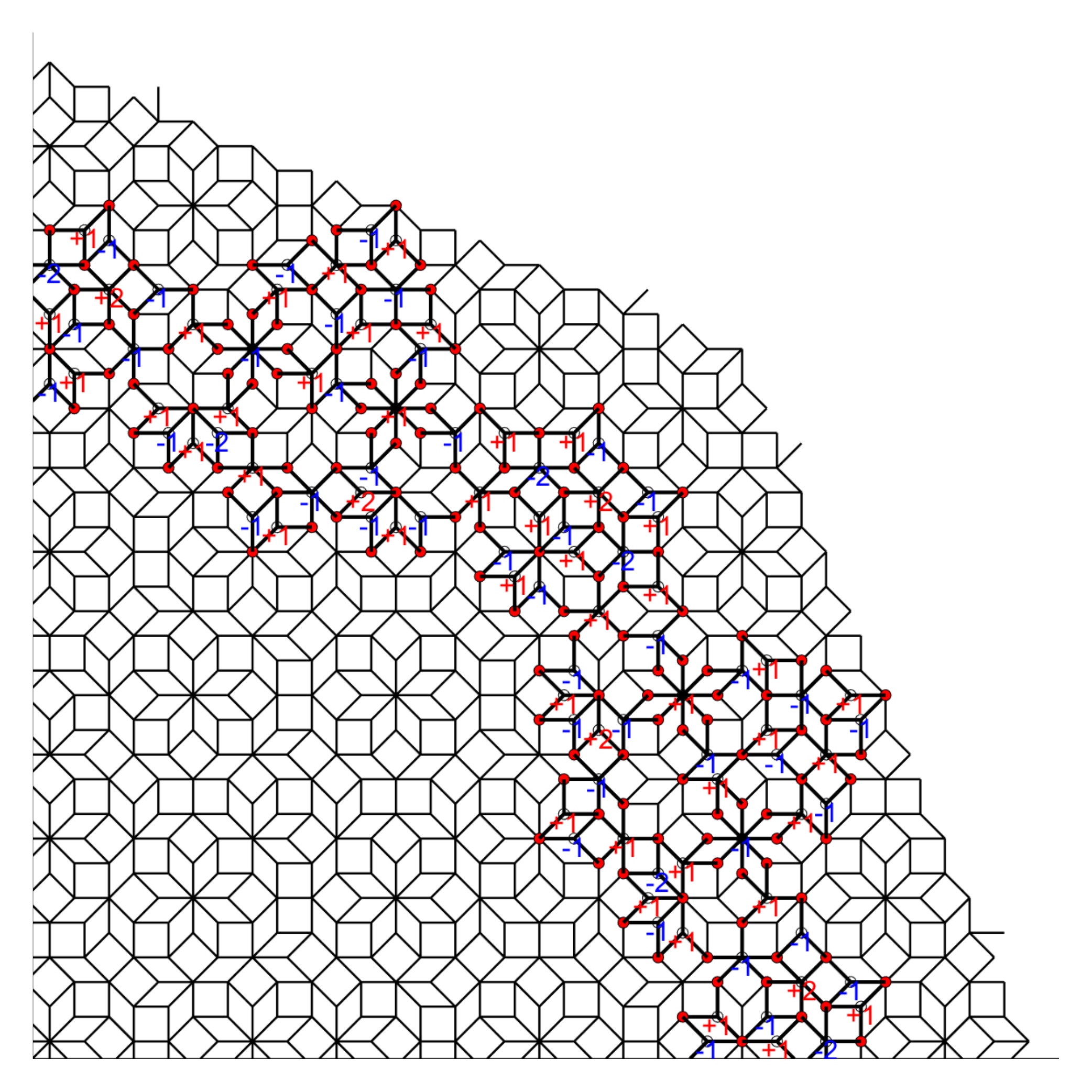}
    \includegraphics[trim=8mm 8mm 8mm 8mm,clip,width=0.45\textwidth]{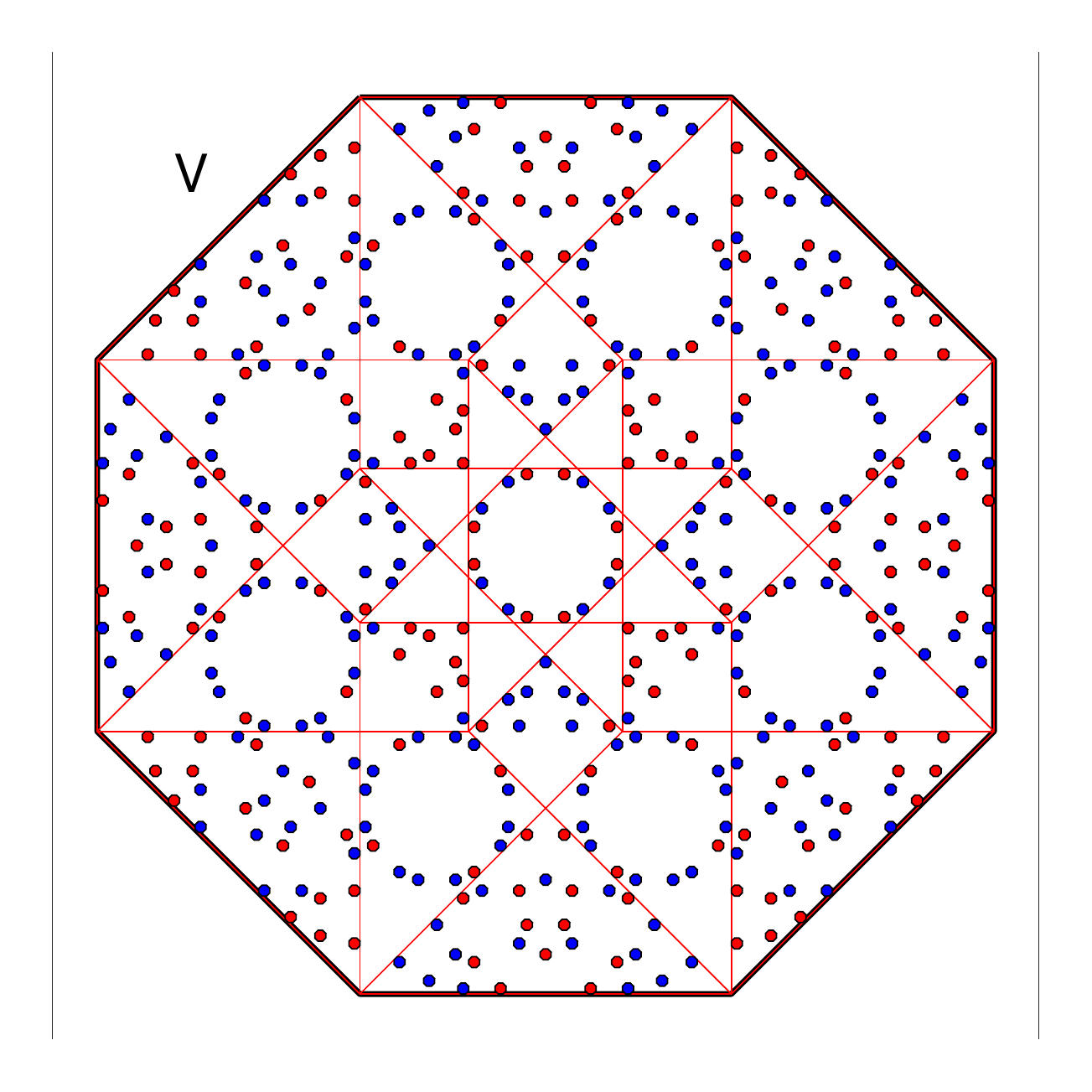}
    \caption{Type-R LS wavefunction and perpendicular space allowed regions.Type-Q and Type-R cover new 8-edge regions and are both independent as they are orthogonal.}
    \label{fig:TR_RealSpace}
\end{figure*}\begin{figure*}[!h]
    \centering
    \includegraphics[trim=8mm 8mm 8mm 8mm,clip,width=0.45\textwidth]{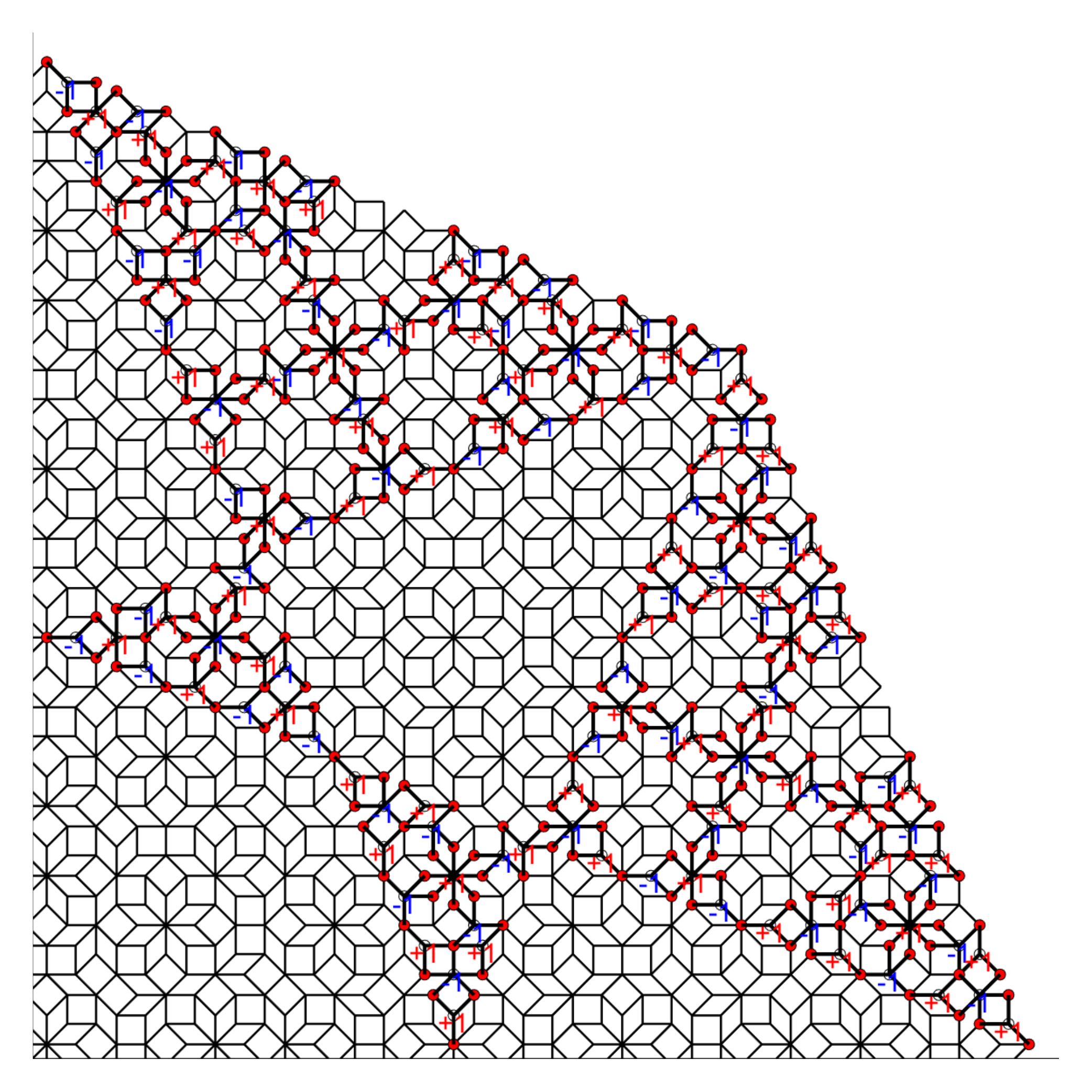}
    \includegraphics[trim=8mm 8mm 8mm 8mm,clip,width=0.45\textwidth]{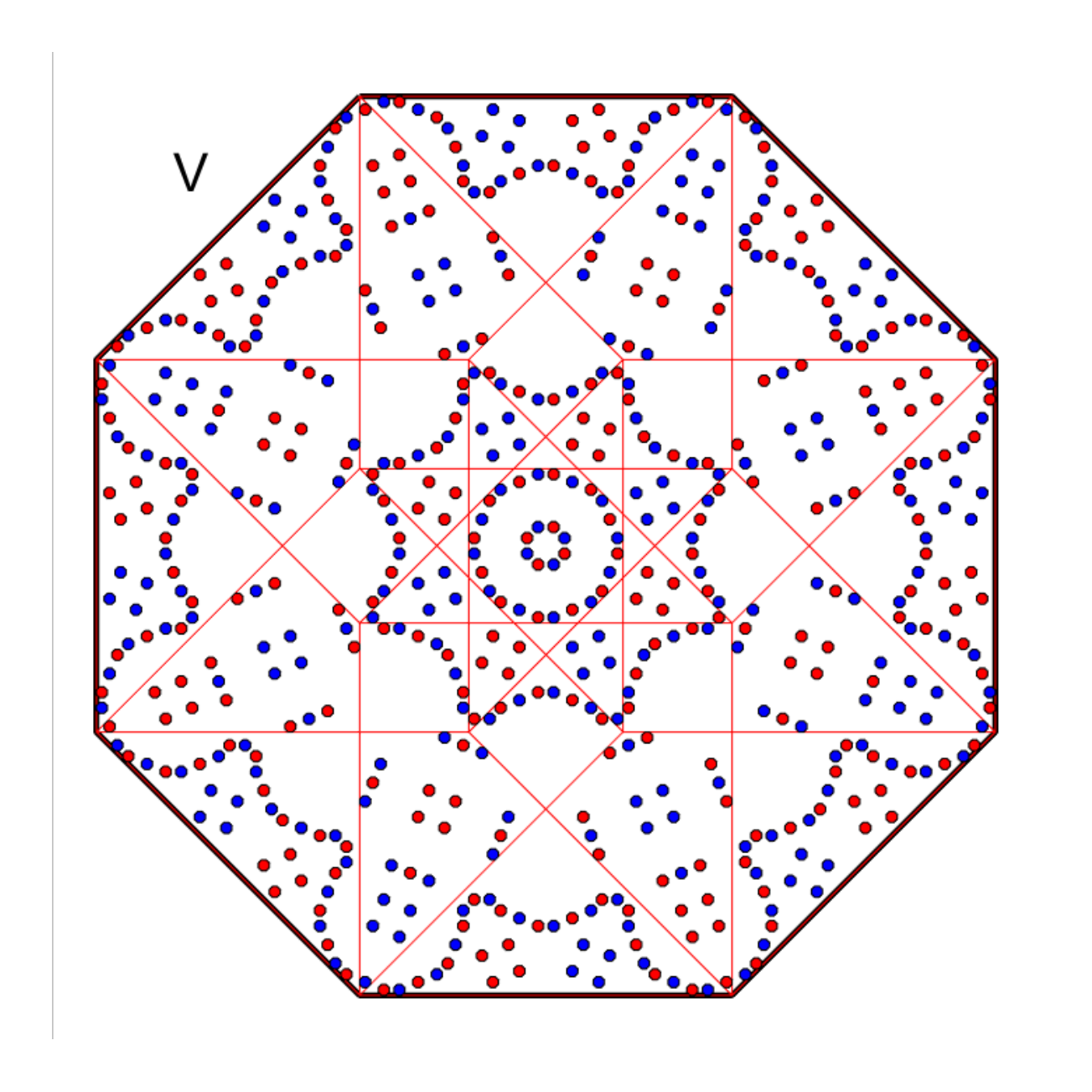}
    \caption{Type-S LS wavefunction and perpendicular space allowed regions.Type-S and Type-T cover new 4-edge regions (regions furthest away from the center of $V$) and are both independent as they are orthogonal.}
    \label{fig:TS_RealSpace}
\end{figure*}

\begin{figure*}[!h]
    \centering
    \includegraphics[trim=8mm 8mm 8mm 8mm,clip,width=0.45\textwidth]{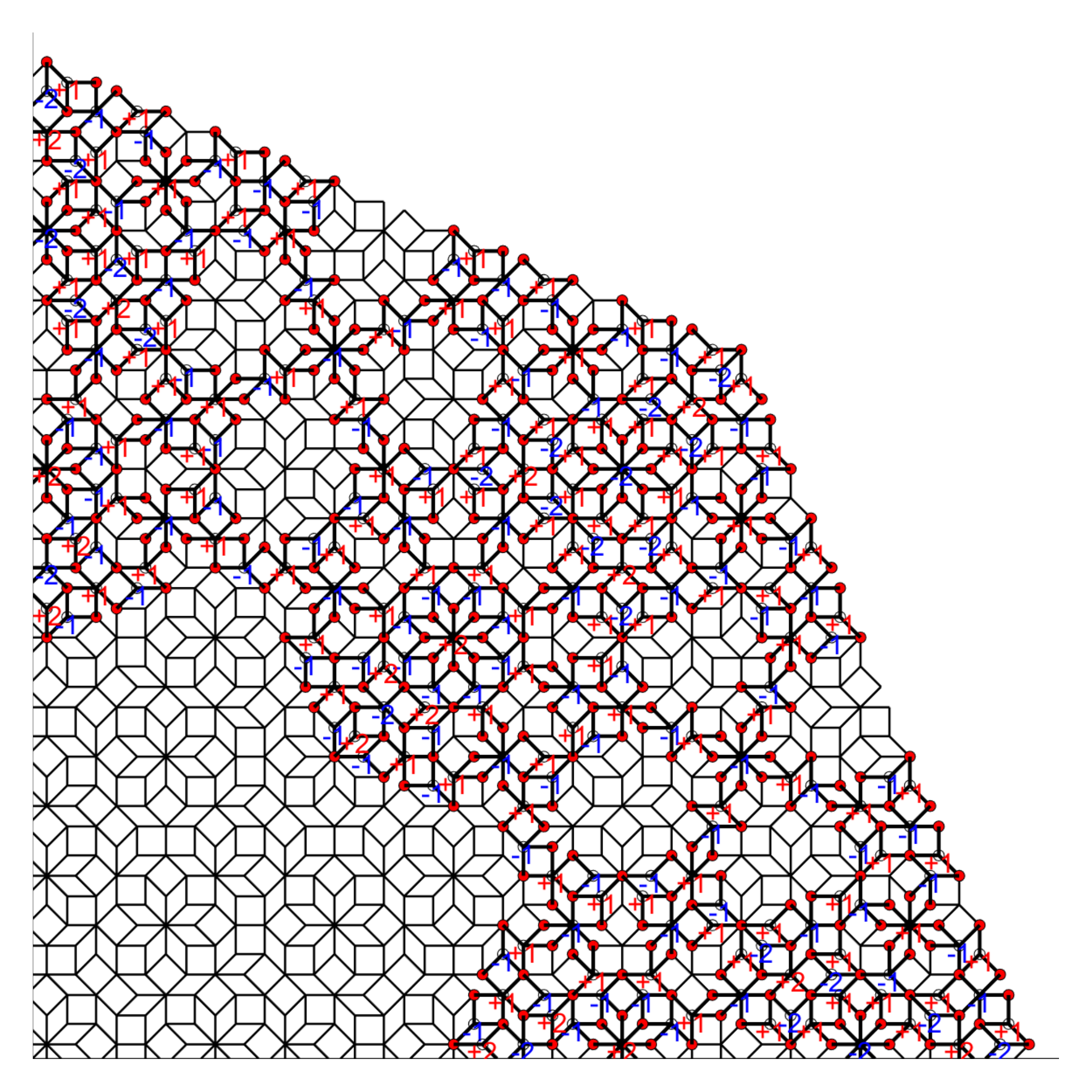}
    \includegraphics[trim=8mm 8mm 8mm 8mm,clip,width=0.45\textwidth]{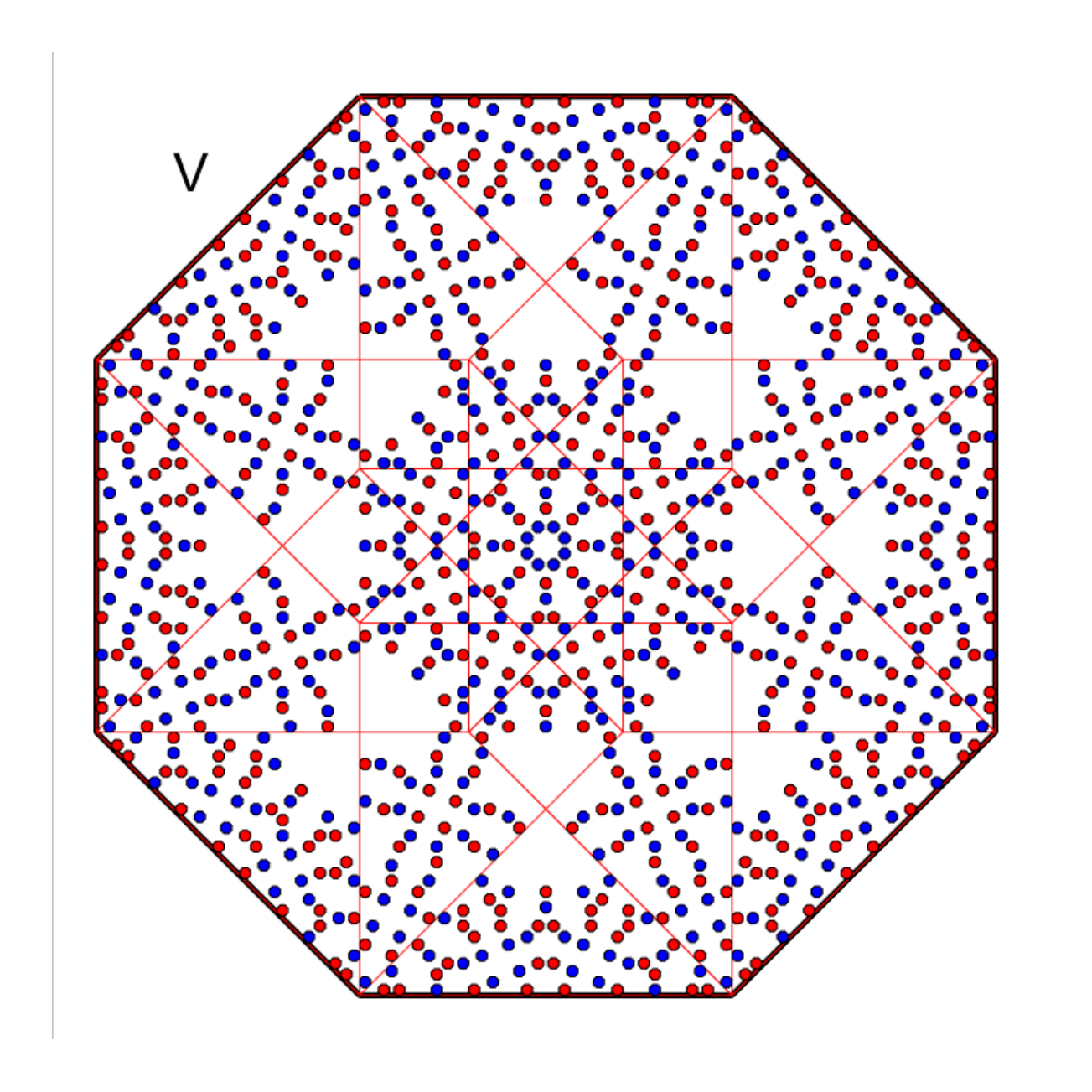}
    \caption{Type-T LS wavefunction and perpendicular space allowed regions. Type-S and Type-T cover new 4-edge regions (regions furthest away from the center of $V$) and are both independent as they are orthogonal.}
    \label{fig:TT_RealSpace}
\end{figure*}

\begin{acknowledgements}
We would like to thank F. Pi\'{e}chon for valuable correspondence. 

\end{acknowledgements}

%

\end{document}